\documentclass[final,1p,sort&compress]{elsarticle}
\usepackage{mathrsfs}
\usepackage{amsmath}
\usepackage{amssymb}
\usepackage{multirow}
\usepackage{array}
\usepackage{tikz}
\usepackage{color}
\usepackage{graphicx}
\usepackage{epstopdf}
\usepackage{leftidx}
\usepackage{mleftright} 

\newcommand*{\colorboxed}{}
\def\colorboxed#1#{%
	\colorboxedAux{#1}%
}
\newcommand*{\colorboxedAux}[3]{%
	\begingroup
	\colorlet{cb@saved}{.}%
	\color#1{#2}%
	\boxed{%
		\color{cb@saved}%
		#3%
	}%
	\endgroup
}

\definecolor{princetonorange}{rgb}{1.0, 0.56, 0.0}

\usepackage{empheq}

\newcommand\widecolourbox[1]{{\setlength\fboxrule{1pt}\setlength\fboxsep{4pt}\colorboxed{princetonorange}{\enspace#1\enspace}}}

\journal{Physics Reports}

\usepackage{units}

\usepackage{enumerate}

\usepackage{url}
\usepackage[colorlinks]{hyperref}
\hypersetup{%
	plainpages=true,
	breaklinks=true,
	hypertexnames=false,
	pageanchor=true,
	colorlinks=true,
	linkcolor={blue},
	citecolor={red},
	urlcolor={blue},
	anchorcolor={black}
}

\newcommand{\ket}[1]{\left|{#1}\right\rangle}
\newcommand{\braket}[2]{\left\langle{#1}|{#2}\right\rangle}
\newcommand{\ketbra}[2]{\left|{#1}\rangle \langle{#2}\right|}
\newcommand{\brakket}[3]{\langle{#1}|{#2}|{#3}\rangle}
\newcommand{\expec}[1]{\left\langle{#1}\right\rangle}

\newcommand{\sz}{\sigma_z}
\newcommand{\sx}{\sigma_x}

\newcommand{\sm}{\sigma_-}
\renewcommand{\sp}{\sigma_+}

\newcommand{\abs}[1]{\left|#1\right|}

\newcommand{\nn}{\nonumber}

\newcommand{\figref}[1]{\mbox{Fig.~\ref{#1}}}

\newcommand{\secref}[1]{\mbox{Sec.~\ref{#1}}}

\newcommand{\appref}[1]{\mbox{Appendix~\ref{#1}}}
\renewcommand{\eqref}[1]{\mbox{Eq.~(\ref{#1})}}

\newcommand{\figpanel}[2]{Fig.~\hyperref[#1]{\ref*{#1}(#2)}}
\newcommand{\figpanels}[3]{Fig.~\hyperref[#1]{\ref*{#1}(#2)-(#3)}}
\newcommand{\figpanelNoPrefix}[2]{\hyperref[#1]{\ref*{#1}(#2)}}

\newcommand{\be}{\begin{equation}}
\newcommand{\ee}{\end{equation}}
\newcommand{\bea}{\begin{eqnarray}}
\newcommand{\eea}{\end{eqnarray}}
\def\87rb{$^{87}$Rb}

\begin{document}
	
\begin{frontmatter}
		
\title{Quantum amplification and simulation
			of strong and ultrastrong coupling of light and matter}
		
\author[FirstAddress,SecondAddress,8thAddress]{Wei Qin\corref{cor1}}
\cortext[cor1]{Corresponding authors}
\ead{qin.wei@tju.edu.cn}
		
\author[ThirdAddress]{Anton Frisk Kockum}
		
		
\author[FourthAddress]{Carlos S\'anchez Mu\~noz}
		
\author[SecondAddress,FifthAddress]{Adam Miranowicz}
		
\author[SecondAddress,SixthAddress,SeventhAddress]{Franco Nori\corref{cor1}}
\ead{fnori@riken.jp}

\address[FirstAddress]{Center for Joint Quantum Studies and Department of Physics, School of Science, Tianjin University, Tianjin 300350, China}
\address[SecondAddress]{Theoretical Quantum Physics Laboratory, RIKEN, Saitama 351-0198, Japan}
\address[ThirdAddress]{Department of Microtechnology and Nanoscience, Chalmers University of Technology, 412 96 Gothenburg, Sweden}
\address[FourthAddress]{Departamento de Física Teórica de la Materia Condensada and Condensed
Matter Physics Center (IFIMAC), Universidad Autónoma de Madrid, 28049 Madrid,
Spain}
\address[FifthAddress]{Institute of Spintronics and Quantum Information, Faculty of Physics, Adam Mickiewicz University, 61-614 Pozna\'n, Poland}
\address[SixthAddress]{Center for Quantum Computing, RIKEN, Wako-shi, Saitama 351-0198, Japan}
\address[SeventhAddress]{Physics Department, The University of Michigan, Ann Arbor, Michigan 48109-1040, USA}
\address[8thAddress]{Tianjin Key Laboratory of Low Dimensional Materials Physics and Preparing Technology, Tianjin University, Tianjin 300350, China}
		
\date{\today}
		
\begin{abstract}
			
The interaction of light and matter at the single-photon level is of central importance in various fields of physics, including, e.g., condensed matter physics, astronomy, quantum optics, and quantum information. Amplification of such quantum light-matter interaction can be highly beneficial to, e.g., improve device performance, explore novel phenomena, and understand fundamental physics, and has therefore been a long-standing goal. Furthermore, simulation of light-matter interaction in the regime of ultrastrong coupling, where the interaction strength is comparable to the bare frequencies of the uncoupled systems, has also become a hot research topic, and considerable progress has been made both theoretically and experimentally in the past decade. In this review, we provide a detailed introduction of recent advances in amplification of quantum light-matter interaction and simulation of ultrastrong light-matter interaction, particularly in cavity and circuit quantum electrodynamics and in cavity optomechanics.
			
\end{abstract}
		
		
\end{frontmatter}


\newpage
\tableofcontents
\newpage

\begin{table}[!ht]
\begin{center}
	\caption{Main abbreviations and basic notations used in this review.} \label{table}
	\begin{tabular}{*{1}{p{2.5cm}<{\centering}}*{1}{p{8cm}<{\centering}}}
		\hline \hline
		Notation & Meaning \\
		\hline
		QED & quantum electrodynamics \\
		QIP & quantum information processing \\
		RWA & rotating-wave approximation \\
		JC & Jaynes--Cummings \\
		USC & ultrastrong coupling \\
		DSC & deep strong coupling \\
		BEC & Bose-Einstein condensate \\
		VQS & variational quantum simulation \\
		VQA & variational quantum algorithm \\
        SQUID & superconducting quantum interference device \\
		$D(\alpha)$ & displacement operator with a complex amplitude $\alpha$ \\
		$S\left(\xi\right)$ & squeezing operator with a complex parameter $\xi$ \\
		$\sigma_{x}$, $\sigma_{y}$, $\sigma_{z}$, $\sigma_{\pm}$ & Pauli operators \\
		$g$ & atom-cavity coupling strength \\
		$J$ & inter-cavity coupling strength \\
		$\kappa$ & photon loss rate \\
		$\gamma$ & atomic spontaneous emission rate \\
		$C$ & $C=g^{2}/\kappa\gamma$ single-photon cooperativity \\
		$r$ & squeezing degree \\
		$\omega_{\rm cav}$ & cavity frequency \\
		$\omega_{q}$ & atomic transition frequency \\
		\hline
	\end{tabular}
 \end{center}
\end{table}

\section{Introduction}


\subsection{The importance of light-matter interactions}

The quantum interaction of light and matter is a fundamental area of research in physics that spans across its various fields, encompassing: quantum optics~(see, e.g., Refs.~\cite{scully1997book, AgarwalBook, PerinaBook}), photonics~\cite{AndrewsBook}, cavity quantum electrodynamics (QED)~\cite{haroche2006book}, circuit QED~\cite{you2005superconducting,schoelkopf2008wiring,clarke2008superconducting,you2011atomic,gu2017microwave,Krantz2019,kockum2019quantum,kjaergaard2020superconducting,circuit2021blais}, atom optics~\cite{MeystreBook,GardinerBook,BudkerBook}, quantum sensing~\cite{Degen2017} and quantum metrology~\cite{NawrockiBook}, as well as quantum optical technologies~\cite{ByrnesBook}, including quantum cryptography, quantum communications, and optical quantum information processing (QIP)~\cite{KokBook}. Moreover, light-matter interactions have been actively investigated in condensed matter physics, both at the fundamental level concerning, e.g., cavity quantum materials~\cite{Schlawin2022}, non-equilibrium condensed matter physics with light~\cite{LeHur2016}, light-induced effects in three- and lower-dimensional materials or topological materials~\cite{Bao2022}; but also at the applied level to understand the principles behind devices like lasers, light-emitting diodes, photodetectors, and solar cells. Moreover, we should also mention the importance of light-matter interactions in astronomy (ranging from analyzing emission and absorption spectra to understanding, e.g, black holes, neutron stars, as well as other stellar and interstellar structures and their evolution), quantum chemistry (especially photochemistry)~\cite{levine2009quantum}, and quantum biology (ranging from understanding and simulating photosynthesis to controlling the activity of neurons or other cell types with light using the methods of optogenetics)~\cite{lambert2013quantum}.


\subsection{The importance of amplifying light-matter interactions}

The strength of the quantum interaction between light and matter has a significant impact on various aspects of science and technology. A stronger quantum interaction can lead to several advantages and opportunities for exploring new phenomena, improving technologies, and gaining deeper insights into the fundamental nature of the universe.

Here are some reasons why a stronger quantum interaction between light and matter is beneficial:

\begin{enumerate}

\item Greater sensitivity in quantum measurements: 
In fields like quantum sensing and quantum metrology, a stronger quantum interaction allows for more sensitive measurements. This is especially important in applications such as gravitational wave detection, where precise measurements of tiny perturbations are crucial.

\item  Enhanced control in QIP: 
Stronger interactions can lead to better control over quantum systems, enabling more efficient and reliable QIP, such as quantum computing, quantum communication, and quantum cryptography.

\item  Emergence of novel phenomena: 
In many cases, stronger interactions can lead to the emergence of new and unexpected phenomena, providing opportunities for discovery and innovation.

\item  Exploring exotic quantum states: 
Stronger interactions can facilitate the creation and study of exotic quantum states that can be harnessed for various applications.

\item  Better understanding of fundamental physics: 
Stronger interactions enable exploring the boundary between classical and quantum behavior, providing insights into the fundamental principles of quantum mechanics and potentially revealing new physics beyond our current understanding.

\item  Advanced materials and nanotechnology: 
Stronger interactions between light and matter can lead to the development of novel materials and technologies. This includes the creation of metamaterials with unprecedented optical properties and the design of more efficient photonic devices.

\item  Improved imaging and spectroscopy: 
Stronger interactions improve the resolution and sensitivity of imaging techniques and spectroscopic measurements. This is valuable for various fields, including medical imaging, materials characterization, and remote sensing.

\item  Exploring quantum phase transitions: 
In condensed matter physics, stronger interactions can enable studying quantum phase transitions, where a material's properties change dramatically due to quantum effects, especially at low temperatures.

\item  Testing fundamental principles: 
Stronger interactions can lead to more accurate tests of fundamental principles, such as the equivalence principle or Lorentz invariance, potentially revealing deviations from these principles that could point to new physics.

\item Faster dynamics: 
Increasing the interaction between light and matter enhances the speed of energy exchange and dynamic processes in a system. Stronger interaction allows for faster energy transfer, coherent dynamics, and efficient information exchange between the two components.

\item Increasing system nonlinearities: 
By increasing the interaction strength between light and matter, the energy levels of the matter can be significantly perturbed, leading to enhanced nonlinear responses in the system.

\item  Quantum simulation: 
Strong interactions can be used to simulate complex quantum systems that are difficult to study directly. This has applications in understanding condensed matter physics, simulating chemical reactions, and exploring fundamental physical phenomena.

\end{enumerate}



\subsection{Prototype models for studying light-matter interactions}

The most popular description of the interaction between a two-level quantum system (such as a real or artificial atom or a qubit) and a single-mode quantized electromagnetic field (a cavity mode) without the rotating-wave approximation (RWA) --- a simplification that disregards non-resonant components in light-matter interaction Hamiltonians --- is the quantum Rabi model~\cite{Rabi1937}. This model simplifies to the standard Jaynes--Cummings (JC) model~\cite{jaynes1963comparison, Shore1993} under the RWA, i.e., when the counter-rotating terms in the Rabi interaction Hamiltonian are negligible. The multi-atom generalizations of the quantum Rabi and JC models are known as the Dicke~\cite{Dicke1954} and Tavis--Cummings~\cite{tavis68a} models, respectively. Several interaction-amplification methods exist with the goal of simulating the Rabi (or Dicke) model by using the JC (or Tavis--Cummings) models under the RWA together with classical or quantum drives, as described in more detail below.

The quantum Rabi model serves as a prototype closely linked to various fundamental models and emerging phenomena. These encompass the Hopfield~\cite{Hopfield1958} and Jahn--Teller~\cite{Jahn1937, Hines2004, Larson2008, Meaney2010, Dereli2012} representations, as well as renormalization-group models, including the spin-boson~\cite{Leggett1987, LeHur2012, Leppakangas2018} and Kondo~\cite{Kondo1964, LeHur2012, Snyman2015} descriptions. Thus, simulating the quantum Rabi model enables simulating these or many other models.

The quantum Rabi model and its generalizations have lead to a discovery of a diverse range of novel physical effects (like the creation of photons from the quantum vacuum~\cite{Ciuti2005,johansson2009dynamical,johansson2010dynamical,wilson2011observation,nation2012colloquium,johansson2013nonclassical}), but they have also brought about notable theoretical complexities. Among these challenges, a prominent one is the breakdown of the RWA. As a result, various aspects of the standard quantum-optical theoretical framework require reconsideration to ensure the precise incorporation of all interaction terms inherent to this regime~\cite{Kockum2019, Forn-Diaz2019, LeBoite2020,de2024nonperturbative}. Note that this breakdown is not unique to quantized fields, and has also led to many interesting phenomena with classical fields, including, e.g., Bloch-Siegert shift~\cite{bloch40a,tuorila2010stark,edbert2017large}, coherent destruction of tunneling~\cite{grossmann1991coherent,wubs2010instantaneous,ivakhnenko2023nonadiabatic}, and driving-driven tunneling oscillations~\cite{Grifoni1998driven,hartmann2000driven,nakamura2001rabi,hausinger2010dissipative}. In order to correctly describe a quantum Rabi-like system when the counter-rotating terms cannot be neglected, the standard formalisms should be generalized to avoid violating various no-go theorems. For example, as discussed in, e.g., Refs.~\cite{salmon2022gauge,Mercurio2022} and references therein:

\begin{enumerate}

\item The conventional master equation used in quantum optics does not accurately capture the way a quantum Rabi-like system interacts with its surrounding environment~\cite{DeLiberato2009, Settineri2018, Macri2022,de2023Relaxation}. 

\item The expected photon output rate is no longer directly linked to the number of photons within the cavity. This means that usual normal-order correlation functions do not correctly describe the photon emission rate in the quantum Rabi and Dicke models. Thus, it is \emph{not} possible to observe, e.g., a continuous flow of photons from the ground state of the Rabi model, which would apparently imply a perpetual-motion behavior~\cite{Stassi2013, Ciuti2006, DiStefano2018}. A direct application of the standard formalism could lead to such unphysical results. 

\item Additionally, these observations seem to contradict the principle of gauge invariance. Thus, one should be very careful in calculating observables to avoid gauge ambiguities~\cite{de2018cavity,De2018breakdown,DiStefano2019,gustin2023gauge,Mercurio2023,hughes2023reconciling}.

\end{enumerate}


\subsection{Ultrastrong and deep strong coupling regimes}

The coupling between light and matter, particularly in the context of quantum systems, is often categorized into four regimes defined by the strength (which can be weak, strong, ultrastrong, and deep strong) of the interaction between the two subsystems~\cite{Kockum2019, Forn-Diaz2019, LeBoite2020}. These regimes have important implications for the behavior and properties of the coupled light-matter systems.

The determination of whether the coupling is strong or weak hinges on the comparison between the value of a coupling strength $g$ and the losses (characterized by some damping rates, say $\kappa_i$) in the system. Thus, the weak-coupling and strong-coupling regimes are $g\le\kappa_i$ and $g>\kappa_i$, respectively. However, ultrastrong coupling (USC) and deep strong coupling (DSC) should not be misunderstood as just stronger coupling, as their characterization does not involve the losses $\kappa_i$, but rather juxtaposes the value of $g$ against the frequency of light (say $\omega_c$ of a cavity mode) and the transition frequency of the matter (say $\omega_a$ of an atom), the uncoupled constituents of the system. In these regimes, the coupling strength becomes comparable to or even larger than the natural frequencies, which implies that the counter-rotating terms in the quantum Rabi Hamiltonian cannot be neglected. Specifically, the USC and DSC regimes are defined by the conditions $\eta\equiv g/\omega_{c,a} > 0.1$ and $\eta > 1$~\cite{Casanova2010}, respectively. Note that this \unit[10]{\%} threshold value for USC is a matter of convention.

The USC regime can lead to dramatic changes in the energy levels and dynamics of the coupled system. New phenomena, such as avoided level crossings and nonperturbative effects, can arise, causing significant deviations from the behavior observed for the weak coupling and the strong coupling. This regime is of particular interest for exploring fundamental quantum effects and potentially enabling new quantum technologies.

In the USC and DSC regimes, the interaction between light and matter is so intense that it can even affect the vacuum state of the electromagnetic field. More specifically: (i) In the strong coupling regime, the ground state of a coupled light-matter system, e.g., a cavity mode and a two-level atom described in the JC model, corresponds to the uncoupled system with the cavity in the vacuum state and the atom in the ground state. However, (ii) in the USC regime, the ground state of the quantum Rabi model is a coherent superposition of all states with an even total number of virtual excitations in the cavity mode and the atom, with the superposition amplitudes decreasing with the increasing number of virtual photonic excitations. Moreover, (iii) the ground state of the quantum Rabi model in the DSC regime corresponds to virtual photonic even and odd Schr\"odinger's cat states entangled with the atomic cat states. These counterintuitive results lead to the emergence of entirely new energy levels and states, which fundamentally alter the system behavior.

Numerous novel effects inherent in the USC and DSC regimes have been theoretically predicted and their various applications have been proposed, including those summarized in Refs.~\cite{Kockum2019, Forn-Diaz2019}. In addition to quantum nonlinear optics, quantum optomechanics, and atom optics, which are described in a greater detail in this review, those proposals encompass also various other fields including: QIP~\cite{Nataf2011, Romero2012, Kyaw2015, Wang2016, Wang2017, Xu2017, Stassi2017, Stassi2018, Stassi2020}, quantum metrology~\cite{Ruggenthaler2018, Garbe2020, Gietka2022, Salado2021}, quantum plasmonics~\cite{tame2013quantum, Benz2016, Todisco2018, Munkhbat2018,SaezBlazquez2023can}, quantum field theory~\cite{Sabin2010, DiStefano2017, DiStefano2019, Savasta2021}, polariton-enhanced superconductivity~\cite{Sentef2018, Schlawin2019}, metamaterials~\cite{Scalari2012, Scalari2013, Bayer2017}, quantum thermodynamics~\cite{Seah2018}, and quantum chemistry (especially chemistry QED)~\cite{Galego2015, Herrera2016, Ebbesen2016, Bennett2016, Kowalewski2016,flick2017atoms,Martinez2018,sedov2020chiral,garcia2021manipulating,genet2021inducing}.

Concerning applications of USC for quantum sensing and quantum metrology, we mention novel high-resolution spectroscopy~\cite{Ruggenthaler2018}, which takes advantage of reduced linewidths and enhanced signal-to-noise ratios achievable in USC setups. Criticality-enhanced metrology in the USC regime has been predicted for the quantum Rabi~\cite{Garbe2020}, Dicke~\cite{Gietka2022}, and Hopfield~\cite{Salado2021} models. In particular, Ref.~\cite{Salado2021} predicted an improved precision of a thermometric quantum sensor operating at a quantum phase transition. Recent experimental observation of a superradiant phase transition with emergent cat states in a controlled quantum Rabi model~\cite{Zheng2023} shows a feasible way of realizing criticality-enhanced metrology in the USC regime. Mechanical states, which enable quantum-enhanced metrology, can be deterministically generated in USC-regime optomechanics, as shown in Ref.~\cite{Macri2016}. The experimental approach developed in Ref.~\cite{Zheng2023} can also lead to realizing noise-biased cat qubits for fault-tolerant quantum computation in the USC regime.

It is worth noting that QIP often relies on the coherent exchange or transfer of excitations between light and matter, and this pivotal aspect finds significant relevance in both the strong coupling and USC regimes. However, USC proves notably more efficient at such transfer processes by using virtual photons. The realm of QIP benefits extensively from the capabilities of USC systems. Proposals of applications encompass: QIP protocols with dramatically improved coherence times and quantum-operation fidelity~\cite{Nataf2011}, ultrafast quantum gate operations~\cite{Romero2012, Wang2017, Xu2017}, long-lasting quantum memories~\cite{Kyaw2015, Stassi2018}, holonomic QIP protocols~\cite{Wang2016}, quantum error-correction codes~\cite{Stassi2017}, and scalable quantum processors~\cite{Stassi2020}. Noteworthy advantages span beyond mere reduction in operation times and improved coherence times and gate fidelity: for example, USC also empowers simpler protocols, where the inherent evolution of an USC system supersedes the need for intricate sequences of quantum gates (see, e.g., Ref.~\cite{Stassi2017}). Several of these proposed applications utilize entangled ground states and the underlying parity symmetry.

As mentioned above, it is possible to observe entirely new phenomena in the USC or DSC regimes, e.g., the entangled hybrid light-matter ground state (corresponding to a Schr\"odinger cat of other Schr\"odinger cat states) of the quantum Rabi model in the DSC regime, which can be considered a new stable state of matter observed experimentally in Ref.~\cite{Yoshihara2017}. The most recent experiment of that research group demonstrated another interesting effect in the DSC regime --- an extremely large Lamb shift in a multimode QED system~\cite{Ao2023}.

Various experimental demonstrations of the USC regime (for reviews, see Refs.~\cite{Kockum2019, Forn-Diaz2019}) have been reported in different systems including: intersubband polaritons~\cite{Anappara2009, Gunter2009, Todorov2010, Jouy2011,zaks2011thz, Geiser2012, Delteil2012, Askenazi2014, Askenazi2017}, superconducting quantum circuits~\cite{Niemczyk2010, Forn-Diaz2010, Baust2016, Forn-Diaz2016, Bosman2017, Chen2017, Yoshihara2017, Yoshihara2017a, Yoshihara2018, Vrajitoarea2022, Ao2023}, Landau polaritons~\cite{Muravev2011, Scalari2012, Scalari2013, Maissen2014, Bayer2017, Paravicini2019, Keller2020, Mornhinweg2023}, organic molecules~\cite{Schwartz2011, Kena2013, Gambino2014, Gubbin2014, Mazzeo2014, Barachati2018, Todisco2018, Genco2018, Eizner2018, Wang2022}, optomechanical systems~\cite{George2016, Benz2016, Manjeshwar2023, Dare2023}, magnons~\cite{Flower2019, Bourcin2023, Zhang2023}, quantum dots~\cite{Valmorra2021, Scarlino2022}, and hybrid superconducting-optomechanical systems~\cite{Pirkkalainen2015}. In several experiments, even the DSC regime has been reached~\cite{Yoshihara2017, Yoshihara2018, Bayer2017, Mueller2020, Ao2023}. To date, the highest normalized coupling constant (close to 2) was experimentally achieved in Ref.~\cite{Mueller2020} for plasmon polaritons in three-dimensional nanoparticle metallic crystals.

Despite this impressive experimental progress, it is important to note that probing and controlling the dynamics of such USC systems over a wide range of parameters remains difficult. In general, while stronger quantum interactions offer numerous advantages, they also bring challenges, such as increased complexity and technical difficulties in controlling and manipulating quantum systems. Striking a balance between harnessing the benefits and overcoming the challenges is a key aspect of research and technological development in this field.

The USC regime presents the potential for inducing and observing various classes of higher-order processes and nonlinear optical phenomena involving only two-level systems and virtual photons~\cite{Stassi2017,Kockum2017a,Kockum2017,cirio2019multielectron,Wang2024}, multiphoton quantum Rabi oscillations~\cite{Garziano2015}, nonclassical state preparation~\cite{ashhab2010qubit}, parity symmetry breaking and Higgs-like mechanism~\cite{Garziano2014}, bunched-light emission from individual qubits~\cite{Garziano2017}, conversion of an atomic superposition state into an entangled photonic state~\cite{Macri2018}, as well as simultaneous excitation by a single photon of two or more qubits in a single-mode resonator~\cite{Garziano2016, Zhao2017, Wang2017, macri2020spin, Tomonaga2023} or in different resonators in an array of two or three weakly coupled resonators~\cite{Garziano2020}. Unfortunately, most of these and other interesting effects predicted for the quantum Rabi model in the USC regime have not been experimentally demonstrated yet, because of the technological challenges mentioned above. We believe that it will be much easier to induce and observe them by quantum simulations.


\subsection{The importance of quantum simulations}

To overcome significant experimental problems of reaching and coherently controlling the USC and DSC regimes, various quantum simulation methods have been developed~\cite{Lloyd1996, Buluta2009, Georgescu2014,Jaako2020quantum, Altman2021}. These methods use an easy-to-control quantum system to simulate the properties of a more complex quantum model of interest. More specifically, quantum simulations refer to using controllable quantum systems, such as quantum computers or specialized quantum simulators, to model and understand the behavior of complex quantum systems that are difficult to study using classical computers or analytical methods. These simulations aim to simulate and mimic the behavior of quantum systems in order to gain insights into their properties, dynamics, and interactions.

The need for quantum simulations arises from various reasons. To mention only some of them:

\begin{enumerate}

\item
Complexity of quantum systems, which are often intractable by classical simulations: 
As the number of quantum particles or quantum excitations increases, simulating their interactions using classical computers becomes exponentially difficult. Quantum simulations have the potential to outperform such classical computations by utilizing quantum parallelism, which allows quantum systems to explore multiple possible states simultaneously.

\item
New insights and discoveries: 
Quantum simulations can provide insights into quantum phenomena that are otherwise difficult to observe or understand. They enable the exploration of novel materials, the study of exotic quantum states, and the investigation of fundamental physical principles that govern quantum systems.

\item
Understanding quantum dynamics: 
Quantum simulations enable to study the time evolution of quantum systems, shedding light on processes like chemical reactions, energy transfer, and quantum phase transitions.

\item
Verification of quantum algorithms: 
Quantum computers are still in their early stages of development, and verifying the correctness of quantum algorithms is crucial. Quantum simulations can be used to test and verify these algorithms on small scales before they are scaled up to larger and more complex problems.

\item
Quantum optimization: 
Quantum simulations can be used to tackle quantum and classical optimization problems that arise in various fields including: material science and engineering, energy and resource optimization, cryptanalysis and security, optimization in telecommunications, climate modeling and environmental management, or aerospace and aviation, as well as those fields which are not necessarily directly related to physics, like machine learning and artificial intelligence, drug discovery and development, traffic and transportation optimization, logistics and supply chain management, or even finance and portfolio optimization, among many others. Quantum annealing and other quantum optimization techniques hold the promise of solving these problems more efficiently than classical methods.

\end{enumerate}

In summary, quantum simulations are essential tools for understanding and harnessing the behavior of quantum systems, providing means to explore complex phenomena, discover new materials and properties, and develop and verify quantum algorithms. As quantum technologies continue to advance, quantum simulations are expected to play a pivotal role in various scientific and technological advancements.

In this review, we focus on increasing light-matter interactions via quantum simulations, which covers (at least partially) all the above-mentioned applications. As such, we mainly cover methods for simulating the quantum Rabi model in the USC regime by applying drives to the JC model operating in the strong coupling regime. But it is worth noting that the standard and generalized quantum Rabi models enable further simulating and testing large classes of phenomena or even other theories. These include simulating: deterministic quantum nonlinear optics without real photons, but only with virtual photons and single atoms~\cite{Kockum2017, Kockum2017a, Stassi2017}, supersymmetry (SUSY)~\cite{Tomka2015}, unconventional (polariton-enhanced) superconductivity~\cite{Sentef2018}, the Higgs mechanism~\cite{Garziano2014}, and other effects and theories (for a review, see Ref.~\cite{Kockum2019}).

Finally, we note that standard formalisms of quantum optics can indeed be used to describe light-matter systems in the simulated USC or DSC regime, which can be realized by, e.g., applying quantum or classical drives to a JC-like system. This is another important theoretical advantage of studying the simulated USC regime compared to the true one.


\subsection{Examples of methods for amplifying light-matter interactions}

Among various methods of the light-matter-coupling amplification in JC-type systems, as reviewed in Secs.~\ref{sec:Amplification} and \ref{sec:simulations}, we pay special attention to two approaches, which are based on applying classical and quantum drives.


\subsubsection{Light-matter interactions amplified by classical drives}
\label{ClassicalDrives}

As demonstrated theoretically in Ref.~\cite{Ballester2012}, the quantum Rabi model in the USC regime can be fully simulated by applying two-tone classical drives to the JC model. That method has been demonstrated experimentally in Ref.~\cite{Lv2018} by driving a trapped ion by a pair of counter-propagating Raman laser beams. A similar quantum-simulation method has been experimentally implemented with a superconducting qubit embedded in a cavity-QED setup (a microstrip resonator) and driven by two microwave tones~\cite{Braumuller2017}.

The simulation of the quantum Rabi model by applying strong classical drives (instead of quantum ones) to a JC system enables enhanced-fidelity ultrafast geometric quantum computation~\cite{Chen2022}. Although a single classical drive applied to a JC system cannot simulate a full quantum Rabi model, it is enough to induce numerous effects, which are usually attributed to the USC regime~\cite{Wang2017, SanchezMunoz2020a}. These include the examples described in detail in \appref{sec:examples-one-drive}, i.e., frequency conversion, simultaneous emission of two photons by a single atom, and an analogous effect of the simultaneous excitation of two atoms by a single photon.


\subsubsection{Light-matter interactions amplified by quantum drives}
\label{QuantumDrives}

Another popular method of amplifying light-matter interactions is based on applying {\it parametric amplification} (often referred to as degenerate and non-degenerate parametric amplification, or parametric down-conversion), which can generate squeezed states of light (or other bosonic fields) as an output. When a weak quantum signal interacts with a strong pump beam in a nonlinear medium, the interaction can lead to squeezing of one of the quadratures of the signal beam, which means that the uncertainty in that quadrature is reduced below the vacuum noise level.

Parametric amplification and quadrature squeezing have a significant role in various fields, including quantum optics~\cite{LoudonBook, scully1997book, AgarwalBook, CarmichaelBook, PerinaBook}), atom optics~\cite{MeystreBook}, and even nonrelativistic and relativistic quantum field theories~\cite{ItzyksonBook}. Parametric amplification offers ways to manipulate the quantum properties of light for various purposes, including improving signal-to-noise ratios, enhancing measurement sensitivity, and enabling advanced quantum technologies. We note that the pioneering work of Kennard~\cite{Kennard1927} on nonclassical states (which are now referred to as squeezed states) was published almost a century ago, while the first applications of squeezing in quantum metrology, i.e., for gravitational-wave detectors and interferometers were developed over 40 years ago in Refs.~\cite{Hollenhorst1979, Caves1980, Dodonov1980, Caves1981}. Those works can be considered as the beginning of quantum metrology. To date, the most visible applications of quadrature squeezing are for (i) quantum metrology with squeezed vacuum in the Laser Interferometer Gravitational-Wave Observatory (LIGO)~\cite{Aasi2013, Grote2013, Tse2019} and the Advanced Virgo Detector~\cite{Acernese2019}, and (ii) quantum-optical information processing, e.g., in experimental demonstrations of quantum advantage via boson sampling with squeezed light~\cite{Zhong2020, Zhong2021, Madsen2022}.

Various applications of the USC regime simulated by applying parametric amplification as a quantum drive have been proposed. For example, giant entangled cat states can be generated in a time-dependent quantum Rabi model simulated by applying parametric amplification to a JC system~\cite{Chen2021prl}. Parametric amplification can further enable creating and stabilizing long-lived macroscopic quantum superposition states, not only in a single atom, but also in atomic ensembles~\cite{qin2021generating}, and can even enable ensemble qubits for QIP~\cite{qin2023proposal}. It can also be used for beating the so-called 3-dB limit for intracavity squeezing via quantum reservoir engineering~\cite{qin2022beating}. Other examples can be found in the main part of the review.

Squeezing-enhanced interactions between a boson field and matter are not limited to squeezed optical photons. Actually, the first experimental demonstrations of such amplification schemes were realized with squeezed phonons in Boulder~\cite{burd2020quantum, burd2023experimental} (see also Ref.~\cite{burd2019quantum}) and squeezed microwave photons in Paris~\cite{villiers2022dynamically}. More specifically, the Boulder group reported in Ref.~\cite{burd2020quantum} the 3.25-fold amplification of a phonon-mediated interaction between \textit{two} trapped-ion $^{25}$Mg$^+$ hyperfine qubits by parametric modulation of the trapping potential. Another Boulder experiment~\cite{burd2023experimental} showed the interaction between the motional and spin states of a \textit{single} trapped $^{25}$Mg$^+$ ion amplified by phonon squeezing. Phonons in those experimental implementations correspond to a normal mode of trapped-ion motion. The experiments reported in Refs.~\cite{burd2019quantum, burd2023experimental} are based on sequential and multiple application of proper squeezing and displacement operations in a close analogy to the theoretical proposal for amplifying Kerr interactions~\cite{bartkowiak2014quantum}. Moreover, the Paris-group experiment~\cite{villiers2022dynamically} demonstrated two-fold amplified interactions via microwave-photon squeezing (at \unit[5.5]{dB}) in a superconducting circuit. Specifically, the amplified interaction was observed between a coplanar waveguide resonator capacitively coupled to a transmon qubit. Moreover, a recent theoretical proposal applies the same idea in magnonics, i.e., for amplifying phonon-mediated magnon-spin interactions via virtually-excited squeezed phonons~\cite{Wang2023,nori2023squeezed}.

Other quantum simulation methods of the USC regime include cavity-assisted Raman transitions, digital simulations, and variational methods among others; they are reviewed in \secref{sec:simulations}.


\subsection{Outline of the review}

In summary, this review provides a comprehensive overview of various mechanisms for the amplification of light-matter interactions, especially in cavity and circuit QED, and cavity optomechanics.

The review is focused on describing USC between photons or phonons and qubits, which are realized by real atoms (like trapped ions) or artificial atoms (e.g., superconducting quantum circuits). Nevertheless, many methods reviewed in this paper can be readily generalized to achieve or simulate the USC between other types of quantum excitations, e.g., between phonons and magnons (see, e.g., Ref.~\cite{Wang2023,nori2023squeezed}), photons and magnons~\cite{Golovchanskiy2021}, or photons and plasmons~\cite{Mueller2020}. Thus, the reviewed methods offer new opportunities for quantum technologies also in other fields, like microwave superconducting spintronics or microwave plasmonics.

We discuss different methods which enable amplifying the interactions between light and matter from the strong to ultrastrong, or even deep strong, coupling regimes. These methods include {\it resonant, parametric, and collective amplification}, among others. The amplified photon-mechanical and photon-atom interactions are then explored in detail, with an overview of various amplification mechanisms. Next, simulation methods are discussed, including cavity-assisted Raman transitions, coupled waveguides, and ultracold atoms in optical lattices. Theoretical proposals and experimental implementations are presented, including the application of single or two classical drives in the JC model to simulate the quantum Rabi model or Rabi-like models, and then to nonlinear processes in the USC regime. The review also covers digital simulation methods and variational quantum simulations (VQSs). Finally, we discuss simulation techniques involving coupled waveguides, ultracold atoms in an optical lattice, atomic quantum dots, and a superfluid Bose-Einstein condensate (BEC), as well as the USC between two resonators through three-wave mixing.

For pedagogical reasons, we also present, in the main text and in appendices, detailed derivations of some key results of the applied methods. In particular, we show how the noise induced by squeezing the cavity with a squeezed vacuum reservoir can be effectively eliminated, and how effective Hamiltonians can be intuitively derived within a second-order perturbation theory. We give a few illustrative and detailed examples showing how the effective Hamiltonians, derived for JC-type systems driven by a single classical drive in the strong coupling regime, can enable the simulation of some nonlinear effects characteristic for the USC regime.
\section{Amplification of quantum light-matter interactions}
\label{sec:Amplification}

Below, we first review, in \secref{Amplified photon-mechanical interactions in cavity optomechanic}, amplification methods for optomechanical interactions in cavity optomechanics, including, e.g., amplification via linearization, resonant amplification, parametric amplification, and so on. Then, in \secref{Amplified photon-atom interactions in cavity quantum electrodynamics}, we consider amplification methods of photon-atom interactions in cavity and circuit quantum electrodynamics, including, e.g., parametric amplification, collective amplification, etc. Furthermore, in \secref{Experimental demonstrations of parametrically amplified light-matter interactions}, we introduce recent experimental demonstrations of using parametric squeezing to amplify light-matter interactions in trapped-ion and superconducting-circuit systems. Finally, in \secref{Amplified Kerr-type light-matter interaction via quadrature squeezing}, we introduce the amplification of Kerr-type light-matter interactions with squeezing.


\subsection{Amplified photon-mechanical interactions in cavity optomechanics}
\label{Amplified photon-mechanical interactions in cavity optomechanic}

Cavity optomechanics explores the interaction between electromagnetic radiation and mechanical motion~\cite{Aspelmeyer2014cavity,liu2018cavity}. This optomechanical interaction fundamentally originates from the momentum transfer of cavity photons to mechanical objects, referred to as radiation-pressure forces, and can be described by the Hamiltonian (hereafter we set $\hbar=1$)~\cite{law1995interaction}
\begin{equation}
	H_{\rm OMI} = -G a^{\dag}a x,
\end{equation}
where $G$ is the cavity-frequency dispersive shift per displacement, $a$ is the annihilation operator for the cavity mode, and $x$ is the mechanical displacement, e.g., of the cavity mirror. The radiation-pressure force upon the mechanical object is accordingly given by $F = G a^{\dagger}a$. The mechanical motion can be modelled by a single-mode harmonic oscillator with a Hamiltonian 
\begin{equation}
	H_{m} = \omega_{m} b^{\dag}b, 
\end{equation}
where $\omega_{m}$ is the mechanical frequency and $b$ is the phonon annihilation operator. The displacement $x$ is accordingly expressed as 
\begin{equation}
x = x_{\rm zpf}\mleft(b+b^{\dag}\mright),
\end{equation}
where
\begin{equation}
x_{\rm zpf} = \frac{1}{\sqrt{2m_{\rm eff}\omega_{m}}}
\end{equation}
is the zero-point fluctuation of the mechanical resonator, with $m_{\rm eff}$ being the effective mass of the mechanical resonator. As a result, the optomechanical interaction $H_{\rm OMI}$ becomes 
\begin{equation}\label{eq:optomechanical interaction}
	H_{\rm OMI} = -g_{0}a^{\dag}a\mleft(b+b^{\dag}\mright),
\end{equation}
where $g_{0}=Gx_{\rm zpf}$ is the single-photon optomechanical-coupling strength. A necessary condition for Eq.~(5) to be valid is that $g_{0}$ is much smaller than $2\omega_{\rm cav}\pm\omega_{m}$, such that the mechanically induced creation and annihilation of photon pairs can be neglected. In most experimental situations, this condition is well satisfied. Indeed, $g_{0}$ is extremely weak and, as a result, the ratio $g_{0}/\omega_{m}$ is very small. For these reasons, it is an experimental challenge to observe effects of the detuned optomechanical interaction $H_{\rm OMI}$. Therefore, a large number of methods have been proposed to amplify the optomechanical interaction. Below, we review such methods.


\subsubsection{Amplification via linearization}
\label{subsubsec: Amplification via linearization}

When the cavity mode is driven by a strong coherent drive, the optomechanical interaction $H_{\rm OMI}$ can be linearized approximately and, as a result, the coupling strength $g_{0}$ is enhanced with the average number of intracavity photons. Let us assume that the Hamiltonian of the coherent driving is 
\begin{equation}
H_{\rm dr} = \mathcal{E}\exp\mleft(i\omega_{d}t\mright)a^{\dagger}+{\rm H.c.},
\end{equation}
with complex amplitude $\mathcal{E}$ and frequency $\omega_{d}$. The quantum Langevin equations of motion for the operators $a$ and $b$ in a frame rotating at the cavity frequency $\omega_\mathrm{cav}$ are then given by
\begin{align}\label{eq:equation of motion for cavity}
	\dot{a}=\;&-i\Delta a+ig_{0}a\mleft(b+b^{\dagger}\mright)-i\mathcal{E}-\frac{\kappa}{2}a-\sqrt{\kappa}a_{\rm in}\mleft(t\mright),\\
	\dot{b}=\;&-i\omega_{m}b+ig_{0}a^{\dagger}a-\frac{\gamma_{m}}{2}b-\sqrt{\gamma_{m}}b_{\rm in}\mleft(t\mright),
	\label{eq:equation of motion for motion}
\end{align}
where $\Delta=\omega_{\rm cav}-\omega_{d}$ is the detuning of the cavity resonance $\omega_{\rm cav}$ from the strong driving
field, $\kappa$ is the cavity decay rate, and $\gamma_{m}$ is the mechanical decay rate. Moreover, $a_{\rm in}\mleft(t\mright)$ and $b_{\rm in}\mleft(t\mright)$ are the input-noise annihilation operators for the cavity field and the mechanical motion, respectively. Note that here, to derive the quantum Langevin equations of motion in Eqs.~(\ref{eq:equation of motion for cavity}) and~(\ref{eq:equation of motion for motion}), the Weisskopf-Wigner approximation has been made~\cite{scully1997book}, such that the decay rates $\kappa$ and $\gamma_{m}$ are constant.

Because of the presence of a strong drive $\mathcal{E}$, one can divide the cavity field into the sum of an average amplitude $\alpha$ and a small fluctuation $\delta a$, such that $a=\alpha+\delta a$. Likewise, the mechanical motion is reexpressed as $b=\beta+\delta b$. Indeed, $\delta a$ and $\delta b$ can also be understood as the displaced versions of the operators $a$ and $b$, respectively. Substituting these displaced operators into the equations of motion in Eqs.~(\ref{eq:equation of motion for cavity}) and (\ref{eq:equation of motion for motion}), and then separating the classical and quantum parts, yields
\begin{align}
	\dot{\alpha}=\;&-i\Delta^{\prime}\alpha-i\mathcal{E}-\frac{\kappa}{2}\alpha,\\
	\dot{\beta}=\;&-i\omega_{m}\beta+ig_{0}\mleft|\alpha\mright|^{2}-\frac{\gamma_{m}}{2}\beta,\\
	\label{eq:displaced cavity mode}
	\delta\dot{a}=\;&i\Delta^{\prime}\delta a+ig_{0}\mleft(\alpha+\delta a\mright)\mleft(\delta b+\delta b^{\dagger}\mright)-\frac{\kappa}{2}\delta a-\sqrt{\kappa}a_{\rm in}\mleft(t\mright),\\
	\label{eq:displaced mechanical mode}
	\delta\dot{b}=\;&-i\omega_{m}\delta b+ig_{0}\mleft(a^{\dagger}a+\alpha a^{\dagger}+\alpha^{*}a\mright)-\frac{\gamma_{m}}{2}\delta b-\sqrt{\gamma_{m}}b_{\rm in}\mleft(t\mright),
\end{align}
where $\Delta^{\prime}=\Delta-g_{0}\mleft(\beta+\beta^{*}\mright)$ is a new detuning induced by the optomechanical interaction. By setting $\dot{\alpha}=\dot{\beta}=0$, the average amplitudes $\alpha$ and $\beta$ are found to be
\begin{align}
	\alpha=\;&-\frac{i\mathcal{E}}{i\Delta^{\prime}+\kappa/2},\\
	\beta=\;&\frac{ig_{0}\mleft|\alpha\mright|^{2}}{i\omega_{m}+\gamma_{m}/2}\approx\;\frac{g_{0}\mleft|\alpha\mright|^{2}}{\omega_{m}}.
\end{align}

Neglecting the weak nonlinear coupling terms in Eqs.~(\ref{eq:displaced cavity mode}) and~(\ref{eq:displaced mechanical mode}), the equations of motion for the displaced operators $\delta a$ and $\delta b$ are given by
\begin{align}
	\delta\dot{a}=\;&-i\Delta^{\prime}\delta a+i\alpha g_{0}\mleft(\delta b+\delta b^{\dagger}\mright)-\frac{\kappa}{2}\delta a-\sqrt{\kappa}a_{\rm in},\\
	\delta\dot{b}=\;&-i\omega_{m}\delta b+ig_{0}\mleft(\alpha \delta a^{\dagger}+\alpha^{*}\delta a\mright)-\frac{\gamma_{m}}{2}\delta b-\sqrt{\gamma_{m}}b_{\rm in},
\end{align}
both of which correspond to an effective optomechanical interaction 
\begin{equation}\label{eq:linearized Hamiltonian}
	H_{\rm OMI}^{\rm L}=-g_{c}\mleft[\exp\mleft(-i\theta\mright)\delta a+\exp\mleft(i\theta\mright)\delta a^{\dag}\mright]\mleft(\delta b+\delta b^{\dagger}\mright),
\end{equation}
where 
\begin{empheq}[box =\widecolourbox]{equation}
	g_{c}=g_{0}\sqrt{\bar{n}_{c}},
\end{empheq}
is referred to as the linearized optomechanical interaction strength, and $\theta$ is the phase of the amplitude $\alpha$. Here, $\bar{n}_{c}=\mleft|\alpha\mright|^2$ is the average number of intracavity photons. Thus $g_{c}$ is enhanced by a factor of $\sqrt{\bar{n}_{c}}$, compared to the bare single-photon coupling $g_{0}$. It has been experimentally shown that such an enhanced coupling can even reach the regime of the USC~\cite{peterson2019ultrastrong}. 

In the case of $\Delta^{\prime}\approx\omega_{m}$, i.e., in the red-detuned regime, the linearized Hamiltonian $H_{\rm OMI}^{L}$ can, under the RWA, be reduced to 
\begin{equation}\label{eq:Beam splitter Hamiltonian}
	H_{\rm OMI}^{\rm L}\approx H_{-}=-g_{c}\mleft[\exp\mleft(-i\theta\mright)\delta a\delta b^{\dagger}+\exp\mleft(i\theta\mright)\delta a^{\dag}\delta b\mright].
\end{equation}
In the displaced frame, $H_{-}$ acts as a beam-splitter-like interaction and results in the exchange of a single excitation between the cavity and the mechanical resonator. In the original frame, this exchange corresponds to an anti-Stokes scattering process where a single photon of the driving field is scattered into the cavity resonance, while simultaneously absorbing a mechanical phonon. The Hamiltonian $H_{-}$ has already been widely used for, e.g., sideband cooling of mechanical motion~\cite{marquardt2007quantum, wilson2007theory, teufel2011sideband, asjad2016suppression, clark2017sideband} and coherent state transfer between the cavity and mechanical modes~\cite{fiore2011storing, verhagen2012quantum, dong2012optomechanical}.

In the case of $\Delta^{\prime}\approx-\omega_{m}$, i.e., in the blue-detuned regime, the linearized Hamiltonian $H_{\rm OMI}^{L}$, under the RWA, reduces to 
\begin{equation}\label{eq:two-mode squeezing Hamiltonian}
	H_{\rm OMI}^{\rm L}\approx H_{+}=-g_{c}\mleft[\exp\mleft(-i\theta\mright)\delta a\delta b+\exp\mleft(i\theta\mright)\delta a^{\dag}\delta b^{\dag}\mright].
\end{equation}
In the displaced frame, $H_{+}$ acts as a two-mode-squeezing-like interaction and results in a simultaneous excitation of a cavity photon and a mechanical phonon. In the original frame, this simultaneous excitation corresponds to a Stokes scattering process, where a single photon of the driving field is scattered into the cavity resonance, while simultaneously exciting a mechanical phonon. The Hamiltonian $H_{+}$ has already been widely used for, e.g., quantum-limited amplification~\cite{massel2011microwave, nunnenkamp2014quantum} and generating entanglement between the cavity and mechanical modes~\cite{vivoli2016proposal, hofer2016proposal, marinkovic2018optomechanical}.


\subsubsection{Resonant amplification}

In \secref{subsubsec: Amplification via linearization}, we introduced a method of amplifying the optomechanical interaction with a strong coherent driving. This is the most commonly used method in cavity optomechanics. However, such a method neglects the intrinsic nonlinearity of the optomechanical interaction $H_{\rm OMI}$ in \eqref{eq:optomechanical interaction}. To make that nonlinearity significant, the single-photon strong coupling regime is required. In this regime, the single-photon coupling strength $g_{0}$ exceeds the cavity loss rate $\kappa$, i.e., $g_{0}\gtrsim\kappa$. However, the optomechanical interaction $H_{\rm OMI}$ in fact describes an off-resonant interaction, and thus its strength strongly depends on the ratio $g_{0}/\omega_{m}$. Unfortunately, that ratio is usually very small, typically of the order of $10^{-5}$ to $10^{-2}$. This strongly suppresses the nonlinearity of the optomechanical interaction, even in the case of $g_{0}\gtrsim\kappa$. For this reason, many proposals to amplify the single-photon nonlinearity in cavity optomechanics focused on how to make the nonlinear interaction resonant or near-resonant. 

\begin{figure}
	\centering
	\includegraphics[width=\linewidth]{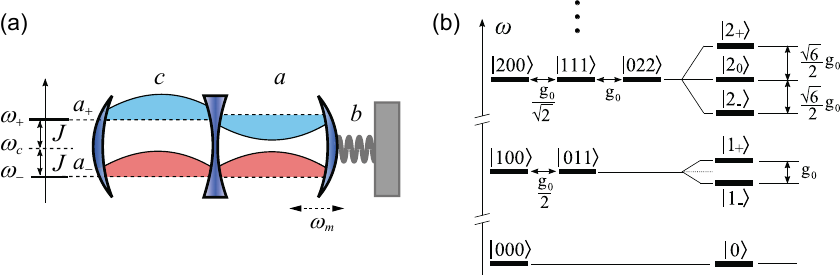}
	\caption{(a) Schematics of the double-cavity approach to enhance the single-photon optomechanical interaction. The two cavities, both with frequency $\omega_{\rm cav}$, are coupled through a beam-splitter interaction of strength $J$, which in turn leads to two normal modes with resonance frequencies $\omega_{\pm}=\omega_{\rm cav}\pm J$. The rightmost mirror is oscillating with frequency $\omega_{m}$.
    (b) Resonant two-mode optomechanical interaction described by $H_{\rm OMI}^{\rm DC}$ in Eq.~(\ref{eq:H_OMI double cavity}). Under the resonance condition $\omega_{m}\approx 2J$, a photon can be scattered, through the optomechanical interaction $g_{0}$, from the mode $a_{-}$ ($a_{+}$) into the mode $a_{+}$ ($a_{-}$) by the absorption (emission) of a mechanical phonon in the mode $b$. Panels (a) and (b) are reproduced with permission from Ref.~\cite{komar2013single}, P.~K\'om\'ar et al., \href{https://link.aps.org/doi/10.1103/PhysRevA.87.013839}{Phys.~Rev.~A \textbf{87}, 013839 (2013)}.}
    \label{fig_resonantly_enhanced_optomechanics}
\end{figure}

A possible approach is to linearly couple the optomechanical cavity to another empty cavity~\cite{stannigel2012optomechanical, ludwig2012enhanced, komar2013single}, as shown in \figref{fig_resonantly_enhanced_optomechanics}. Such a double-cavity setup can be realized with, e.g., a membrane-in-the-middle optomechanical system~\cite{thompson2008strong} or two coupled whispering-gallery-mode microresonators~\cite{grudinin2010phonon, jing2014pt, zhang2018phonon}.

Let us assume that the intercavity coupling is given by 
\begin{equation}\label{eq:intercavity coupling}
	H_{\rm IC} = -J \mleft( a c^{\dag} + a^{\dag} c \mright),
\end{equation}
where $c$ is the annihilation operator for the empty cavity and $J$ is the coupling strength. Note that the presence of the negative sign does not affect the results of interest. In the case when these two cavities have the same frequency $\omega_{\rm cav}$, the coupling $H_{\rm IC}$ leads to the formation of the normal modes (often referred to as supermodes)
\begin{equation}
	a_{+}=\frac{1}{\sqrt{2}}\mleft(a+c\mright) \quad {\rm and} \quad a_{-}=\frac{1}{\sqrt{2}}\mleft(a-c\mright),
\end{equation}
with resonance frequencies $\omega_{\pm}=\omega_{\rm cav}\pm J$, respectively.
When expressed in terms of the normal modes, the optomechanical interaction $H_{\rm OMI}$ in Eq.~(\ref{eq:optomechanical interaction}) becomes
\begin{equation}
	H_{\rm OMI}^{\rm }=-\frac{g_{0}}{2}\mleft(a_{+}^{\dagger}a_{+}+a_{-}^{\dagger}a_{-}\mright)\mleft(b+b^{\dagger}\mright)-\frac{g_{0}}{2}\mleft(a_{+}a_{-}^{\dagger}+a^{\dagger}_{+}a_{-}\mright)\mleft(b+b^{\dagger}\mright).
\end{equation}
By tuning the splitting of the two normal modes to be equal to the mechanical frequency, i.e., $\omega_{m}\approx2J$, one can apply the RWA, yielding
\begin{equation}\label{eq:H_OMI double cavity}
	H_{\rm OMI}^{\rm }\approx H_{\rm OMI}^{\rm DC}=\frac{g_{0}}{2}\mleft(a_{+}a_{-}^{\dagger}b+{\rm H.c.}\mright),
\end{equation}
where the superscript ``DC" refers to the double-cavity optomechanical system and a phase factor of $-1$ has been absorbed into $g_{0}$. The Hamiltonian $H_{\rm OMI}^{\rm DC}$ describes the resonant exchange of photons between the two normal modes by the absorption or emission of a mechanical phonon, as depicted in Fig.~\ref{fig_resonantly_enhanced_optomechanics}(b). Such an exchange process leads to the formation of dressed states, e.g., 
\begin{align}
\ket{1_{\pm}}=\;&\frac{1}{\sqrt{2}}\mleft(\ket{100}\pm\ket{011}\mright),\\
\ket{2_{\pm}}=\;&\frac{1}{\sqrt{6}}\mleft(\ket{200}\pm\sqrt{3}\ket{111}+\sqrt{2}\ket{022}\mright),\\
\ket{2_{0}}=\;&\frac{1}{\sqrt{3}}\mleft(\sqrt{2}\ket{200}-\ket{022}\mright)
\end{align}
for the three lowest energy levels, with the unchanged ground state, i.e., $\ket{0}=\ket{000}$. Here,  $\ket{n_{+}n_{-}n_{m}}$ refers to a state with $n_{\pm}$ photons in the normal modes $a_{\pm}$, and $n_{m}$ phonons in the mechanical mode $b$. The resonant nonlinearity in Eq.~(\ref{eq:H_OMI double cavity}) can enable the double-cavity optomechanical system to enter the single-photon strong coupling regime more easily than the usual single-cavity optomechanical system. 

As demonstrated in Refs.~\cite{liao2014modulated,liao2016macroscopic}, if the coupling strength $J$ in Eq.~(\ref{eq:intercavity coupling}) is assumed to be modulated sinusoidally so that
\begin{equation}
	J\mapsto J\mleft(t\mright)=\zeta\omega_{0}\cos\mleft(\omega_{0}t\mright),
\end{equation}
then 
the optomechanical interaction $H_{\rm OMI}$ becomes 
\begin{equation}
	H_{\rm OMI}\approx H_{\rm OMI}^{\rm M}=g_{\rm M}\mleft(c^{\dagger}c-a^{\dagger}a\mright)\mleft[b\exp\mleft(-i\delta t\mright)+{\rm H.c.}\mright],
\end{equation}
with 
\begin{equation}
g_{\rm M}=\frac{1}{2}g_{0}J_{2n_{0}}\mleft(2\zeta\mright)
\end{equation}
being an effective optomechanical interaction. Here, $\zeta$ is the dimensionless modulation amplitude, $\omega_{0}$ is the modulation frequency, $\delta=\omega_{m}-2n_{0}\omega_{0}$ is the modulation-induced detuning, and $J_{n}\mleft(z\mright)$ is the $n$th-order Bessel function of the first kind. The number $n_{0}$ is a special integer such that the coupling with a detuning $\delta$ is nearly resonant. The superscript ``M" refers to the modulation of the intercavity coupling. 

\begin{figure}
	\centering
	\includegraphics[width=\linewidth]{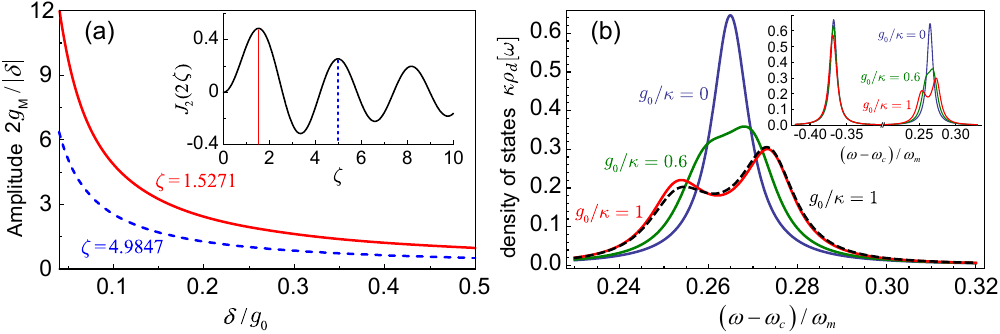}
	\caption{(a) Maximum mechanical displacement, $2g/\mleft|\delta\mright|$, in a double-cavity optomechanical system with a periodically modulated intercavity coupling. Inset: the second-order Bessel function of the first kind, $J_{2}\mleft(2\zeta\mright)$, plotted as a function of the modulation amplitude $\zeta$.
    (b) Modified cavity density of states by the residual optomechanical interaction [described by $H_{\delta}$ in Eq.~(\ref{eq_residual_optomechanical_coupling})] for the ``$+$" polariton mode. The inset shows the asymmetrical density of states for the ``$-$" (left) and ``$+$" (right) polariton modes. The solid curves are analytical results; the dashed curve is obtained from a numerical simulation of the master equation. Panels (a) and (b) are adapted with permission, respectively, from Ref.~\cite{liao2016macroscopic}, J.-Q.~Liao et al., \href{https://link.aps.org/doi/10.1103/PhysRevLett.116.163602}{Phys.~Rev.~Lett.~\textbf{116}, 163602 (2016)} and Ref.~\cite{nonlinear2013lemonde}, M.-A.~Lemonde et al., \href{https://link.aps.org/doi/10.1103/PhysRevLett.111.053602}{Phys.~Rev.~Lett.~\textbf{111}, 053602 (2013)}.}
    \label{fig_double_cavity_plus_polaritons}
\end{figure}

The Hamiltonian $H_{\rm OMI}^{\rm M}$ essentially describes an effective force $\propto\langle c^{\dagger}c-a^{\dagger}a\rangle$ acting on a mechanical resonator with an effective frequency $\delta$. Compared to the natural optomechanical interaction $H_{\rm OMI}$, the near-resonant coupling $H_{\rm OMI}^{\rm M}$ can induce a single-photon mechanical displacement proportional to $g_{0}/\mleft|\delta\mright|$ rather than to $g_{0}/\omega_{m}$. This indicates that, as shown in \figpanel{fig_double_cavity_plus_polaritons}{a}, the optomechanical nonlinearity is strongly amplified. The physical reason for this amplification is that the single-photon hopping between the two cavities at the
proper times accumulates the displacement effect when the driving force is in phase with the mechanical oscillation. A similar enhancement of the nonlinearity can also be obtained by modulating the cavity frequencies~\cite{liao2015enhancement}. The resonant amplification based on modulation has been studied for the generation of mechanical Schr\"{o}dinger cat states via flipping a qubit repeatedly~\cite{tian2005entanglement}, and even for the detection of the virtual radiation pressure arising from atom-field USC~\cite{cirio2017amplified}. 

To obtain the linearized Hamiltonian $H_{\rm OMI}^{\rm L}$ in Eq.~(\ref{eq:linearized Hamiltonian}), the residual optomechanical interaction
\begin{equation}\label{eq_residual_optomechanical_coupling}
	H_{\delta}=-g_{0}\delta a^{\dagger} \delta a\mleft(\delta b+\delta b^{\dagger}\mright) 
\end{equation}
is neglected. This coupling is of the same form as the standard optomechanical interaction and, as suggested in Refs.~\cite{nonlinear2013lemonde, borkje2013signatures, kronwald2013optomechanically}, can be amplified when the coupling $H_{\delta}$ is treated as a perturbation to the linearized Hamiltonian $H_{\rm OMI}^{\rm L}$ in Eq.~(\ref{eq:linearized Hamiltonian}). In the absence of the coupling $H_{\delta}$, the Hamiltonian $H_{\rm OMI}^{\rm L}$ can be diagonalized, yielding
\begin{equation}
	H_{\rm OMI}^{\rm L}=E_{+}p_{+}^{\dag}p_{+}+E_{-}p_{-}^{\dag}p_{-},
\end{equation}
where the operators $p_{\pm}$ represent two normal modes with eigenenergies
\begin{equation}
	E_{\pm}=\frac{1}{\sqrt{2}}\mleft(\Delta^{2}+\omega_{m}^{2}\pm\sqrt{\mleft(\Delta^{2}-\omega_{m}^{2}\mright)^{2}-16\Delta g_{c}^{2}\omega_{m}}\mright)^{1/2},
\end{equation}
respectively. These two normal modes hybridize the photonic and mechanical degrees of freedom and can thus be considered as polaritons. Expressed in terms of the polariton modes, the Hamiltonian $H_{\delta}$ is transformed to
\begin{equation}	H_{\delta}\approx\tilde{H}_{\delta}=\tilde{g}_{0}\mleft(p_{+}^{\dagger}p_{-}^{2}+{\rm H.c.}\mright),
\end{equation}
where $\tilde{g}_{0}$ is an effective coupling strength proportional to the single-photon coupling $g_{0}$. Here, we have assumed that $E_{+}=2E_{-}$, such that the coupling $\tilde{H}_{\delta}$ becomes resonant and, thus, dominant. Other off-resonant couplings have been neglected under the RWA. The Hamiltonian $\tilde{H}_{\delta}$ describes a process where a ``$+$" polariton is created and simultaneously two ``$-$" polaritons are destroyed, and vice versa. This process can strongly modify the cavity density of states even with a weak single-photon coupling $g_{0}$, as depicted in \figpanel{fig_double_cavity_plus_polaritons}{b}. It could be further exploited for observing optomechanically induced transparency and also measuring the 
average number of mechanical phonons.


\subsubsection{Parametric amplification}
\label{Parametric amplification}

In this section, we introduce another method, which uses parametric amplification to enhance the nonlinear optomechanical interaction~\cite{bartkowiak2014quantum, lu2015squeezed}. Before discussing this method, let us first recall degenerate parametric amplification, which is a very common nonlinear process in quantum optics.

Parametric amplification essentially describes a nonlinear interaction between three distinct light fields, usually referred as
to the pump, signal, and idler, respectively. This nonlinear interaction down-converts a pump photon into a correlated photon pair (i.e., a signal photon and an idler photon) under energy and momentum conservation. Here, we focus our attention mainly on {\it degenerate parametric amplification} (DPA), in which the signal and idler photons in each pair are identical, i.e., have the same frequency and the same polarization. There are similar results for non-degenerate parametric amplification. The Hamiltonian describing DPA inside a cavity is
\begin{equation}\label{eq:parametric driving Hamiltonian}
	H_{\rm DPA}=\Delta_{\rm 2ph} a^{\dag}a+\frac{1}{2}\Omega_{\rm 2ph}\mleft[\exp\mleft(-i\theta_{\rm 2ph}\mright)a^{2}+{\rm H.c.}\mright],
\end{equation}
where $\Omega_{\rm 2ph}$ and $\theta_{\rm 2ph}$ are the amplitude and phase of the parametric (or two-photon) driving, and $\Delta_{\rm 2ph}=\omega_{\rm cav}-\omega_{\rm 2ph}/2$ is the detuning between the cavity frequency $\omega_{\rm cav}$ and half the parametric driving frequency $\omega_{\rm 2ph}$.

The dynamics described by $H_{\rm DPA}$ squeezes the cavity field. To proceed, we now introduce the squeezing operator defined by
\begin{equation}\label{eq:squeezing operator}
	S\mleft(\xi\mright)=\exp\mleft[\frac{1}{2}\mleft(\xi^{*}a^{2}-\xi a^{\dagger 2}\mright)\mright],
\end{equation}
where $\xi=r\exp\mleft(i\theta_{\rm 2ph}\mright)$ is an arbitrary complex number. Here, $r$ is the squeezing parameter, which determines the degree of squeezing. The squeezing operator acting on the cavity mode $a$ causes a squeezed cavity mode $a_{\rm sq}$, which is given by the Bogoliubov transformation
\begin{equation}
	a_{\rm sq}\equiv S\mleft(\xi\mright)aS^{\dagger}\mleft(\xi\mright)=a\cosh\mleft(r\mright)+a^{\dagger}\exp\mleft(i\theta_{\rm 2ph}\mright)\sinh\mleft(r\mright).
\end{equation}
It is easily found, after a straightforward calculation, that
\begin{equation}\label{eq:Bogoliubov transformation}
	a=a_{\rm sq}\cosh\mleft(r\mright)-a^{\dagger}_{\rm sq}\exp\mleft(i\theta_{\rm 2ph}\mright)\sinh\mleft(r\mright).
\end{equation}
By expressing $H_{\rm DPA}$ in terms of the $a_{s}$ mode and then choosing a proper squeezing parameter, i.e.,
\begin{equation}
	r=\frac{1}{4}\ln\frac{\Delta_{\rm 2ph}+\Omega_{\rm 2ph}}{\Delta_{\rm 2ph}-\Omega_{\rm 2ph}},
\end{equation}
the Hamiltonian $H_{\rm DPA}$ becomes diagonal,
\begin{equation}\label{eq:diagonalized DPA}
	H_{\rm DPA}^{\rm sq}=\omega_{\rm sq}a^{\dagger}_{\rm sq}a_{\rm sq},
\end{equation}
where $\omega_{\rm sq}=\sqrt{\Delta^{2}_{\rm 2ph}-\Omega_{\rm 2ph}^{2}}$ is the frequency of the squeezed cavity mode.  Here, we have assumed that $\Delta_{\rm 2ph}>\Omega_{\rm 2ph}$, such that the system is stable. Note that the ground state of the Hamiltonian $H_{\rm sq}$ is the vacuum state in the squeezed frame, which corresponds to the squeezed vacuum in the lab frame.

\begin{figure}
	\centering
	\includegraphics[width=\linewidth]{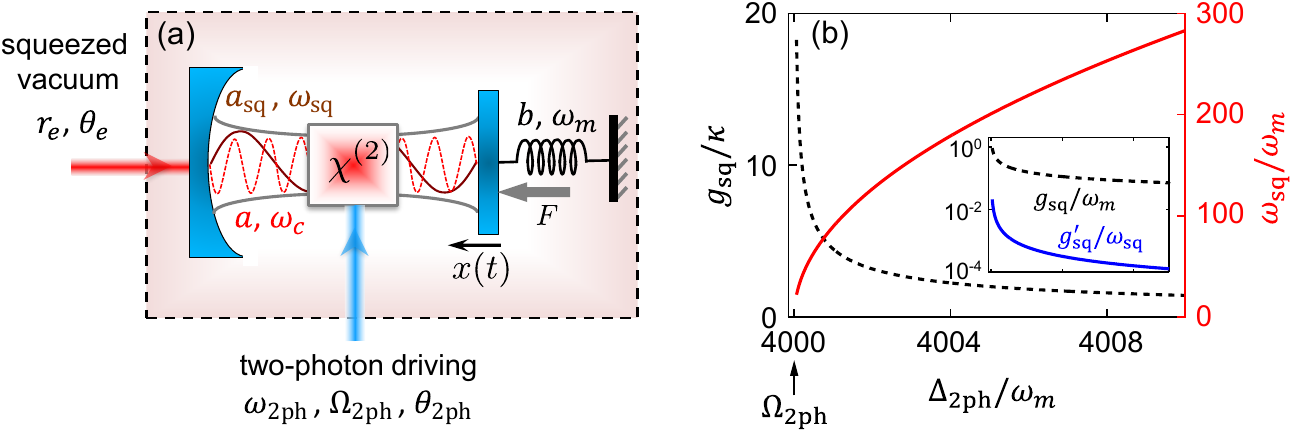}
	\caption{(a) Schematics of using parametric amplification to enhance the optomechanical interaction. A cavity, with frequency $\omega_{\rm cav}$, contains a $\chi^{\mleft(2\mright)}$ nonlinear crystal, which is driven by a two-photon driving of frequency $\omega_{\rm 2ph}$, amplitude $\Omega_{\rm 2ph}$, and phase $\theta_{\rm 2ph}$. The right mirror of the cavity is oscillating with frequency $\omega_{m}$ and position  $x\mleft(t\mright)$.
    (b) Enhanced optomechanical interaction $g_{\rm sq0}$ (left axis) and resonance frequency of the squeezed cavity mode $\omega_{\rm sq}$ (right axis) as a function of the detuning $\Delta_{\rm 2ph}$ for $\Omega_{\rm 2ph}=4000\omega_{m}$. The inset shows the ratios $g_{\rm sq0}/\omega_{m}$ and $g_{\rm sq0}^{\prime}/\omega_{\rm sq}$. Panels (a) and (b) are reproduced with permission from Ref.~\cite{lu2015squeezed}, X.-Y.~L\"{u} et al., \href{https://link.aps.org/doi/10.1103/PhysRevLett.114.093602}{Phys.~Rev.~Lett.~\textbf{114}, 093602 (2015)}.}
    \label{fig_parametrically_enhanced_optomechanics}
\end{figure}

The nonlinear transformation in Eq.~(\ref{eq:Bogoliubov transformation}) shows a Bogoliubov coefficient of $\cosh\mleft(r\mright)$, which {\it increases exponentially with the squeezing parameter $r$. This indicates that in the squeezed frame, the coupling of the cavity mode to other degrees of freedom can be enhanced exponentially}. Such a mechanism has been used for enhancing the optomechanical interaction in cavity-optomechanical systems~\cite{lu2015squeezed}. A schematic setup is illustrated in \figpanel{fig_parametrically_enhanced_optomechanics}{a}. Crucially, a nonlinear $\chi^{\mleft(2\mright)}$ crystal is placed inside a cavity, such that the cavity mode is subject to a detuned two-photon driving. As mentioned above, the cavity mode $a$ is squeezed by the two-photon driving and, as a result, becomes the squeezed mode $a_{\rm sq}$. Expressed in terms of the mode $a_{\rm sq}$, the optomechanical interaction Hamiltonian in Eq.~(\ref{eq:optomechanical interaction}) is transformed to
\begin{equation}\label{eq:optomechanical interaction in terms of a_s}
	H_{\rm OMI}=-g_{\rm sq}^{\rm om}a_{\rm sq}^{\dagger}a_{\rm sq}\mleft(b+b^{\dagger}\mright)+\frac{1}{2}g_{\rm sq}^{\rm 2ph}\mleft(a_{\rm sq}^{2}+a_{\rm sq}^{\dagger2}\mright)\mleft(b+b^{\dagger}\mright),
\end{equation}
where 
\begin{equation}
	g_{\rm sq}^{\rm om}=g_{0}\cosh\mleft(2r\mright)
\end{equation}
describes the strength of the optomechanical interaction between the squeezed cavity field and the mechanical resonator, and 
\begin{equation}
g_{\rm sq}^{\rm 2ph}=g_{0}\sinh\mleft(2r\mright)
\end{equation}
is the strength of a two-photon process in the squeezed frame.

We now consider the case of $\omega_{\rm sq}\gg\mleft\{\omega_{m},g_{\rm sq}^{\rm 2ph}\mright\}$, as shown in \figpanel{fig_parametrically_enhanced_optomechanics}{b}. Here, the RWA is allowed, such that the second term in \eqref{eq:optomechanical interaction in terms of a_s} is negligible, yielding a standard optomechanical interaction Hamiltonian in the squeezed frame,
\begin{equation}\label{eq:enhanced optomechanical interaction}
	H_{\rm OMI} \approx H_{\rm OMI}^{\rm sq}=-g_{\rm sq}^{\rm om}a_{\rm sq}^{\dag}a_{\rm sq}\mleft(b+b^{\dag}\mright).
\end{equation}
When $r\geq1$, the coupling strength, $g_{\rm sq}^{\rm om}$, can be approximated by
\begin{empheq}[box =\widecolourbox]{equation}
	g_{\rm sq}^{\rm om}\approx g_{0}\exp\mleft(2r\mright)=g_{0}\sqrt{\frac{\Delta_{\rm 2ph}+\Omega_{\rm 2ph}}{\Delta_{\rm 2ph}-\Omega_{\rm 2ph}}}.
\end{empheq}
This implies that, compared to the single-photon coupling $g_{0}$ in the original laboratory frame, an {\it exponential enhancement} of $g_{\rm sq}^{\rm om}$ can be achieved with increasing the squeezing parameter $r$, i.e., when $\Delta_{\rm 2ph}$ approaches $\Omega_{\rm 2ph}$ from the right, as seen in Fig.~\ref{fig_parametrically_enhanced_optomechanics}(b). Under a similar mechanism, the quadratic optomechanical interaction
can also be exponentially enhanced~\cite{gu2018enhanced,zhang2019enhancing}.

The parametric driving, while squeezing the cavity mode, also introduces thermal noise and two-photon-correlation noise into the cavity. These two types of noise are generally considered detrimental in strong-squeezing processes. However, by injecting a squeezed vacuum field into the cavity~[see \figpanel{fig_parametrically_enhanced_optomechanics}{a}], one can eliminate both types of noise (see \appref{app:Derivations of eliminating the noise induced by squeezing the cavity with a squeezed vacuum reservoir}), and as a result the dynamics of the optomechanical system can be described by the master equation
\begin{equation}\label{eq:simplified master equation for optomechanical systems}
	\dot{\rho} = -i\mleft[H_{\rm OM}^{\rm sq},\rho\mright]+\kappa\mathcal{L}\mleft(a_{\rm sq}\mright)\rho+\gamma_{m}\mleft(n_{\rm th}+1\mright)\mathcal{L}\mleft(b\mright)\rho+\gamma_{m}n_{\rm th}\mathcal{L}\mleft(b^{\dagger}\mright)\rho,
\end{equation}
where 
\begin{equation}
H_{\rm OM}^{\rm sq}=\omega_{\rm sq}a^{\dagger}_{\rm sq}a_{\rm sq}+\omega_{m}b^{\dagger}b+H^{\rm sq}_{\rm OMI},
\end{equation}
while $\gamma_{m}$ is the mechanical decay rate and $n_{\rm th}$ is the thermal phonon number of the mechanical mode. It is seen that by increasing the squeezing parameter $r$, the optomechanical system can be driven into the single-photon strong coupling regime (i.e., $g_{\rm sq}^{\rm om}>\kappa$) even from the single-photon WC regime ($g_{0}<\kappa$). As a direct result, such a parametric enhancement of the optomechanical interaction improves the conversion~\cite{luo2020squeezing}, entanglement~\cite{xiong2017improve}, and even  cross-Kerr nonlinearity~\cite{yin2018enhanced} between the optical and microwave fields. 
The two-mode squeezing of the cavity field is also capable of exponentially enhancing the optomechanical interaction, but at the same time a two-mode, rather than single-mode, squeezed vacuum field is needed to suppress the noise induced by this two-mode squeezing~\cite{li2016enhanced}.

In the case of $\omega_{m}\approx2\omega_{\rm sq}$, the Hamiltonian in Eq.~(\ref{eq:optomechanical interaction in terms of a_s}) reduces to
\begin{equation}
	H_{\rm OMI}\approx H_{-}^{\rm sq}=\frac{1}{2}g_{\rm sq}^{\rm 2ph}\mleft(a_{\rm sq}^{2}b^{\dagger}+{\rm H.c.}\mright),
\end{equation}
which can be used to simulate the dynamical Casimir effect~\cite{qin2019emission,wilson2011observation}.
In fact, the dynamics described by $H_{-}^{\rm sq}$ can be interpreted in the laboratory
frame as mechanically induced two-photon hyper-Raman
scattering, i.e., an anti-Stokes scattering of a driving photon
pair, rather than a single photon, into a higher-energy mode by absorbing a phonon. 

Furthermore, by squeezing the cavity field, one can also eliminate the quantum backaction heating even in the unresolved sideband regime, such that ground-state cooling~\cite{lau2020ground} and mechanical squeezing~\cite{xiong2020strong} can be implemented.
Instead, squeezing the mechanical mode provides another way to enhance the optomechanical interaction. For examples, along this line, photon blockade~\cite{lemonde2016enhanced,yin2017nonlinear}, mechanical squeezing~\cite{zhang2018quantum}, superradiant quantum phase transition~\cite{lu2018entanglement}, and enhancing quadratic optomechanical interaction~\cite{gu2018enhanced} have been studied.

Recently, Ref.~\cite{arenz2020amplification} showed that by combining parametric amplification processes and dynamical-decoupling techniques, one can amplify the desired interaction, and at the same time suppress an undesired interaction, so as to speed up the dynamical evolution of the system.


\subsubsection{Other amplification mechanisms}

\paragraph{Parity-time symmetry} In contrast to conventional Hermitian Hamiltonians, non-Hermitian parity-time ($\mathcal{PT}$) symmetric Hamiltonians~\cite{bender1998real, el2018non, ozdemir2019parity} exhibit a phase transition from the unbroken to broken $\mathcal{PT}$ phases at an exceptional point, where the eigenvalues are changed from real to complex numbers. By coupling two different optical or microwave cavities, one with passive loss and the other with active gain, $\mathcal{PT}$-symmetric systems can be created~\cite{peng2014parity, hodaei2014parity, chang2014parity}. In such double-cavity systems, many counterintuitive aspects occur, particularly in the vicinity of exceptional points, which can be explored to implement a strong nonlinearity for cavity optomechanics. By manipulating the gain-to-loss ratio, a nonlinear regime for the intracavity-photon
intensity can emerge, such that the optical pressure and the mechanical gain are enhanced simultaneously, resulting in an ultralow-threshold phonon laser~\cite{jing2014pt, zhang2018phonon}. A $\mathcal{PT}$-symmetry-induced enhancement mechanism has also been used for demonstrating optomechanical chaos~\cite{lv2015pt}, optomechanically-induced transparency~\cite{jing2015optomechanically}, and high-precision metrology~\cite{liu2016metrology}.

\paragraph{Collective effects} By coupling a cavity field to an array of mechanical resonators, a collectively enhanced optomechanical interaction can be obtained~\cite{gartner2018integrated,piergentili2018two}. It has been predicted in Refs.~\cite{xuereb2012strong, xuereb2013collectively} that  the single-photon coupling strength of the cavity mode to a collective mechanical mode can scale as $N^{3/2}$, where $N$ is the number of mechanical resonators. Compared to the case of a single mechanical resonator, the resulting collective coupling can be made up to several orders of magnitude stronger, which, as a direct consequence, enables exploiting the long-range interactions between distant mechanical resonators. Optimal configurations for these optomechanical arrays are given in Ref.~\cite{newsom2020optimal}, such that the optomechanical interaction strength scales exponentially with $N$ before the saturation. Furthermore, the collective enhancement of the single-photon optomechanical interaction has also been shown in the cases where the collective motion of ensembles of ultracold atoms, serving as a mechanical resonator, is coupled to the cavity field~\cite{brennecke2008cavity,murch2008observation,brooks2012non,cold2013ritsch}.

\paragraph{Nonlinear Josephson junctions} For microwave-regime superconducting cavities, the mechanical motion, which modulates the capacitance of a cavity, can
couple to the cavity mode via the radiation-pressure interaction~\cite{regal2008measuring, teufel2011sideband, massel2011microwave}. When a Cooper-pair transistor is embedded inside a superconducting cavity~\cite{enhancing2014heikkila}, the single-photon optomechanical interaction strength can be enhanced by several orders of magnitude. Such a giant enhancement arises due to the presence of the nonlinearity of the Josephson junctions, and has been experimentally demonstrated in Ref.~\cite{pirkkalainen2015cavity}. Similarly, the nonlinearity of Coulomb blockade can also be used to significantly enhance the single-photon optomechanical interaction~\cite{blien2020quantum}. It has been suggested that, with a Cooper-pair transistor~\cite{rimberg2014cavity} or a Cooper-pair box~\cite{Manninen2022}, even the USC between cavity photons and a mechanical resonator can be reached. Moreover, an ultrastrong single-photon optomechanical interaction can also be obtained by embedding a dc superconducting quantum interference device (SQUID), with a suspended arm as a mechanical resonator, into a superconducting cavity~\cite{shevchuk2017strong, nation2016ultrastong, kounalakis2020flux}.


\subsection{Amplified photon-atom interactions in cavity quantum electrodynamics}
\label{Amplified photon-atom interactions in cavity quantum electrodynamics}

Cavity QED is the field of studying the fundamental interactions of atoms with photons in high-Q cavities, in which photons are confined for a long time and, thus, can repeatedly interact with the atoms. Cavity QED has been considered to be a promising platform to explore various applications ranging from fundamental tests of quantum theory to powerful quantum technologies. To understand this platform, let us first recall the interaction of a single-electron atom and a single-mode radiation field.

The atom can be considered as a dipole with a dipole momentum $\boldsymbol{d}=e\boldsymbol{r}$, where $e$ is the electronic charge and $\boldsymbol{r}$ is the position vector of the electron. Typically, the wavelength of an electromagnetic field is much larger than the size of the atom, such that the dipole approximation can be applied. Under this approximation, the interaction between the atom and the field is modelled by
\begin{equation}\label{eq:interaction under dipole approx}
	\mathcal{H}_{\rm int}=-\boldsymbol{d}\cdot\boldsymbol{E},
\end{equation}
where
\begin{equation}\label{eq:electronic field}
	\boldsymbol{E}=\boldsymbol{\epsilon}\mleft(a+a^{\dagger}\mright),
\end{equation}
represents the field at the location of the atomic nucleus, with $\boldsymbol{\epsilon}$ being a vector with the dimension of the electric field, and $a$ ($a^{\dagger}$) is the annihilation (creation) operator for the field mode. When expressed in terms of atomic energy eigenstates $\ket{i}$, the dipole momentum $\boldsymbol{d}$ is  
\begin{equation}\label{eq:atomic dipole}
	\boldsymbol{d}=\sum_{i,j}\boldsymbol{d}_{ij}\sigma_{ij},
\end{equation}
where $\boldsymbol{d}_{ij}=e\brakket{i}{\boldsymbol{r}}{j}$ is the dipole transition matrix element between the levels $\ket{i}$ and $\ket{j}$, and $\sigma_{ij}=\ketbra{i}{j}$ denotes the corresponding atomic transition operator. It follows, by substituting Eqs.~(\ref{eq:electronic field}) and (\ref{eq:atomic dipole}) into Eq.~(\ref{eq:interaction under dipole approx}), that
\begin{equation}
	\mathcal{H}_{\rm int}=-\sum_{i,j}\mleft(\boldsymbol{d}_{ij}\cdot\boldsymbol{\epsilon}\mright)\mleft(a+a^{\dagger}\mright)\sigma_{ij}.
\end{equation}
For a two-level atom with the ground state $\ket{g}$ and the excited state $\ket{e}$, $\boldsymbol{d}_{gg}=\boldsymbol{d}_{ee}=0$ because the electronic states $\ket{g}$ and $\ket{e}$ are of a definite parity, $g_{gg}=g_{ee}=0$. Moreover, we assume that $\boldsymbol{d}_{ge}=\boldsymbol{d}_{eg}$, such that the interaction Hamiltonian $\mathcal{H}_{\rm int}$ now becomes the well-known quantum Rabi interaction Hamiltonian, i.e.,
\begin{equation}\label{eq:rabi Hamiltonian 00}
	H_{\rm Rabi}^{\rm int}=g\mleft(a+a^{\dagger}\mright)\mleft(\sigma^{-}+\sigma^{+}\mright),
\end{equation}
where $g=-\boldsymbol{d}_{ge}\cdot\boldsymbol{\epsilon}=-\boldsymbol{d}_{eg}\cdot\boldsymbol{\epsilon}$ is the coupling strength, while $\sigma^{-}=\ketbra{g}{e}$ and $\sigma^{+}=\ketbra{e}{g}$ are the ladder operators of the atom. Equation~(\ref{eq:rabi Hamiltonian 00}) can describe the atom-field interaction, in particular, when the coupling strength $g$ is comparable to the atomic transition frequency or the cavity frequency. The quantum Rabi model is discussed in the following sections. Here, we focus on the case when $g$ is much smaller than these two characteristic frequencies. In this case, one can apply the RWA so as to neglect the counter-rotating components, yielding the JC interaction Hamiltonian~\cite{jaynes1963comparison}, i.e., 
\begin{equation}\label{eq:atomcavity_JC_coupling}
	H_{\rm Rabi}^{\rm int}\approx H_{\rm JC}^{\rm int}=g\mleft(a\sigma^{+}+a^{\dagger}\sigma^{-}\mright),
\end{equation}
describing an important and basic type of atom-field interactions. 

In cavity QED, the interaction $H_{\rm JC}^{\rm int}$ leads to a coherent exchange of energy between the atom and the cavity and, thus, has been widely used, particularly for QIP. However, owing to the presence of decoherence, exploiting such an atom-cavity system for QIP often requires the strong coupling regime, where the strength $g$ exceeds both the atomic spontaneous emission rate $\gamma$ and the cavity decay rate $\kappa$. Within the strong coupling regime, a single excitation can be coherently exchanged between the atom and the cavity before their coherence is lost. A typical parameter quantifying this property is the cooperativity, defined as
\begin{equation}
	C=\frac{g^{2}}{\kappa\gamma},
\end{equation} 
which shows that the ability, e.g., to process quantum information, increases as the coupling strength $g$. The first demonstration of the strong coupling with single atoms was reported in Ref.~\cite{meschede1985one}. Below we review several methods for amplifying the coupling strength $g$ and the cooperativity $C$. 


\subsubsection{Parametric amplification}
\label{subsubsec:Parametric amplification}

\begin{figure}[b!]
	\centering
	\includegraphics[width=\linewidth]{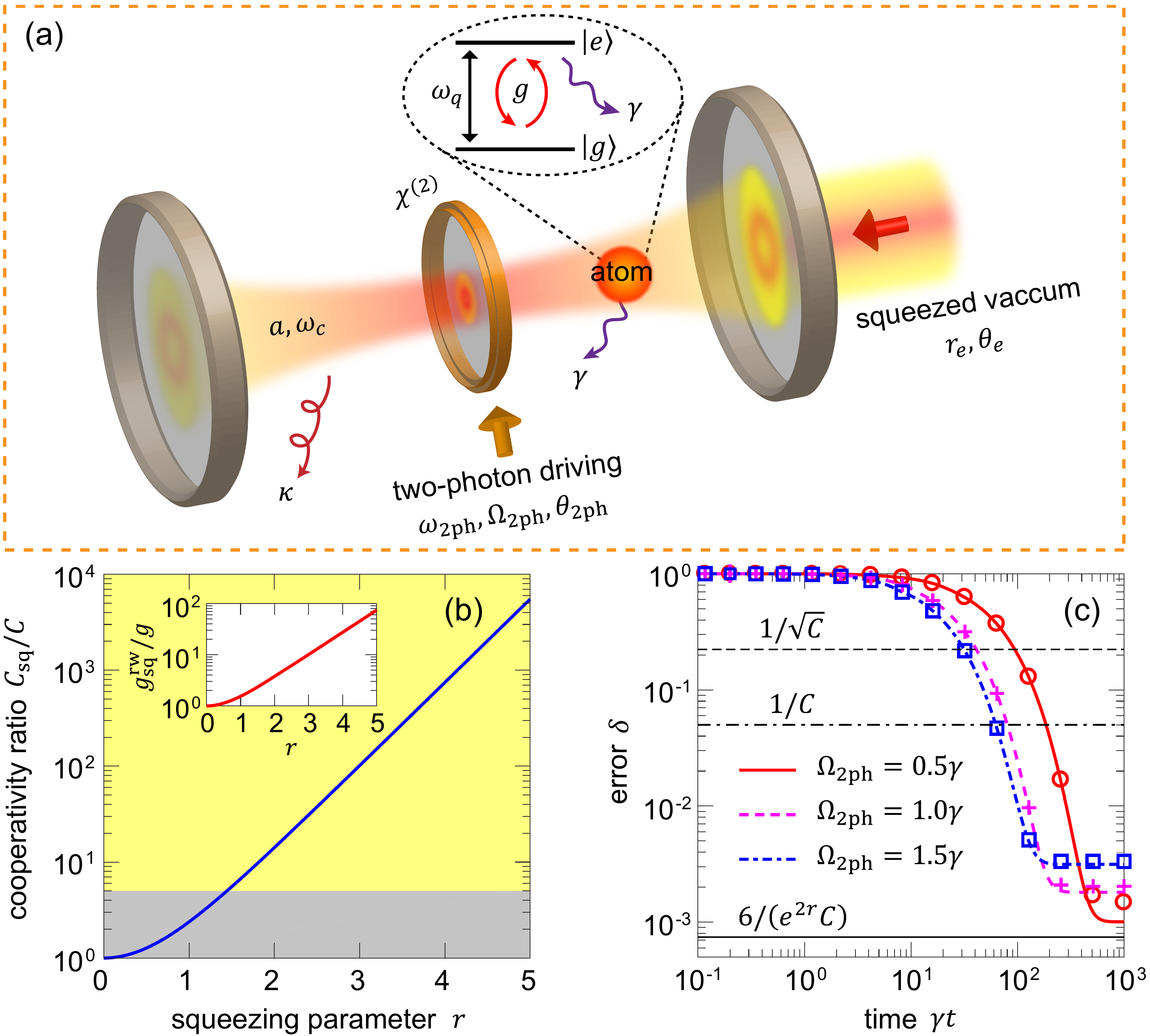}
	\caption{(a) Schematics of using parametric amplification to enhance the photon-atom interaction $g$ and thus the cooperativity $C$. The single-mode cavity contains a $\chi^{\mleft(2\mright)}$ nonlinear medium, which is strongly pumped at amplitude $\Omega_{\rm 2ph}$, frequency $\omega_{\rm 2ph}$, and phase $\theta_{\rm 2ph}$. A two-level atom, with the ground state $\ket{g}$ and the excited state $\ket{e}$, is coupled to the cavity mode with a strength $g$; $\omega_{\rm cav}$ and $\omega_{q}$ are the resonance frequencies of the cavity and the atom, respectively, $\kappa$ is the cavity decay rate, and $\gamma$ is the atomic spontaneous emission rate.
    (b) Exponentially enhanced cooperativity $C_{\rm sq}$ and coupling $g_{\rm sq}^{\rm rw}$ (inset) between the squeezed cavity mode and the atom. The gray and yellow shaded areas refer to the weak ($C_{\rm sq}<1$) and strong ($C_{\rm sq}>1$) coupling regimes under the assumption of $C=0.2$.
    (c) Exponentially enhanced entanglement of two three-level $\Lambda$-type atoms mediated by the squeezed cavity mode. With squeezing, the lower bound of the entanglement error $\delta=1-F$ is lowered to $\approx6/\mleft(e^{2r}C\mright)$, clearly far below $1/C$ and $1/\sqrt{C}$, which are the lower bounds imposed, respectively, by unitary and dissipative state preparation without squeezing. Curves show effective predictions and symbols indicate exact results. Panels (a)-(c) are adapted with permission from Ref.~\cite{qin2018exponentially}, W.~Qin et al., \href{https://link.aps.org/doi/10.1103/PhysRevLett.120.093601}{Phys.~Rev.~Lett.~\textbf{120}, 093601 (2018)}.}
    \label{fig_parametrically_enhanced_cavity_QED}
\end{figure}

The coupling in Eq.~(\ref{eq:atomcavity_JC_coupling}) essentially originates from the fluctuations of the electromagnetic vacuum~\cite{haroche2006book}, and thus amplifying these fluctuations via antisqueezing could induce an enhancement of the coupling strength $g$. The basic idea underlying this amplification method, essentially the same with the idea presented in Sec.~\ref{Parametric amplification}, is shown schematically in \figpanel{fig_parametrically_enhanced_cavity_QED}{a}.

When the cavity field is parametrically driven (i.e., is squeezed), as described by the Hamiltonian in Eq.~(\ref{eq:parametric driving Hamiltonian}), the coupling strength $g$ can be exponentially enhanced~\cite{qin2018exponentially,leroux2018enhancing,zeytinouglu2017engineering}. Upon substituting the Bogoliubov transformation in Eq.~(\ref{eq:Bogoliubov transformation}) into Eq.~(\ref{eq:atomcavity_JC_coupling}), the atom-cavity coupling Hamiltonian $H_{\rm JC}^{\rm int}$ is transformed to
\begin{equation}\label{eq:full Hamiltonian in squeezed mode}
	H_{\rm JC}^{\rm int}=g_{\rm sq}^{\rm rw}\mleft(a_{\rm sq}\sigma^{+}+{\rm H.c.}\mright)+g_{\rm sq}^{\rm cr}\mleft[\exp\mleft(i\theta_{\rm 2ph}\mright)a_{\rm sq}\sigma^{-}+{\rm H.c.}\mright],
\end{equation}  
where 
\begin{align}
g_{\rm sq}^{\rm rw}=\;g\cosh\mleft(r\mright),\\
g_{\rm sq}^{\rm cr}=\;-g\sinh\mleft(r\mright)
\end{align}
characterize the strengths of the rotating-wave and counter-rotating interactions, respectively, between the squeezed cavity mode $a_{\rm sq}$ and the atom. The counter-rotating interaction describes the processes not conserving excitation number, and thus can be neglected in the large-detuning regime $\mleft|g_{\rm sq}^{\prime}\mright|/\mleft(\omega_{\rm sq}+\Delta_{q}\mright)\ll1$. Here, $\Delta_{q}=\omega_{q}-\omega_{\rm 2ph}/2$, where $\omega_{q}$ is the atomic transition frequency. As a result, the interaction Hamiltonian $H_{\rm JC}^{\rm int}$ in Eq.~(\ref{eq:full Hamiltonian in squeezed mode}) becomes approximated by
\begin{equation}
	H_{\rm JC}^{\rm int, sq}=g_{\rm sq}^{\rm rw}\mleft(a_{\rm sq}\sigma^{+}+{\rm H.c.}\mright),
\end{equation} 
given in terms of the coupling strength $g_{\rm sq}^{\rm rw}$. Therefore, for $r\geq1$, an exponentially-enhanced atom-cavity coupling,
\begin{empheq}[box =\widecolourbox]{equation}\label{eq:atom_cavity_coupling_enhancement}
	g_{\rm sq}\approx\frac{1}{2}g\exp\mleft(r\mright),
\end{empheq}
can be predicted, as plotted in the inset of \figpanel{fig_parametrically_enhanced_cavity_QED}{b}. Since there are $\sim\exp\mleft(2r\mright)$ photons converted into a squeezed single-photon state, the exponential enhancement, given in Eq.~(\ref{eq:atom_cavity_coupling_enhancement}), can also be understood as a collective enhancement. This mechanism is to some degree similar to {\it a collective enhancement of the coupling of a single photon to an atomic ensemble}.

As demonstrated in \appref{app:Derivations of eliminating the noise induced by squeezing the cavity with a squeezed vacuum reservoir}, in order to suppress the noise induced by the parametric driving (i.e., by squeezing the cavity mode), one can couple a squeezed vacuum reservoir to the cavity mode. By properly tuning the relevant parameters, the dynamics of the atom-cavity system can be described by a standard master equation, 
\begin{equation}\label{eq:simplified master equation for atom cavity}
	\dot{\rho}=-i\mleft[H_{\rm JC}^{\rm sq},\rho\mright]+\kappa\mathcal{L}\mleft(a_{\rm sq}\mright)\rho+\gamma\mathcal{L}\mleft(\sigma^{-}\mright)\rho,
\end{equation}
where 
\begin{equation}
H_{\rm JC}^{\rm sq}=\omega_{\rm sq}a_{\rm sq}^{\dagger}a_{\rm sq}+\frac{1}{2}\Delta_{q}\sigma_{z}+H_{\rm JC}^{\rm int, sq}
\end{equation}
is the system Hamiltonian in terms of the squeezed mode $a_{\rm sq}$, and $\rho$ is the density matrix of the system. This master equation indicates that one can define an effective cooperativity 
\begin{equation}
	C_{\rm sq}=\frac{g_{\rm sq}^{2}}{\kappa\gamma},
\end{equation}
in the squeezed frame. It is found that, as shown in \figpanel{fig_parametrically_enhanced_cavity_QED}{b}, an exponential enhancement in the cooperativity for $r\geq1$ can occur, i.e., 
\begin{empheq}[box =\widecolourbox]{equation}\label{seq: enhancement in cooperativity}
	C_{\rm sq}\approx\frac{1}{4}C\exp\mleft(2r\mright).
\end{empheq}
A typical application of such a giant enhancement is to improve quantum entanglement or gate operations between separated atoms in the same cavity. Note that in this type of applications, the interaction between these atoms is mediated by a cavity mode in the squeezed vacuum, rather than in the usual vacuum. However, this is not a problem because the cavity mode, often serving as a quantum data bus, can be made effectively decoupled from the atoms of interest, or disentangled from the already entangled atoms at the end of the state preparation or the gate operation.

When entangled states of separated atoms are prepared in optical or microwave cavities, the state error $\delta=1-F$ scales as $\delta\propto1/\sqrt{C}$ for the preparation approaches based on unitary gates, and as $\propto1/C$ for dissipative state preparation. Here, $F$ is the fidelity between the actual and ideal states. Thus, the cooperativity enhancement given in Eq.~(\ref{seq: enhancement in cooperativity}), when applied to dissipative state preparation, can enable an exponential improvement in the state error~\cite{qin2018exponentially}, i.e., 
\begin{empheq}[box =\widecolourbox]{equation}
\delta\propto\frac{1}{e^{2r}C},
\end{empheq}
as shown in \figpanel{fig_parametrically_enhanced_cavity_QED}{c}. There, two three-level $\Lambda$-type atoms are considered, and the desired state is an entanglement of the ground states of these two atoms. The role of the parametrically enhanced coupling between the atoms and the squeezed cavity mode is to {\it exponentially suppress the transitions from the desired state to some decaying states} and, as a result, suppress the decay out of the desired state. Note here that the cavity degree of freedom is always effectively decoupled from the atoms and, thus, the state of the cavity mode (e.g., the vacuum or squeezed vacuum) is not important. 

So far, parametrically amplified photon-atom interactions, and thus mediated atom-atom interactions, have been widely studied. This mechanism has been explored to generate lasing into a squeezed cavity mode~\cite{Sanchez2021}. It has been also shown that when adiabatically eliminating the degree of freedom, an exponential enhancement of the dipole-dipole coupling between atoms can be observed~\cite{wang2019enhancement}. If the amplitude of the atom driving is modulated in time, then a fast and high-fidelity generation of steady-state entanglement~\cite{chen2019fast}, and similarly, a high-fidelity implementation of arbitrary phase gates~\cite{wang2020noise} can be achieved by a parametric driving of the cavity. In the case of two coupled cavities, it is possible to parametrically drive a single cavity, which is coupled to an atom, to enhance the coupling of this atom to another cavity~\cite{wang2019squeezing}. When, furthermore, a multiphoton coupling between the atom and the cavity is taken into account, their single-photon or two-photon coupling can be greatly enhanced even for a small squeezing parameter~\cite{wang2020enhancing}. Enhanced spin-phonon, and in turn spin-spin, interactions via squeezing a mechanical resonator have been theoretically demonstrated in hybrid systems in Ref.~\cite{li2020enhancing}. Very recently, a theoretical proposal that can amplify magnon-spin interactions via virtually-excited squeezed phonons has also been put forward in magnonics~\cite{Wang2023,nori2023squeezed}. 


\subsubsection{Collective amplification}

In Sec.~\ref{subsubsec:Parametric amplification}, we showed that a detuned parametric driving of a cavity enables the coupling between an atom and the squeezed cavity mode to be enhanced exponentially. Because exponentially many photons are converted into a single-photon state of the squeezed cavity mode, this parametric enhancement of the atom-field coupling can thus be understood as originating from the coupling of a single atom to many photons. 

In this section, we introduce an opposite enhancement mechanism, which is based on the coupling of a single photon to an ensemble containing many atoms or spins. To proceed, let us assume that the ensemble contains $N$ identical two-level atoms. The coupling between the ensemble and a cavity mode is described by the interaction Hamiltonian
\begin{equation}
	H_{\rm ens}^{\rm int} = \sum_{j=1}^{N}g_{j}\mleft(a^{\dagger}\sigma_{j}^{-}+a\sigma_{j}^{+}\mright),
\end{equation}
where $g_{j}$ is the single atom-cavity coupling strength, and $\sigma_{j}^{\pm}$ are the ladder operators of the $j$th atom. The atomic
ensemble can be considered as a large collective pseudospin with $S=N/2$, such that it can be described with collective spin operators
\begin{equation}
	S_{\pm}=\;\frac{1}{g}\sum_{j=1}^{N}g_{j}\sigma^{\pm}_{j},\quad {\rm and} \quad 
	S_{z}=\;\frac{1}{2}\sum_{j=1}^{N}\sigma_{j}^{z},
\end{equation}
where $g^{2}=\frac{1}{N}\sum_{j=1}^{N}g_{j}^{2}$. In the special case where $g_{j}$ is a constant, i.e., $g_{j}=g$, one has
$S_{\pm}=\sum_{j=1}^{N}\sigma^{\pm}_{j}$. The interaction Hamiltonian $H_{\rm ens}^{\rm int}$ is accordingly transformed into
\begin{equation}
	H_{\rm ens}^{\rm int}=g\mleft(aS_{+}+a^{\dagger}S_{-}\mright).
\end{equation}

Let us now apply the Holstein-Primakoff transformation~\cite{holstein1940field}
\begin{equation}
	S_{-}=\;\sqrt{N-s^{\dagger}s}s, \quad
	S_{+}=\;s^{\dagger}\sqrt{N-s^{\dagger}s}, \quad {\rm and} \quad
	S_{z}=-N/2+s^{\dagger}s,
\end{equation}
where $s$ and $s^{\dagger}$ are the bosonic annihilation and creation operators, which satisfy the commutation relation $\mleft[s,s^{\dagger}\mright]=1$.
In the low-excitation regime, where the average number of excited atoms is much smaller than the total number of atoms (i.e., $\langle s^{\dagger}s\rangle\ll N$), the operators $S_{-}$ and $S_{+}$ are further simplified to 
\begin{align}
	S_{-}\approx\;\sqrt{N}s \quad {\rm and} \quad S_{+}\approx\;\sqrt{N}s^{\dagger},
\end{align}
respectively. It is seen that the collective spin behaves as a quantum harmonic oscillator. In this case, the Hamiltonian $H_{\rm ens}^{\rm int}$ becomes
\begin{equation}
	H_{\rm ens}^{\rm int}=g_{\rm col}\mleft(as^{\dagger}+a^{\dagger}s\mright),
\end{equation}
where
\begin{empheq}[box =\widecolourbox]{equation}\label{seq: collective enhancement in coupling}
	g_{\rm col}=\sqrt{N}g.
\end{empheq}
This means that the collective coupling $g_{\rm col}$ is enhanced by the square root of the number of atoms, compared to the single-atom coupling strength $g$.
Such a {\it collective enhancement} comes at the expense of reducing the nonlinearity of the coupled atom-cavity system. 

The collective strong coupling between light and matter was first demonstrated experimentally with an ensemble of Rydberg atoms in Ref.~\cite{kaluzny1983observation}, and has been a fundamental building block of quantum repeaters for long-distance quantum communication~\cite{duan2001long, kuzmich2003generation, van2003atomic, chou2005measurement, chaneliere2005storage, yuan2008experimental, hammerer2010quantum, sangouard2011quantum, grezes2016towards, jing2019entanglement, yu2020entanglement, pu2021experimental}. Moreover, a collective enhancement of light-matter interactions allows one to generate various nonclassical states in large ensembles, e.g., spin-squeezed states~\cite{kitagawa1993squeezed, hald1999spin, sorensen2002entangling, leroux2010implementation, schleier2010states, ma2011quantum, dalla2013dissipative, facilitating2020tucker, macri2020spin, qin2020strong, groszkowski2020heisenberg,liu2023generation} and atomic Schr\"{o}dinger cat states~\cite{agarwal1997atomic, generating2003massar, qin2021generating}. Thus, it also forms an essential ingredient of quantum metrology for high-precision measurements~\cite{pezze2018quantum}.

\begin{figure}[t]
	\centering
	\includegraphics[width=\linewidth]{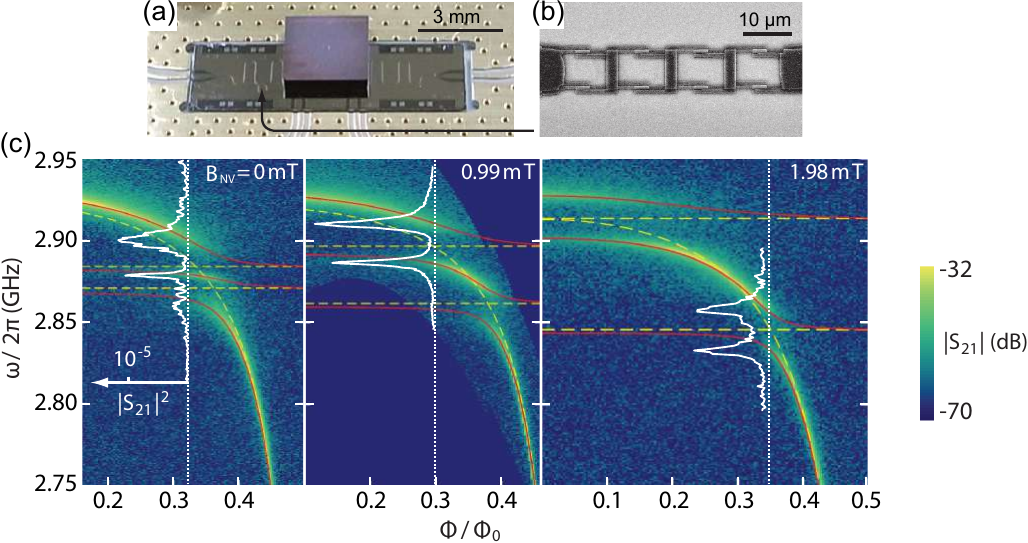}
	\caption{Experimental demonstration of collectively enhanced photon-atom interaction.
    (a)~Photograph of the experimental sample: a nitrogen-vacancy (NV) electronic-spin ensemble is placed on top of a superconducting coplanar waveguide resonator, whose resonance frequency is tuned by modulating the magnetic flux $\Phi$ threading a SQUID array shown in (b).
    (c) Transmission spectrum $\mleft|S_{21}\mleft(\omega\mright)\mright|$ of the resonator. Two avoided crossings, induced by the strong coupling of the resonator mode to two microwave transitions between the ground-state spin-triplet sublevels of NV centers, can be observed clearly. Here, $B_{\rm NV}$ is an external magnetic field applied to induce the spin Zeeman splitting and $\Phi_{0}=h/2e$ is the flux quantum. Panels (a)-(c) are adapted with permission from Ref.~\cite{kubo2010strong}, Y.~Kubo et al., \href{https://link.aps.org/doi/10.1103/PhysRevLett.105.140502}{Phys.~Rev.~Lett.~\textbf{105}, 140502 (2010)}.}\label{fig_collectively_enhanced_cavity_QED}
\end{figure}

Recently, much attention has been focused on coupling nitrogen-vacancy (NV) electronic-spin ensembles to superconducting circuits to simultaneously exploit the complementary advantages of these two different physical systems (e.g., strong nonlinearity and ease of design of superconducting circuits~\cite{you2005superconducting,schoelkopf2008wiring,clarke2008superconducting,you2011atomic,gu2017microwave,Krantz2019,kockum2019quantum,kjaergaard2020superconducting,circuit2021blais}, and extremely long coherence times of NV spins~\cite{bar2013solid}), therefore building an important type of hybrid quantum system~\cite{xiang2013hybrid, clerk2020hybrid}. However, the coupling of a single NV spin to a superconducting cavity usually is $g/2\pi \approx \unit[10]{Hz}$~\cite{verdu2009strong}, which is too weak to allow for a coherent exchange of quantum information between the NV spin and the cavity field. However, the NV spin ensembles typically contains $N \approx 10^{12}$ NV spins, and therefore a collective coupling of $g_{\rm col}/2\pi \approx \unit[10]{MHz}$ can be achieved according to Eq.~(\ref{seq: collective enhancement in coupling}). Such a collective enhancement has been widely demonstrated experimentally~\cite{kubo2010strong, amsuss2011cavity, kubo2011hybrid, PhysRevA.85.012333, putz2014protecting, grezes2014multimode, astner2017coherent}. The most common setup of such experiments is sketched in \figpanel{fig_collectively_enhanced_cavity_QED}{a}~\cite{kubo2010strong}. A diamond single crystal containing a number of NV centers is placed on top of a superconducting coplanar resonator, whose resonance frequency is tunable with an array of four SQUIDs [see \figpanel{fig_collectively_enhanced_cavity_QED}{b}]. The collectively enhanced ensemble-cavity coupling can, as seen from \figpanel{fig_collectively_enhanced_cavity_QED}{c}, cause two clear anticrossings, an indication that the ensemble-cavity system is in the strong coupling regime. Note that here these two anticrossings arise from the fact that there are two distinct microwave transitions between the ground-state spin-triplet sublevels of NV centers, and both can be coupled to the resonator mode. In addition, the collectively enhanced coupling of ensembles of ion spins~\cite{PhysRevLett.105.140501, PhysRevLett.110.157001, wisby2014coupling} or $^{87}$Rb atoms~\cite{hattermann2017coupling} to a superconducting cavity has also been reported in experiments.

The dissipative dynamics of the atomic ensemble can be described with the standard Lindblad operator 
\begin{equation}
	\frac{\gamma}{2}\sum_{j=1}^{N}\mathcal{L}\mleft(\sigma_{j}^{-}\mright)\rho\,=\frac{\gamma}{2}\sum_{j=1}^{N}\mleft(2\sigma_{j}^{-}\rho\sigma_{j}^{+}-\sigma_{j}^{+}\sigma_{j}^{-}\rho-\rho \sigma_{j}^{+}\sigma_{j}^{-}\mright).
\end{equation}
Here, we have assumed the atoms to be fully independent. It follows, on performing the Fourier transform
\begin{equation}
	\widetilde{\sigma}_{k}^{-}=\frac{1}{\sqrt{N}}\sum_{j}\exp\mleft(-ikj\mright)\sigma_{j}^{-}
\end{equation}
and then using the relation $\sqrt{N}\widetilde{\sigma}_{k=0}^{\pm}=S_{\pm}$, that 
\begin{equation}\label{seq:atomic emission in momentum space}
	\sum_{j}\mathcal{L}\mleft(\sigma_{j}^{-}\mright)\rho=\frac{1}{N}\mathcal{L}\mleft(S_{-}\mright)\rho+\sum_{k\neq0}\mathcal{L}\mleft(\widetilde{\sigma}_{k}^{-}\mright)\rho,
\end{equation}
where the first and second terms on the right-hand side describe the dissipative processes of the zero and nonzero momentum modes, respectively. If the coherent dynamics only involves the zero-momentum ($k=0$) mode, then one can focus on only that mode~\cite{gelhausen2017many,shammah2018open,macri2020spin}; that is,
\begin{equation}\label{seq:zero momentum mode dissipation}
	\sum_{j}\mathcal{L}\mleft(\sigma_{j}^{-}\mright)\rho=\frac{1}{N}\mathcal{L}\mleft(S_{-}\mright)\rho.
\end{equation}
This is also valid in the steady-state limit or the long-time limit, because the nonzero momentum modes in Eq.~(\ref{seq:atomic emission in momentum space}) only decay. In particular, such a reduction can exactly describe the dissipative dynamics of an atomic ensemble initially in the ground state. Therefore in the low-excitation regime, 
the dissipative dynamics of the atomic ensemble is determined by
\begin{equation}
	\sum_{j=1}^{N}\mathcal{L}\mleft(\sigma_{j}^{-}\mright)\rho=\mathcal{L}\mleft(s\mright)\rho,
\end{equation}
which corresponds to a damped quantum harmonic oscillator. It is found from that the local dissipation can be described by the collective dissipation in some cases, such that the collective cooperativity is given by
\begin{empheq}[box =\widecolourbox]{equation}\label{seq: collective enhancement in cooperativity}
	C_{\rm col}=NC,
\end{empheq}
which increases proportionally to the number, $N$, of atoms in the ensemble. It has been shown that even an ensemble weakly coupled to the cavity can induce strong coupling of a single atom to the cavity~\cite{schutz2020ensemble}. 

The collective cooperativity $C_{\rm col}$ in Eq.~(\ref{seq: collective enhancement in cooperativity}) is only valid for ensembles of independent atoms. When well separated, the atoms can be considered as independent. But when the spacing of these atoms is very small, their dipole-dipole interaction and their collective dissipative dynamics need to be taken into account. In this case, it has been suggested that the collective cooperativity can be further enhanced~\cite{plankensteiner2017cavity,plankensteiner2019enhanced}, with an effective collective cooperativity given by
\begin{equation}
	C_{\rm col}^{\rm eff}=\frac{\mathbf{G}^\intercal\mathbf{G}}{\kappa\gamma_{\rm eff}},
\end{equation}
where $\mathbf{G}=\mleft(g_{1},\ldots,g_{N}\mright)^\intercal$ and $\gamma_{\rm eff}$ is an effective collective decay rate. By optimizing the amplitude profile of the transverse cavity field, the maximum value of $\mathbf{G}^\intercal\mathbf{G}$ can be obtained. At the same time, for collective subradiant states, the decay rate $\gamma_{\rm eff}$ is strongly suppressed such that $\gamma_{\rm eff}\ll\gamma$. Thus, the effective collective cooperativity $C_{\rm col}^{\rm eff}$ is significantly enhanced. Compared to $C_{\rm col}$ in the cases of independent atoms, which scales linearly with the number of atoms $N$, this subradiant enhancement results in a nonlinear scaling of $C_{\rm col}^{\rm eff}$ with $N$ (e.g., $\propto N^{4}$). 


\subsubsection{Other amplification mechanisms}
\paragraph{Plasmonic cavities} 

Usually, improving both the quality factor $Q$ and the mode volume $V$ for the same cavity remains challenging, due to the diffraction limit. This means that it is difficult to achieve simultaneously a small cavity decay rate $\kappa$ and a strong photon-atom interaction $g$, which limits the cooperativity $C$. Plasmonic particles (i.e., metal nanoparticles) driven by an external field can produce intense localized fields near them~\cite{tame2013quantum}. Many cavities have already utilized such a mechanism to localize an electromagnetic field into a region of nanometer scale and, as a result, to significantly decrease the mode volume of these cavities~\cite{seo2009full, kuttge2010ultrasmall, hu2016design, santhosh2016vacuum, chikkaraddy2016single}. This yields an enhancement in the coupling of an atom to the cavity field when the atom is placed closed to the plasmonic particles, and in turn drives the system to the strong coupling regime. For example, for whispering-gallery-mode cavities with an ultrahigh $Q$ but relatively large $V$, the resulting cooperativity is enhanced by approximately two orders of magnitude compared to that obtained in the case of a bare cavity~\cite{xiao2012strongly, peng2017enhancing}. This type of enhancement has been used, e.g., to generate  indistinguishable single photons~\cite{wein2018feasibility} and quantum entanglement~\cite{liu2013generation}.

\paragraph{Hybrid cavities} In addition, it has been demonstrated that coupling two different cavities, one with a low $Q$ (i.e., with a large cavity loss rate) and the other with a high $Q$ (i.e., with a small cavity loss rate) can effectively realize a high-$Q$ and small-$V$ cavity~\cite{coherent2014liu}. The atom is assumed to be coupled to the low-$Q$ cavity. Due to a large detuning and, thus, a low photon occupation, the low-$Q$ cavity mode can be eliminated adiabatically, yielding an effective interaction between the atom and the high-$Q$ cavity mode. This effective atom-cavity system combines the respective advantages of these two cavities (i.e., a high $Q$ and a small $V$). The strong coupling regime, characterized by $C>1$, can then be reached, as long as the detuning of the low-$Q$ cavity is large enough. 


\subsection{Amplified Kerr-type light-matter interactions via quadrature squeezing}
\label{Amplified Kerr-type light-matter interaction via quadrature squeezing}

Phase shifts induced by Kerr-type effects are typically very small when dealing with single photons~\cite{Venkataraman2013}. A number of experiments have demonstrated the feasibility to generate and observe cross-Kerr phase shifts even on the order of a few tens of degrees per photon, which can enable performing at least limited quantum logic gates. These include experiments based on cavity QED using single atoms~\cite{Turchette1995}, quantum dots~\cite{Fushman2008}, or atomic ensembles~\cite{Beck2016}, circuit QED~\cite{Hoi2013}, and photonics using optical fibers~\cite{Matsuda2009}. Anyway, much larger phase shifts at the single-photon level are in high demand in order to harness the full power of the Kerr or Kerr-type effects for quantum computing.

As demonstrated in Ref.~\cite{bartkowiak2014quantum}, it is possible, at least theoretically, to boost a cross-Kerr phase shift to an
arbitrary value by employing sequentially either one- or two-mode quadrature-squeezing operations. Utilizing such Kerr amplification
techniques may prove valuable in implementing quantum nondemolition (QND) measurements~\cite{Imoto1985} or practical quantum-optical entangling gates, like a deterministic Fredkin gate~\cite{Milburn1989} or a conditional phase (CPHASE)
gate~\cite{Chuang1995,Brod2016}, which rely on giant Kerr nonlinear interactions at the level of individual photons. 

There
is yet another fundamental advantage of the approach proposed in Ref.~\cite{bartkowiak2014quantum}; namely, this method shows the
feasibility of enhancing (at least some types of) higher-order nonlinear interactions by applying lower-order nonlinear effects.
Indeed, the Kerr effect generated in an nonlinear medium is proportional to its third-order susceptibility $\chi^{(3)}$, while
squeezing generation depends on the second-order susceptibility $\chi^{(2)}$.

In this section, we recall the amplification method and circuits proposed in Ref.~\cite{bartkowiak2014quantum}. The method was developed based on a vector coherent-state theory and applied physical operations that adhere to the commutation relations associated with the SU(1,1) generators. Thus, let us consider cross-Kerr-type nonlinear interaction between a two-level atom $a$ and an optical mode $b$, as described by the effective Hamiltonian ($\hbar=1$)
\begin{equation}
	H^{ab}_{\rm Kerr}= g \sigma_+ \sigma_- b^\dag b = g n_a n_b,
	\label{HamiltonianKerr}
\end{equation}
where ${g}$ is the strength of the Kerr interaction, which is proportional to the third-order susceptibility, $\chi^{(3)}$, of the nonlinear medium; $\sigma_+$ ($\sigma_-$) is the atomic raising (lowering) operator; $n_{a}=\sigma_+ \sigma_-$ is the atomic excitation number operator; $b$ ($b^{\dagger}$) is the annihilation (creation) operator for the optical mode $b$; and $n_{b}= b^{\dagger} b$ is the photon number operator in the mode $b$.

\begin{figure}
	\centering
	\includegraphics[width=9cm]{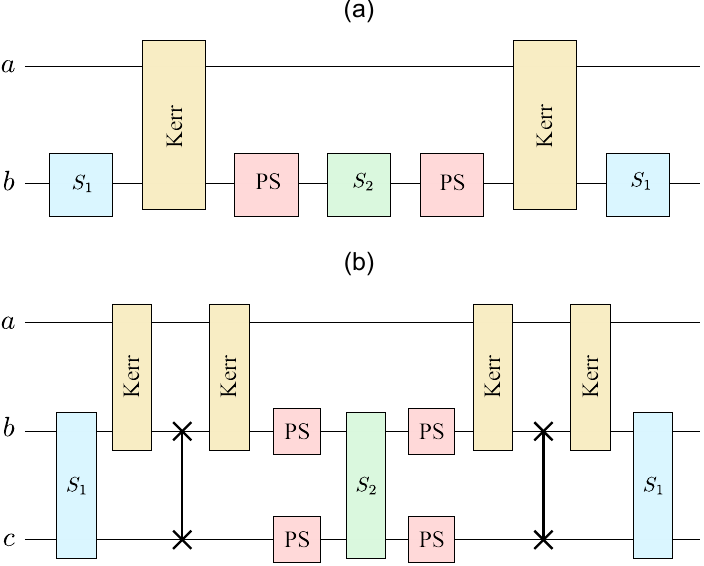}
	\caption{Circuits of Ref.~\cite{bartkowiak2014quantum} for amplifying the cross-Kerr interaction between a two-level atom (in path $a$) and	optical modes (in paths $b$ and $c$) by sequentially applying quadrature squeezing: 
    (a) single-mode squeezing  $S_{1} = S_b(\theta_1)$ and $S_{2}= S_b(\theta_2)$, according to Eq.~(\ref{KerrAmpGates1}); or 
    (b) two-mode squeezing $S_{1} = S_{bc}(\theta_1)$ and $S_{2}= S_{bc}(\theta_2)$, according to Eq.~(\ref{KerrAmpGates2}), separated with proper phase shifts (PSs). The SWAP gates are represented by the lines connecting the $\times$ symbols.}
	\label{fig:KerrCircuits}
\end{figure}

The main purpose of this method is to amplify $g$ by applying either standard single-mode quadrature-squeezing operators ($k = 1, 2$)
\begin{equation}
	S_k\equiv S_{b}(\theta_k) = \exp \mleft[ -\frac{\theta_k}{2} \mleft( b^2 - b^{\dag 2} \mright) \mright]
	\label{squeezing1}
\end{equation}
to the mode $b$, or the two-mode squeezing operators
\begin{equation}
	S_{bc}(\theta_1) = \exp \mleft[ -\theta_1 \mleft( b c - b^\dag c^\dag \mright) \mright]
    \label{squeezing2}
\end{equation}
to the mode $b$ and an auxiliary mode (say $c$). In these definitions, $\theta_k$ is a real squeezing parameter; the extra
minus indicates that the squeezing angle is $\pi$; the subscripts $b$ and $c$ indicate the modes on which the squeezing operations
are applied; and $c$ ($c^{\dagger}$) is the annihilation (creation) operator for the mode $c$. The circuit shown in \figpanel{fig:KerrCircuits}{a} enables enhancing the Kerr nonlinearity according to the relations
\begin{equation}
	\overbrace{S_b(\theta_1)}^{\rm squeezing}
    \overbrace{e^{i \frac{g}{2} (2 n_a n_b - n_b)}}^{\rm Kerr\&PS}
    \overbrace{S_b(\theta_2)}^{\rm squeezing}
    \overbrace{e^{i\frac{g}{2} (2 n_a n_b - n_b)}}^{\rm Kerr\&PS}
    \overbrace{S_b(\theta_1)}^{\rm squeezing}
	=
    \overbrace{e^{\frac{i}{2} (g_\gamma - g) (2 n_a - 1)}}^{\rm PS}
	\overbrace{e^{i g_\gamma (2 n_a n_b - n_b)}}^{\rm amplified\, Kerr\&PS},
    \label{KerrAmpGates1}
\end{equation}
where PS stands for a linear phase shift (or shifter) in the mode $b$. The parameters $g$ and $\theta_1$ determine the squeezing parameter
\begin{equation}
	\theta_2 = {\rm arctanh} \mleft[ - \cos g \tanh \mleft( 2 \theta_1 \mright) \mright]
	\label{theta2}
\end{equation}
in the gate $S_2$ in \figpanel{fig:KerrCircuits}{a}, and the amplified Kerr interaction strength
\begin{equation}
	g_\gamma = \arctan \mleft[ \tan g \cosh \mleft( 2 \theta_1 \mright) \mright].
    \label{gamma}
\end{equation}

By defining the Kerr unitary operator as $K(\Delta\phi) = \exp[i\Delta\phi\, n_a n_b]$, the left-hand side of \eqref{KerrAmpGates1} can be rewritten in a form with the operations clearly corresponding to the gates in \figpanel{fig:KerrCircuits}{a}, i.e.,
\begin{equation}
	K(\Delta\phi_{\rm amp}) = P' S_1 K(\Delta\phi_{\rm in}) P S_2 P K(\Delta\phi_{\rm in}) S_1,
	\label{circuit1}
\end{equation}
where $P = \exp(-i\beta)$, with $\beta = g n_b/2$; and $P' = \exp(-i\beta')$, with $\beta' = (g_\gamma - g) (n_a - 1/2) - g_\gamma n_b$, are linear phase shifts. For brevity, the less important phase shift $P'$ is not shown in \figpanel{fig:KerrCircuits}{a}. Moreover, $\Delta\phi_{\rm in} = g$ and $\Delta\phi_{\rm amp} = 2 g_\gamma$ are, respectively, the initial and amplified Kerr interaction strengths. The cross-Kerr amplification factor can be defined by the ratio of these final and initial phase shifts,
\begin{equation}
	\kappa_{\rm amp} = \frac{\Delta\phi_{\rm amp}}{\Delta\phi_{\rm in}} = \frac{2g_\gamma}{g}.
    \label{r1}
\end{equation}

Let us assume typical experimental conditions, for which $\Delta\phi_{\rm in}\ll 1$ and also the squeezing parameter $\theta_1$ is relatively small, to guarantee that $\tan(2\Delta\phi_{\rm in})\cosh (2\theta_1)\ll 1$. Then it is easy to show, by expanding $g_\gamma$ in \eqref{gamma} in power series of $g = 2 \Delta\phi_{\rm in}$, that the method enables {\it increasing the initial Kerr interaction strength by an exponential} factor determined by the squeezing parameter $\theta_1$, i.e.,
\begin{equation}
	\Delta\phi_{\rm amp} \approx 4\Delta\phi_{\rm in}\cosh(2\theta_1),
\end{equation}
and so
\begin{equation}
	\kappa_{\rm amp} \approx 2\cosh(2\theta_1),
    \label{r2}
\end{equation}
which is independent of the initial Kerr interaction strength $\Delta\phi_{\rm in}$.

The single-optical-mode amplification method, described by \eqref{KerrAmpGates1}, was generalized in Ref.~\cite{bartkowiak2014quantum} to the two-optical-mode case. Specifically, the cross-Kerr interaction, given in \eqref{HamiltonianKerr}, is also assumed between the atom $a$ and another optical mode $c$, i.e.,
\begin{equation}
	H^{ac}_{\rm Kerr} = g \sigma_+ \sigma_- c^\dag c = g n_a n_c,
	\label{HamiltonianKerrPrime}
\end{equation}
where $n_c = c^\dag c$ is the photon-number operator in the mode $c$. This generalized amplification method is described by the relation
\begin{eqnarray}
	&\overbrace{S_{bc}(\theta_1)}^{\rm squeezing}
	\overbrace{e^{i\frac{g}{2} (2 n_a n_b - n_b)}}^{\rm Kerr_{ab}\&PS}
    \overbrace{e^{i\frac{g}{2} (2 n_a n_c - n_c)}}^{\rm Kerr_{ac}\&PS}
	\overbrace{S_{bc}(\theta_2)}^{\rm squeezing}
	\overbrace{e^{i\frac{g}{2} (2 n_a n_b - n_b)}}^{\rm Kerr_{ab}\&PS}
    \overbrace{e^{i\frac{g}{2} (2 n_a n_c - n_c)}}^{\rm Kerr_{ac}\&PS}
	\overbrace{S_{bc}(\theta_1)}^{\rm squeezing}
    \nonumber \\
	&=
    \underbrace{e^{i (g_\gamma-g) (2 \hat n_a - 1)}}_{\rm PS}
	\underbrace{e^{i g_\gamma (2 n_a n_b - n_b)} e^{i g_\gamma (2 n_a n_c - n_c)}}_{\rm amplified\ Kerr\&PS},
    \label{KerrAmpGates2}
\end{eqnarray}
where the squeezing parameter $\theta_2$ is given by \eqref{theta2}, while the Kerr amplification factor $g_\gamma$ is given by \eqref{gamma} multiplied by a factor two.

The operations on the left-hand side of \eqref{KerrAmpGates2} correspond to the gates shown in \figpanel{fig:KerrCircuits}{b}. This correspondence can be even better seen by rewriting \eqref{KerrAmpGates2} in the form of \eqref{circuit1}, but for the combined Kerr operators and phase shifts, as defined by (for $j=b,c$)
\begin{eqnarray}
	K = K_{ab} K_{ac}, \quad &{\rm where}& \quad K_{aj} = \exp(i g n_a n_j), \\
	P = P_b P_c, \quad &{\rm where}& \quad P_j = \exp[-i (g/2) n_j], \\
	P' = \exp(-i\beta'), \quad &{\rm where}& \quad \beta' = 2 (g_\gamma-g) (n_a - 1/2) - g_\gamma (n_b + n_c),
	\label{KPP}
\end{eqnarray}
respectively, and for the two-mode squeezing operators instead of the single-mode ones. Note that the phase shift $P'$, which does not affect the Kerr amplification factor, is, for brevity, not shown in \figpanel{fig:KerrCircuits}{b}, analogously as it was omitted in \figpanel{fig:KerrCircuits}{a}. It might be surprising that \eqref{KerrAmpGates2} contains the terms describing the Kerr interactions of the atom $a$ with both optical modes, while the circuit shown in \figpanel{fig:KerrCircuits}{b} includes only the Kerr gates between the atom $a$ and the mode $b$. This modification of the circuit is explained by the relation
\begin{equation}
	K = K_{ab} U^{bc}_{\rm SWAP} K_{ab} U^{bc}_{\rm SWAP},
	\label{Kerr_swap}
\end{equation}
using the SWAP gate, $U^{bc}_{\rm SWAP}$, between the optical modes $b$ and $c$.


\begin{figure}[t]
	\centering
	\includegraphics[width=\linewidth]{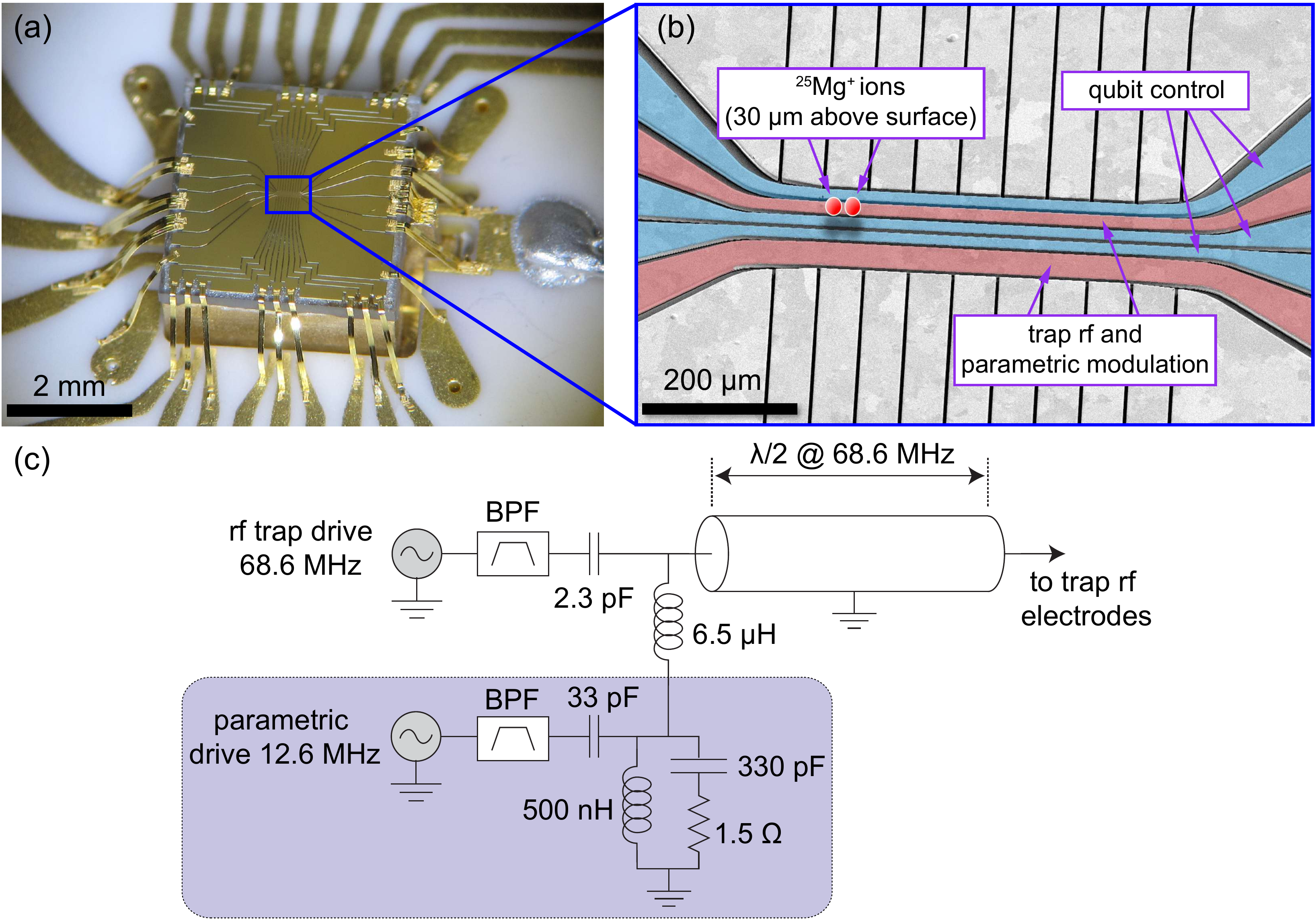}
	\caption{Experimental setup and circuits of Ref.~\cite{burd2020quantum} for demonstrating parametrically amplified light-matter interactions.
    (a) Photograph of the surface-electrode ion trap.
    (b) False-color SEM image of the central region, boxed in (a), of the ion trap. Two $^{25}{\rm Mg}^{+}$ ions are trapped $\sim\unit[30]{\mu m}$ above the trap surface. The blue electrodes provide the qubit control, while the red electrodes provide the trapping rf and the parametric modulation for the motional mode.
    (c) Circuit diagram for parametric modulation of the ion motional frequency. BPF refers to a band-pass filter. Panels (a)-(c) were provided by Dr.~Daniel H.~Slichter from NIST.}
    \label{fig-trapped-ions-setup}
\end{figure}

\subsection{Experimental demonstrations of parametrically amplified light-matter interactions}
\label{Experimental demonstrations of parametrically amplified light-matter interactions}

In 2021, an experimental demonstration of parametrically squeezing a bosonic mode to enhance the generation of quantum entanglement between qubits was reported in a trapped-ion system in Ref.~\cite{burd2020quantum} by a group at the National Institute of Standards and Technology (NIST). It has further been experimentally shown, by the same group in Ref.~\cite{burd2023experimental} in 2023, that the use of parametric squeezing can realize the amplification of the system Hamiltonian, even without the precise knowledge of this Hamiltonian~\cite{arenz2020amplification}. Recently, the parametrically amplified dispersive interaction between an atom and a squeezed microwave cavity mode was also demonstrated in a superconducting-circuit experiment in Ref.~\cite{villiers2022dynamically}.  Below, we introduce these experiments in more detail.

\subsubsection{Trapped ions}

We start with the trapped-ion experiment reported in Ref.~\cite{burd2020quantum}; the experimental setup is shown in Fig.~\ref{fig-trapped-ions-setup}. The experiment used two $^{25}{\rm Mg}^{+}$ ions, which were trapped $\sim\unit[30]{\mu m}$ above a linear surface-electrode ion trap [see Figs.~\figpanelNoPrefix{fig-trapped-ions-setup}{a} and \figpanelNoPrefix{fig-trapped-ions-setup}{b}]. Furthermore, an out-of-phase radial motional mode, shared by the two $^{25}{\rm Mg}^{+}$ ions and cooled to near the ground state using resolved-sideband cooling from oscillating magnetic field gradients, was used as a bosonic harmonic oscillator degree of freedom, and therefore played the role of light in light-matter interactions. Its parametric modulation, and in turn its squeezing, was implemented by applying an oscillating potential at or close to twice the motional frequency to the rf electrodes of the ion trap [see Figs.~\figpanelNoPrefix{fig-trapped-ions-setup}{b} and \figpanelNoPrefix{fig-trapped-ions-setup}{c}]. The states $\ket{\downarrow}\equiv\ket{F=3,m_{F}=1}$ and $\ket{\uparrow}\equiv\ket{F=2,m_{F}=1}$ in the $^{2}S_{1/2}$ electronic ground state hyperfine manifold of the trapped $^{25}{\rm Mg}^{+}$ ions were used as qubit states. Here, $F$ refers to the total angular momentum, and $m_{F}$ refers to the projection of the total angular momentum along a quantization axis defined by an external magnetic field.

\begin{figure}[t]
	\centering
	\includegraphics[width=\linewidth]{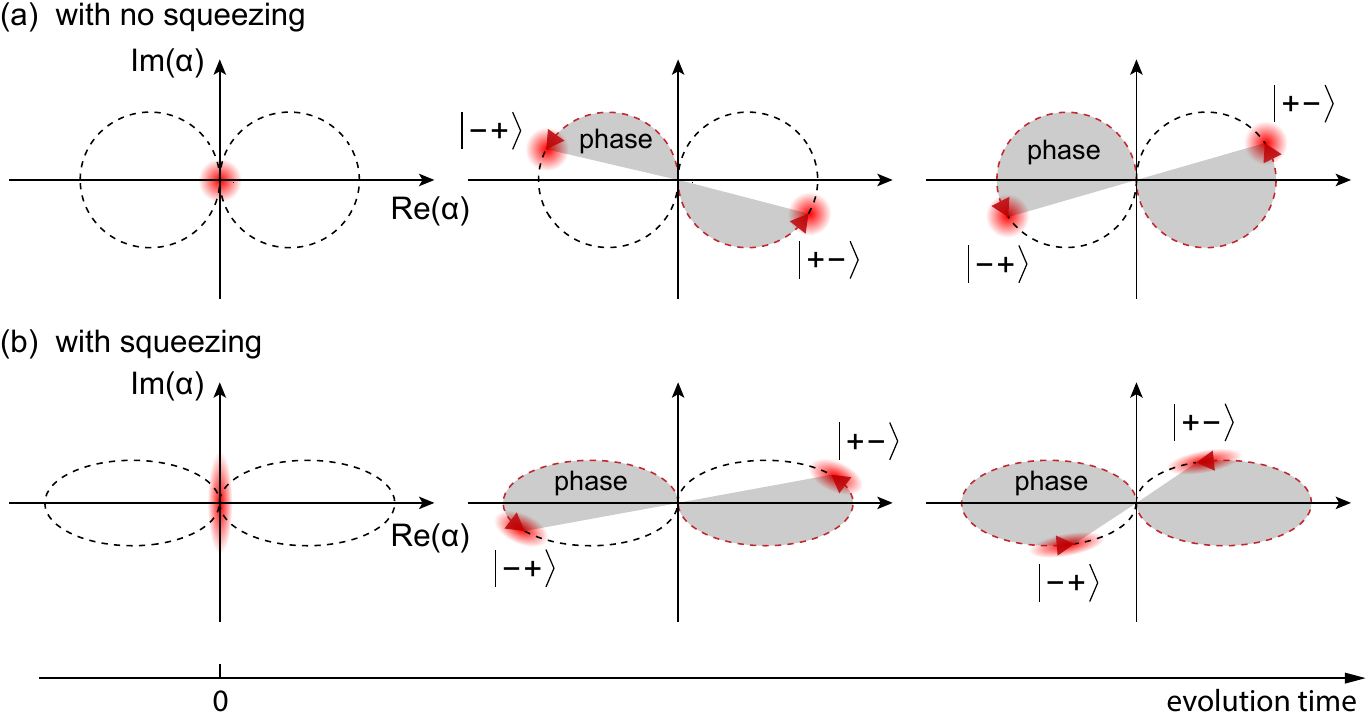}
	\caption{Phase-space representation of the geometric phase accumulation demonstrated in Ref.~\cite{burd2020quantum} for the cases (a) without and (b) with squeezing the motional mode of trapped-ion qubits. The horizontal arrow at the bottom represents the evolution time, with the initial time assumed to be 0. In the case of no squeezing, the geometrical phases acquired by the states $\ket{+-}$ and $\ket{-+}$ are accumulated along circular trajectories, and are equal to the (gray shaded) areas enclosed by their respective trajectories. When the motional mode is squeezed (in fact, periodically), the circular trajectories become elliptical with larger-amplitude displacements. The squeezing speeds up the geometric phase accumulation and, in turn, leads to a faster generation of quantum entanglement. For both cases, the motional mode is disentangled from the already entangled trapped-ion qubits at the end of the state preparation and, thus, the quantum state of the motional mode is not important.}
    \label{fig_phase_space_illustration_of_boson_mediated_interactions}
\end{figure}

The two trapped-ion qubits are coupled to the shared motional mode through the M\o{}lmer-S\o{}rensen interaction, making the phase-space displacement of the motional mode conditioned on the qubit state. In the case of no squeezing, the time evolution drives the states $\ket{+-}$ and $\ket{-+}$ to traverse circular trajectories in phase space, as depicted in \figpanel{fig_phase_space_illustration_of_boson_mediated_interactions}{a}. Here, $\ket{\pm}=\frac{1}{\sqrt{2}}\mleft(\ket{\uparrow}\pm\ket{\downarrow}\mright)$. Consequently, the states $\ket{+-}$ and $\ket{-+}$ accumulate their geometric phases, which are given by the areas enclosed by the respective trajectories and thus are of the same amount [see \figpanel{fig_phase_space_illustration_of_boson_mediated_interactions}{a}]. The motional mode, after some time, returns to its starting point in phase space, i.e., its initial state, and is disentangled from the qubits. At the same time, the states $\ket{+-}$ and $\ket{-+}$ acquire a geometric phase of $\pi/2$ for a single phase-space loop, which in turn induces quantum entanglement between the qubits. 

If a detuned parametric driving is applied to the motional mode, the time evolution rotates the direction of squeezing of the motional mode, and also changes the degree of squeezing periodically. Correspondingly, as shown in Fig.~\ref{fig_phase_space_illustration_of_boson_mediated_interactions}(b), the phase space trajectories, along which the geometric phases that are accumulated for the states $\ket{+-}$ and $\ket{-+}$, become elliptical~\cite{ge2019trapped,ge2019stroboscopic}, with larger-amplitude displacements originating from the parametrically amplified ion-motion coupling. This indicates that compared to the no-squeezing case, a larger geometric phase can be accumulated for a given evolution time, thus leading to a faster generation of quantum entanglement. On the other hand, in analogy to the no-squeezing case, the squeezed motional mode is also disentangled from the qubits after a single phase-space loop. Therefore, whether the motional mode is in a squeezed vacuum or the usual vacuum is not important. 

In the experiment in Ref.~\cite{burd2020quantum}, the observed results demonstrated that with parametric amplification, the entanglement generation is sped up by a factor of $\simeq3.74$, from which an increase of up to $3.25$-fold in the strength of the ion-motion coupling can be estimated. That experiment also showed the dependence of the interaction amplification on the phase $\theta_{\rm 2ph}$, i.e., on the direction of squeezing. This is different from the $\theta_{\rm 2ph}$-independent amplification process proposed in Ref.~\cite{qin2018exponentially}.

In proposals~\cite{lu2015squeezed, lemonde2016enhanced, qin2018exponentially, leroux2018enhancing, zeytinouglu2017engineering} of {\it using parametric squeezing to enhance light-matter interactions}, precise knowledge of the system Hamiltonian is needed in advance, e.g., the relative phase between the squeezing operation and the rest of the system dynamics. To clarify this issue, the use of parametric squeezing to amplify a coherent displacement of a bosonic field can be taken as an example. A direct method is to apply a squeezing operation and then an antisqueezing operation to the bosonic field, yielding
\begin{equation}\label{eq-phase-sensitive-amplification}
D(\alpha_{\rm sq})=S^{\dagger}(\xi)D(\alpha)S(\xi).
\end{equation}
Here, $D(\alpha)=\exp(\alpha a^{\dagger}-\alpha^{*}a)$ is a displacement operator and $S(\xi)$ is the squeezing operator defined in Eq.~(\ref{eq:squeezing operator}). A straightforward calculation gives
\begin{empheq}[box =\widecolourbox]{equation}
\label{eq-alpha-insensitive}
\alpha_{\rm sq}=\mleft[\cosh(r)+e^{i(\theta_{\rm 2ph}-2\phi_{\alpha})}\sinh(r)\mright]\alpha,
\end{empheq}
where $\phi_{\alpha}={\rm arg}(\alpha)$. Clearly, choosing $\theta_{\rm 2ph}-2\phi_{\alpha}=2n\pi$ ($n=0, \pm1, \pm2, \cdots$) can lead to {\it an exponential amplification in the coherent amplitude $\alpha$}, such that $\alpha_{\rm sq}=e^{r}\alpha$. This kind of amplification was experimentally reported in Ref.~\cite{burd2019quantum}, using the same setup as shown in Fig.~\ref{fig-trapped-ions-setup}, but with a single trapped $^{25}{\rm Mg}^{+}$ ion. The displacement operation of motion was implemented by applying a resonant oscillating potential to an electrode of the ion trap. An increase in $\alpha$ by a factor of $\approx9.17$ was demonstrated in that experimental work. However, it is also clear that the amplification given in Eq.~(\ref{eq-alpha-insensitive}) depends strongly on the phase difference $\theta_{\rm 2ph}-2\phi_{\alpha}$. Indeed, for some values of $\theta_{\rm 2ph}-2\phi_{\alpha}$, the coherent amplitude $\alpha$ even becomes deamplified; e.g., the phase difference of $\theta_{\rm 2ph}-2\phi_{\alpha}=(2n+1)\pi$ ($n=0, \pm1, \pm2, \cdots$) leads to $\alpha_{\rm sq}=e^{-r}\alpha$, i.e., {\it an exponential decrease in $\alpha$}. Thus, in order to achieve the desired amplification, the accurate value of the phase $\phi_{\alpha}$ is needed. However, in some cases, the phase $\phi_{\alpha}$ may be unknown or even fluctuate in time. To address this kind of issue, a proposal, which, in the absence of the precise knowledge of the system parameters, can amplify or speed up quantum dynamics using parametric squeezing, was put forward in Ref.~\cite{arenz2020amplification}. 

\begin{figure}[t]
	\centering
	\includegraphics[width=0.8\linewidth]{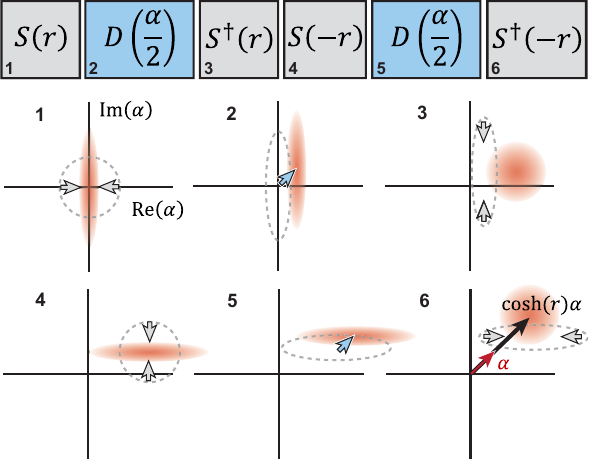}
	\caption{Phase-space representation of squeezing-induced amplification of a coherent displacement of motion, demonstrated in Ref.~\cite{burd2023experimental}. The gray dotted outlines and the orange blobs represent the phase-space distributions before and after, respectively, the operations shown in the corresponding boxes at the top are applied. The gray and blue arrows refer to the squeezing and displacement operations, respectively, while the red and black arrows in the final panel refer to the unamplified and amplified displacements, respectively. All panels are reproduced from Ref.~\cite{burd2023experimental}, S.~C.~Burd et al., \href{https://journals.aps.org/prxquantum/abstract/10.1103/PRXQuantum.5.020314}{PRX Quantum 5, 020314 (2024)}, with permission.}
    \label{fig-HA-phase-space}
\end{figure}

In order to amplify the coherent amplitude $\alpha$ of a bosonic field, whose full knowledge is unknown, the expression in Eq.~(\ref{eq-phase-sensitive-amplification}) needs to be modified, according to the formalism of Ref.~\cite{arenz2020amplification}, to become
\begin{align}
D(\alpha_{\rm sq})=S^{\dagger}(-r)D(\alpha/2)S(-r)S^{\dagger}(r)D(\alpha/2)S(r),
\end{align}
In this case, $\alpha_{\rm sq}$ is then found to be
\begin{empheq}[box =\widecolourbox]{equation}
\label{eq-alpha-sq}
\alpha_{\rm sq}=\cosh(r)\alpha,
\end{empheq}
indicating that the coherent amplitude $\alpha$ can be amplified by a factor $\cosh(r)$, but independently of its phase. This amplification process can be understood better in phase space as shown in Fig.~\ref{fig-HA-phase-space}.

More generally, the method presented in Ref.~\cite{arenz2020amplification} can also amplify a qubit-field interaction without precise knowledge. Consider, as an example, the JC interaction $H_{\rm JC}^{\rm int}$ given in Eq.~(\ref{eq:atomcavity_JC_coupling}). The evolution operator $U(t)$ under $H_{\rm JC}^{\rm int}$ for an evolution time $t$ is Trotterized, such that
\begin{equation}
U(t) = \prod_{n=1}^{N} \mathcal{U} \mleft( \frac{t}{N} \mright).
\end{equation}
Here,  
\begin{equation}
\mathcal{U} \mleft( \frac{t}{N} \mright) = S^{\dag}(-r) U_{0} \mleft( \frac{t}{2N} \mright) S(-r) S^{\dag}(r) U_{0} \mleft( \frac{t}{2N} \mright) S(r),
\end{equation}
where $U_{0}(t/2N)=\exp(-iH_{\rm JC}^{\rm int}t/2N)$ accounts for the evolution operator under $H_{\rm JC}^{\rm int}$ for a sufficiently small time $t/2N$~\cite{burd2023experimental,arenz2020amplification}. According to the Trotter formula~\cite{arenz2020amplification}, a sufficiently large $N$ can enable
\begin{equation}
U(t)\approx\exp\mleft(-iH_{\rm JC}^{\rm int, amp}t\mright),
\end{equation}
where
\begin{empheq}[box =\widecolourbox]{equation}
\label{eq-H-sq}
H^{\rm int, amp}_{\rm JC}=\cosh(r)H_{\rm JC}^{\rm int}.
\end{empheq}
It is easily seen that this is an interaction amplification by a factor $\cosh(r)$, but without the need to know precisely the exact form of the interaction. Note that the JC interaction $H_{\rm JC}^{\rm int}$ taken as an example here does not commute with the squeezing operators and thus Trotterization is needed as mentioned above; however, for some other interaction Hamiltonian that commutes with the squeezing operators (up to a phase), as is the case for a simple displacement operation, Trotterization is not necessary. 

Recently, this kind of squeezing-induced amplification in the absence of a full knowledge of the details of the system parameters has been experimentally reported in trapped-ion systems in Ref.~\cite{burd2023experimental}. The experimental setup was the same as given in Fig.~\ref{fig-trapped-ions-setup}, but a single trapped $^{25}{\rm Mg}^{+}$ ion was used there. The theoretical predictions given in Eqs.~(\ref{eq-alpha-sq}) and (\ref{eq-H-sq}) were demonstrated in that experimental work. The experiment showed that the coherent amplitude $\alpha$ can be amplified by a mean gain of $\approx1.77$, in good agreement with the theoretical prediction of $\approx2.11$ (corresponding to a squeezing parameter of $r\approx1.38$). 

In order to demonstrate the amplification of the qubit-field interaction given in Eq.~(\ref{eq-H-sq}), the experiment used the hyperfine states  $\ket{\downarrow}\equiv\ket{F=3,m_{F}=3}$ and $\ket{\uparrow}\equiv\ket{F=2,m_{F}=2}$ of the trapped $^{25}{\rm Mg}^{+}$ ion as qubit states. This qubit is coupled to a radial motional mode of the ion through a laser-driven stimulated Raman transition, which can be described by a JC Hamiltonian. Note that the phase of such a laser-driven JC interaction is not stabilized with respect to the squeezing phase, representing an example of strengthening the qubit-motion interaction where the exact phase (and thus the form) of $H_{\rm quad}$ is not known. Experimentally, an increase in the ion-motion coupling strength by a factor of $\approx1.56$ was observed for $N=6$, in good agreement with the theoretical prediction of $\approx1.7$ (corresponding to $r\approx1.1$).


\subsubsection{Superconducting circuits}

\begin{figure}
	\centering
	\includegraphics[width=\linewidth]{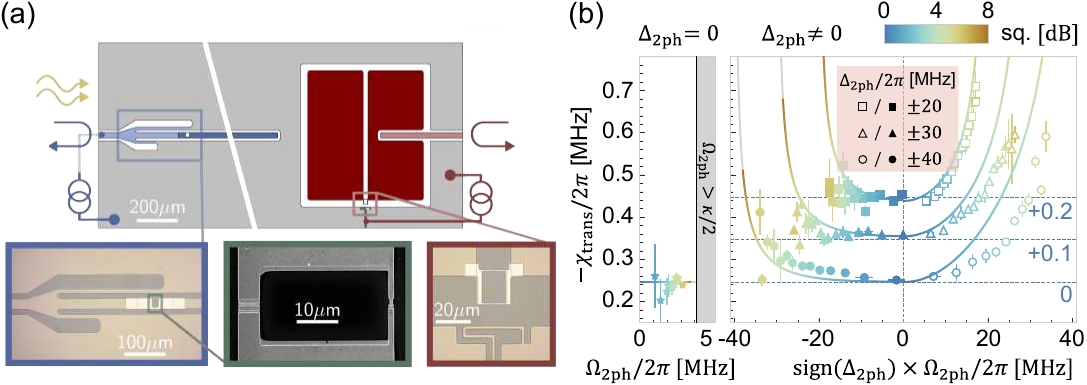}
	\caption{Experimental demonstration of the squeezing-enhanced dispersive atom-cavity coupling reported in Ref.~\cite{villiers2022dynamically}.
    (a) Experimental setup with a diagonal break. A quarter wavelength coplanar waveguide resonator (blue), as a single-mode cavity, is shunted to ground through a SNAIL element (boxed in green), such that it can act as a degenerate parametric amplifier, whose reflection spectrum is measured through an inductive coupler (boxed in blue). This SNAIL resonator is capacitively coupled to a transmon qubit (red) with a SQUID loop (boxed in red) threaded by a flux.
    (b) Dispersive coupling $\chi_{\rm trans}$ of the qubit and the cavity versus the two-photon driving $\Omega_{\rm 2ph}$ for $\Delta_{\rm 2ph}/2\pi=0$, $\pm20$, $\pm30$, and $\pm\unit[40]{MHz}$, and with offsets $0.2$, $0.1$, and $\unit[0]{MHz}$ for clarity. The colormap represents the squeezing in decibels, and the gray shaded area refers to the instability regime. The curves are theoretical predictions, and the symbols are experimental data; moreover, in the case of $\Delta_{\rm 2ph}\neq0$, the open and solid symbols correspond to $\Delta_{\rm 2ph}>0$ and $\Delta_{\rm 2ph}<0$, respectively. Clearly, there is no improvement in $\chi_{\rm trans}$ for $\Delta_{\rm 2ph}=0$, but a two-fold increase in $\chi_{\rm trans}$ for $\Delta_{\rm 2ph}/2\pi = \unit[20]{MHz}$ at a squeezing of $\simeq\unit[5.5]{dB}$. Figures, with some modifications, from Ref.~\cite{villiers2022dynamically}, M.~Villiers et al., \href{https://journals.aps.org/prxquantum/abstract/10.1103/PRXQuantum.5.020306}{PRX Quantum 5, 020306 (2024)}, with permission.}\label{fig_experimental_illustration_squeezing_2022}
\end{figure}

In the superconducting-circuit experiment in Ref.~\cite{villiers2022dynamically}, a quarter-wavelength coplanar-waveguide resonator interrupted by a superconducting nonlinear asymmetric inductive element (SNAIL) is capacitively coupled to a transmon qubit, as shown in \figpanel{fig_experimental_illustration_squeezing_2022}{a}. This SNAIL resonator, when pumped close to twice the resonator frequency, can behave as a degenerate parametric amplifier, and as a result can induce a significant squeezing but with a very weak Kerr nonlinearity. 

The experiment was operated in a dispersive regime to provide spectroscopic features able to decouple the effect of the undesired enhanced noise (i.e., thermal noise and two-photon correlation noise) from the expected enhanced coupling (as opposed to the resonant case). In this regime, the interaction given in Eq.~(\ref{eq:full Hamiltonian in squeezed mode}) leads to a dispersive coupling between the qubit and the squeezed cavity mode, with a strength
\begin{equation}\label{eq-enhanced-dispersive-coupling}
\chi=\frac{2g^2}{\Delta_q-\omega_{\rm sq}}\cosh^2(r)+\frac{2g^2}{\Delta_q+\omega_{\rm sq}}\sinh^2(r).
\end{equation}
Here, the physical parameters are defined in Sec.~\ref{subsubsec:Parametric amplification}. This expression is only for a pure two-level system, and for a transmon qubit with an anharmonicity $\chi_{\rm anh}$, it needs to be modified to 
\begin{equation}\label{eq-enhanced-dispersive-coupling-transmon}
\chi_{\rm trans}=\frac{2g^2}{\Delta_{q}-\omega_{\rm sq}}\frac{\chi_{\rm anh}}{\chi_{\rm anh}+\Delta_{q}-\omega_{\rm sq}}\cosh^2(r)+\frac{2g^2}{\Delta_{q}+\omega_{\rm sq}}\frac{\chi_{\rm anh}}{\chi_{\rm anh}+\Delta_{q}+\omega_{\rm sq}}\sinh^2(r).
\end{equation}
It can be seen that as $\chi_{\rm anh}\rightarrow\infty$, $\chi_{\rm trans}\rightarrow\chi$. Clearly, the dispersive coupling $\chi$ in Eq.~(\ref{eq-enhanced-dispersive-coupling}) and its transmon version $\chi_{\rm trans}$ in Eq.~(\ref{eq-enhanced-dispersive-coupling-transmon}) can increase (even exponentially) with the squeezing parameter $r$.

As shown in \figpanel{fig_experimental_illustration_squeezing_2022}{b}, a two-fold increase in $\chi_{\rm trans}$ can be experimentally observed for $\Delta_{\rm 2ph}/2\pi = \unit[20]{MHz}$ and $\Omega_{\rm 2ph}/2\pi = \unit[17]{MHz}$ (corresponding to a squeezing of $\simeq\unit[5.5]{dB}$). Note that there is no improvement in $\chi_{\rm trans}$ when $\Delta_{\rm 2ph}=0$. An asymmetric improvement of $\chi_{\rm trans}$ for the two cases of $\Delta_{\rm 2ph}>0$ and $\Delta_{\rm 2ph}<0$ was also demonstrated in the experiment. Such an asymmetry arises due mainly to the fact that for $\Delta_{\rm 2ph}>0$ (or $\Delta_{\rm 2ph}<0$), the resonance of the squeezed cavity mode shifts towards (or away from) the qubit resonance, thus decreasing (increasing) the detuning between the qubit and the squeezed cavity mode.

In the case of $\Delta_{\rm 2ph}=0$, the amplification bandwidth decreases as the gain increases. This is because of the fundamental gain-bandwidth constraint; that is, the product of the gain and the amplification bandwidth is approximately equal to the single-photon loss rate $\kappa$. In sharp contrast, in the detuned case of $\Delta_{\rm 2ph}\neq0$, the amplification bandwidth becomes constant and thus independent of the gain~\cite{metelmann2022quantum}. This interesting feature was also observed in the experiment in Ref.~\cite{villiers2022dynamically}. 
\section{Simulation of ultrastrong and deep-strong light-matter interactions}
\label{sec:simulations}

In this section, we switch from considering schemes for \textit{amplifying} existing interactions to ways of \textit{simulating} interactions that are significant compared to the bare transition frequencies in the system. There are many ways to perform such quantum simulations~\cite{Buluta2009, Georgescu2014}, and a plethora of experimental systems in which it can be implemented. Here, we first consider, in Secs.~\ref{sec:CavityAssistedRaman}--\ref{sec:SimulationSingleDrive}, analog simulation schemes where one or two drives are added to a system with a low interaction strength, such that a reference frame can be found where the renormalized parameters can be tuned by the external drives to reach the USC or DSC regimes. Then, in \secref{sec:DigitalSimulation}, we review digital protocols for quantum simulation. Finally, in \secref{sec:OtherSimulation}, we review a number of additional simulation methods and setups, including VQS, two analog simulation setups with cold atoms, and ways to simulate the USC between resonators or in optomechanical systems.


\subsection{Cavity-assisted Raman transitions}
\label{sec:CavityAssistedRaman}


\subsubsection{Derivation}

Some of the most important instances of simulating light-matter systems in the USC regime are based on the use of cavity-assisted Raman transitions to obtain an effective Dicke Hamiltonian. Crucially, the effective frequencies of the involved subsystems in the simulated Hamiltonian are strongly reduced, thereby providing a way to explore regimes where the light-matter coupling can be very strong in comparison. Such a simulation was originally proposed in Ref.~\cite{Dimer2007} as a way to observe the superradiant phase transition of the Dicke model, and has been followed by many proposals for particular systems. For example, in Ref.~\cite{Grimsmo2013}, a method was proposed to simulate the quantum Rabi model in cavity QED with a single rubidium atom in a cavity.

\begin{figure}
	\centering
    \includegraphics[width=0.9\textwidth]{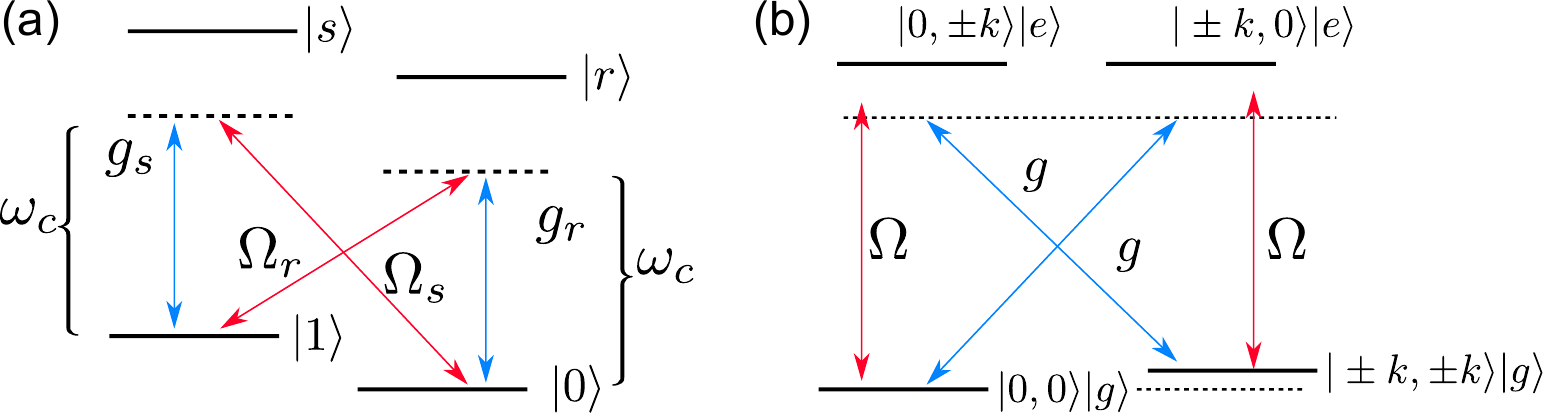}
	\caption{Level structures of simulations of the quantum Rabi model with cavity-assisted Raman transitions. 
    (a) Original proposal with the atomic electronic states, and 
    (b) alternative proposal with the atomic motional states in a BEC. See also Refs.~\cite{Dimer2007} and \cite{Baumann2010}, respectively.}
    \label{fig_cavity_raman}
\end{figure}

This proposed scheme is shown in \figpanel{fig_cavity_raman}{a}. The atomic part consists of two ground states, $\ket{0}$ and $\ket{1}$, and two excited states $\ket{s}$ and $\ket{r}$. The two ground states constitute an effective two-level system. The cavity mode $a$ of frequency $\omega_{\rm cav}$ is coupled to the transitions $\ket{1} \leftrightarrow \ket{s}$ and $\ket{0} \leftrightarrow \ket{r}$. Two external lasers, of frequencies $\omega_{\mathrm{l}r}$ and $\omega_{\mathrm{l}s}$, drive the transitions $\ket{1} \leftrightarrow \ket{r}$ and $\ket{0} \leftrightarrow \ket{s}$, respectively, with Rabi frequencies $\Omega_r$ and $\Omega_s$.
By defining the lowering operators $\sigma_{1s}^-\equiv \ketbra{1}{s}$, $\sigma_{1r}^-\equiv \ketbra{1}{r}$, $\sigma_{0s}^-\equiv \ketbra{0}{s}$, and $\sigma_{0r}^- \equiv \ketbra{0}{r}$, and the number operators
$\sigma_i \equiv \ketbra{i}{i}$, $i \in\{0,1,s,r\}$, the Hamiltonian of the system reads
\begin{equation}
	H = H_0 + H_{\mathrm{int}}+ H_\mathrm{dr}, 
\end{equation}
with 
\begin{align}
	H_0 &= \omega_{\rm cav} a^\dag a + \omega_1 \sigma_1 + \omega_r \sigma_r + \omega_s \sigma_s , \\
	H_\mathrm{int} &= g_s \mleft( a^\dag \sigma_{1s}^- + \text{H.c.} \mright) + g_r \mleft( a^\dag \sigma_{0r}^- + \text{H.c.} \mright) ,\\
	H_\mathrm{\mathrm{dr}} &= \frac{\Omega_s}{2} \mleft( e^{i\omega_{\mathrm{l} s} t}\, \sigma_{0s}^- + \mathrm{H.c.} \mright) + \frac{\Omega_r}{2} \mleft( e^{i\omega_{\mathrm{l} r} t}\, \sigma_{1r}^- + \mathrm{H.c.} \mright).
\end{align}
The time dependence of the Hamiltonian can be removed via a unitary transformation
\begin{equation}
	U=\exp\mleft\{i\mleft[ \omega_1'\sigma_1  + \omega_{\mathrm{l} s} \sigma_s +  (\omega_{\mathrm{l} r}+\omega_1') \sigma_r + (\omega_{\mathrm{l}s}-\omega_1')a^\dagger a\mright]t\mright\}.
\end{equation}
The time independence is enforced by fixing the parameter $\omega_1'$ to be
\begin{equation}
	\omega_1' = (\omega_{\mathrm{l}s}-\omega_{\mathrm{l}r})/2,
\end{equation}
which implies $\omega_1' \approx \omega_1$, since $\omega_{\mathrm{l}r}\approx \omega_{\rm cav} - \omega_1$ and $\omega_{\mathrm{l}s}\approx \omega_{\rm cav} + \omega_1$. In the new rotating frame, the time evolution is governed by the Hamiltonian $H_R = U H U + i (\partial_t U)U^\dagger$, which is now time-independent and reads $H_R = H_{R,0} + V$, where $H_{R,0}$ is the new bare Hamiltonian, given by
\begin{equation}
	H_{R,0} = \delta_c a^\dagger a + \Delta_1 \sigma_1 + \Delta_r \sigma_r + \Delta_s \sigma_s ,
\end{equation}
with $\delta_c = \omega_{\rm cav} - (\omega_{\mathrm{l}s}+\omega_{\mathrm{l}r})/2$, $\Delta_1 = \omega_1-\omega_1'$, $\Delta_r = \omega_r - \frac{1}{2}(\omega_{\mathrm{l}s}+\omega_{\mathrm{l}r})$, $\Delta_s = \omega_s - \omega_{\mathrm{l}s}$; and $V$ is the interaction term that couples the ground states to the excited states: 
\begin{equation}
	V = \frac{\Omega_s}{2} \mleft( \sigma_{0s}^- + \sigma_{0s}^+ \mright) + \frac{\Omega_r}{2} \mleft( \sigma_{1r}^- + \sigma_{1r}^+ \mright) + g_s \mleft( a^\dag \sigma_{1s}^- + a \sigma_{1s}^+ \mright) + g_r \mleft( a^\dag \sigma_{0r}^- + a \sigma_{0r}^+ \mright).
\end{equation}

Under the assumption $\mleft| \Delta_{r/s} \mright| \gg \{\delta_c, \Delta_1, \Omega_{r/s},  g_{r/s}\}$, the excited states can then be adiabatically eliminated. This assumption enables a separation between a manifold of ``fast'' degrees of freedom (involving the excited states), and a manifold of ``slow'' degrees of freedom (involving the ground states).  After adiabatically eliminating the fast degrees of freedom, an effective Hamiltonian for the slow degrees of freedom can be obtained: the states $\ket{1}$ and $\ket{0}$ are connected via the second-order processes, which involve the absorption-emission processes of a cavity photon and a laser photon. As an intuitive example, one can see that the excitation $\ket{0} \rightarrow \ket{r}$ by the absorption of a cavity photon, followed by the transition $\ket{r} \rightarrow \ket{1}$ via the emission of a laser photon, yields an effective dynamics described by a Hamiltonian term $\propto a \sigma_{01}^+$. In order to derive the effective Hamiltonian, one can follow the approach based on the Schrieffer-Wolff transformation described in Ref.~\cite{CohenTannoudji1998}. Up to second order in the interaction term $V$ that couples fast and slow degrees of freedom, the effective Hamiltonian is expressed as
\begin{equation}
	\brakket{i}{H_\mathrm{eff}}{j} = E_i \delta_{ij} + \frac{1}{2} \sum_{\alpha} \brakket{i}{V}{\alpha} \brakket{\alpha}{V}{j} \mleft[ \frac{1}{E_i-E_\alpha} + \frac{1}{E_j-E_\alpha} \mright],
	\label{eq:Heff-second-order}
\end{equation}
where $E_i$ are the energies of the bare Hamiltonian $E_i = \brakket{i}{H_{R,0}}{i}$. The indices $i,j$ label the states in the slow manifold, while $\alpha$ labels the states in the fast manifold. We label the eigenstates of $H_{R,0}$ as $\ket{k, n} \equiv \ket{k} \otimes \ket{n}$, where $k \in \{0,1,s,r\}$ refers to the atomic state, and $\ket{n}$ denotes the Fock state with $n$ cavity photons. Then, we find:
\begin{align}
	\brakket{1,n}{H_\mathrm{eff}}{0,n-1} &\approx -\sqrt{n}\frac{g_s \Omega_s}{2\Delta_s}, \\
	\brakket{1,n}{H_\mathrm{eff}}{0,n+1} &\approx -\sqrt{n+1}\frac{g_r \Omega_r}{2\Delta_r}, \\
	\brakket{1,n}{H_\mathrm{eff}}{1,n} &\approx n \delta_c + \Delta_1 -\frac{\Omega_r^2}{4\Delta_r}-n\frac{g_s^2}{\Delta_s}, \\
	\brakket{0,n}{H_\mathrm{eff}}{0,n} &\approx n\delta_c -\frac{\Omega_s^2}{4\Delta_s}-n\frac{g_r^2}{\Delta_r}.
\end{align}
From these matrix elements, we can, by defining the spin operators $S_- = \sigma_{01}^-$ and $S_z = \frac{1}{2}(2\sigma_1 -1)$, write the effective Hamiltonian as
\begin{equation}
	H_\mathrm{eff} = \Delta_c a^\dag a + \Delta_0 s_z + \chi a^\dag a S_z + \lambda_s \mleft( a^\dag S_- + \mathrm{H.c.} \mright) + \lambda_r \mleft( a^\dag S_+ + \mathrm{H.c.} \mright) ,
	\label{eq:effective_H_raman} 
\end{equation}
where
\begin{align}
	\Delta_c &= \delta_c -\frac{1}{2} \mleft( \frac{g_r^2}{\Delta_r} + \frac{g_s^2}{\Delta_s} \mright) , \\
	\Delta_0 &= \frac{\Delta_1}{2} + \frac{1}{4} \mleft( \frac{\Omega_s^2}{\Delta_s} - \frac{\Omega_r^2}{\Delta_r} \mright) , \\
	\chi &= \frac{g_r^2}{\Delta_r} - \frac{g_s^2}{\Delta_s} ,
\end{align}
and the effective coupling strengths are given by
\begin{equation}
	\lambda_{s} \equiv -\frac{g_{s} \Omega_{s}}{2\Delta_{s}}, \quad \lambda_{r} \equiv -\frac{g_{r} \Omega_{r}}{2\Delta_{r}}.
	\label{eq:effective_coupling-Raman} 
\end{equation}

While we here have presented the derivation for the one-atom case, the effective Hamiltonian for a system of $N$ atoms is only slightly modified by including a factor $N$ for the Lamb shift of the cavity resonance, i.e., $\lambda_A \rightarrow N \lambda_A$~\cite{Dimer2007}. It can be seen that one can reach $\chi=0$ by setting $g_r^2/\Delta_r = g_s^2/\Delta_s$. Similarly, we can set $g_s \Omega_s/\Delta_s = g_r \Omega_r/2\Delta_r$, so that the effective Hamiltonian has the form of a quantum Rabi model for $N=1$, and a quantum Dicke model for $N>1$:
\begin{empheq}[box =\widecolourbox]{equation}
	H_\mathrm{eff} = \Delta_c a^\dag a +  \Delta_0 S_z + \lambda \mleft( a^\dag + a \mright) \mleft( S_- + S_+ \mright),
	\label{eq:effective_H_raman_spin} 
\end{empheq}
with 
\begin{equation}
	\lambda = -\frac{g_{s} \Omega_{s}}{2\Delta_{s}} =  -\frac{g_{r} \Omega_{r}}{2\Delta_{r}}.
\end{equation}
Crucially, we see that the USC regime can be easily achieved, since both the detunings $\Delta_c$ and $\Delta_0$ can be made arbitrarily small compared to the coupling strength $\lambda$. 


\subsubsection{Implementations in atomic condensates}

The first experimental implementation of cavity-assisted Raman transitions for the simulation of the Dicke model, and also the observation of a superradiant transition, was realized in an atomic BEC~\cite{Baumann2010}. In this approach, the spin is defined by the atomic \emph{motional} states, rather than the atomic electronic states.
The underlying mechanism behind the observation of this superradiant transition in atomic clouds inside optical cavities is  the self-organization of the atoms into a checkerboard pattern, as demonstrated theoretically in Ref.~\cite{Domokos2002} and shown experimentally in Ref.~\cite{Black2003}. The theoretical mapping of this transition to a dynamical version of the superradiance transition of the Dicke model was put forward in Refs.~\cite{Chen2007,Nagy2010}. The equivalence between them is established by writing the state of an atom as $\ket{k_x,k_z} \ket{a}$, where $\ket{k_x,k_z}$ describes a momentum eigenstate in the $x$-$z$ plane and $\ket{a}$ is an internal, electronic eigenstate. The energy-level structure and the corresponding couplings are shown in \figpanel{fig_cavity_raman}{b}.

One can define an effective two-level spin system composed of the ground state $\ket{0,0} \ket{g}$ and an excited momentum state $\ket{\pm k,\pm k} \ket{g} = \frac{1}{2}\sum_{\alpha,\beta=\pm} \ket{\alpha k,\beta k} \ket{g}$, which is coupled to the cavity mode via the two-photon absorption and emission processes. The JC-type term $a\sigma^\dag$ emerges from the process of the absorption of a cavity photon. Such a photon absorption excites the state $\ket{0,0} \ket{g}$ to $\ket{\pm k,0} \ket{e} = \frac{1}{\sqrt{2}} \sum_{\mu =\pm} \ket{\mu k,0} \ket{e} $, which, in turn, generates the state $\ket{\pm k, \pm k} \ket{g}$ and the simultaneous emission of a photon into the pump field. The excitation of momentum eigenstates is due to the recoil experienced by the atoms by the absorption or emission a photon of momentum $\pm k$, and therefore a small energy difference between the two ground states is given by the photon recoil energy, which is on the order of kHz.  Similarly, a counter-rotating term $\sigma^\dag a^\dag$ is obtained when the ground state is excited to $\ket{0,\pm k} \ket{e}$ by the absorption of a pump photon (so that the recoil momentum is gained in the $z$ axis instead), which is then brought to the state $\ket{\pm k, \pm k} \ket{g}$ by emitting a photon into the cavity mode. 

This idea was implemented experimentally in Ref.~\cite{Baumann2010}, which reported the first Dicke phase transition, using a $^{87}\mathrm{Rb}$ BEC with $N\sim 10^5$ atoms. The Dicke phase transition and the emergence of self-organization are manifested by an abrupt build-up of the cavity field accompanied by the development of momentum components at $(k_x, k_z)$ in the atomic cloud. The symmetry breaking between two possible superradiant states (associated with two atomic density waves shifted by half an optical wavelength) was further explored by the same group in Ref.~\cite{Baumann2011}, and the dynamical properties of this phase transition were then interpreted in terms of a Kibble-Zurek mechanism in Ref.~\cite{Klinder2015}. Several theoretical works have studied more deeply the critical properties and phase diagrams of these models; see, e.g., Refs.~\cite{Bhaseen2012, Kirton2018}.


\subsubsection{Implementations in trapped thermal atoms}

An implementation that followed the original proposal of Ref.~\cite{Dimer2007} more closely, by using the atomic electronic states instead of the atomic motional states to encode a two-level system, was shown for the first time in Ref.~\cite{Baden2014}, where the effective two-level system was encoded by the hyperfine ground states $\ket{F = 1, m_F = 1}$ and $\ket{F=2,m_F = 2}$ of \87rb atoms, with a Zeeman splitting (on the order of MHz) induced by an external magnetic field. 
In Ref.~\cite{Zhiqiang2017}, the same group further explored the rich phase diagram emergent under situations of the asymmetric coupling, i.e., $\lambda_s \neq \lambda_r$, and in Ref.~\cite{Zhang2018}, the differences between co- and counter-propagating structures were discussed in detail.

The cavity-assisted Raman transitions have been investigated for a plethora of effects and their applications, such as the formation of spinor self-ordering and spin textures~\cite{Kroeze2018, Landini2018}, the generation of spin squeezing~\cite{dalla2013dissipative} and steady-state entanglement in solid-state systems~\cite{Gonzalez-Tudela2013}, the simulation of supersymmetry field theories in quantum optics~\cite{Tomka2015}, the driven dissipative dynamics under strong symmetries (dissipative freezing)~\cite{SanchezMunoz2019}, and the engineering of spin-spin interactions mediated by photons~\cite{Masson2017, Davis2019, Mivehvar2019, Norcia2018}. 


\subsection{Simulated ultrastrong interactions in two-mode-driven Jaynes--Cummings systems}

In addition to the protocols described in \secref{sec:CavityAssistedRaman}, there are a few more which also rely on applying two drives to a system that is not in the USC regime, and in particular, a system in the strong coupling regime, which can be described by the JC Hamiltonian or one of its close relatives. Thanks to the two drives, one can engineer a rotating frame where the model parameters are renormalized to be in the USC regime. This type of proposal has already been realized experimentally for both trapped ions and superconducting circuits.


\subsubsection{Theoretical proposals}
\label{sec:TwoDrivesJCTheory}


\paragraph{Basic implementation}

The simulation scheme proposed in Ref.~\cite{Ballester2012} considers a system consisting of a qubit, of frequency $\omega_q$, coupled with a strength $g$ to a cavity mode of frequency $\omega_{\rm cav}$, such that a full description of the system is given by the quantum Rabi Hamiltonian,
\be
H_{\rm Rabi} = \omega_{\rm cav} a^\dag a + \frac{\omega_q}{2} \sz + g \mleft( a + a^\dag \mright) \sx.
\label{eq:HRabi}
\ee
This Hamiltonian can be viewed as an equivalent variant of the quantum Rabi Hamiltonian in the dipole gauge, i.e.,
	\begin{equation}\label{eq-Rabi-dipole-2level}
		H_{\rm Rabi}^{\rm dg}=\omega_{\rm cav}a^{\dagger}a+\frac{\omega_{q}}{2}\sigma_{z}+ig_{\rm dg}(a^{\dagger}-a)\sigma_{x},
	\end{equation}
since the former can be directly obtained by applying a replacement $a\rightarrow ia$ to the latter. Here, $g_{\rm dg}$ refers to the light-matter coupling strength in the dipole gauge. Note that the form of the quantum Rabi Hamiltonian in the USC regime strongly depends on the choice of gauge.
In the Coulomb gauge, the standard quantum Rabi Hamiltonian is given by
\begin{equation}\label{eq-rabi-coulomb-wrong}
	\mathcal{H}_{\rm Rabi}^{\rm cg}=\omega_{\rm cav}a^{\dagger}a+\frac{\omega_{q}}{2}\sigma_{z}+g_{\rm cg}(a+a^{\dagger})\sigma_{y}+D(a+a^{\dagger})^{2},
\end{equation}
where $g_{\rm cl}=g_{\rm dg}\omega_{q}/\omega_{\rm cav}$, and $D$ is the diamagnetic amplitude. According to Eq.~(\ref{eq-rabi-coulomb-wrong}), it seems as though Eq.~(\ref{eq:HRabi}) was also a simpler variant in the Coulomb gauge, with a rotation $\sigma_{y}\rightarrow\sigma_{x}$ and at the same time neglecting the diamagnetic term $\propto(a+a^{\dagger})^2$. However, in fact, such a system Hamiltonian in the Coulomb gauge in Eq.~(\ref{eq-rabi-coulomb-wrong}) is wrong in the USC regime, since it does not produce the correct eigenvalues and, as a result, causes the breakdown of gauge invariance~\cite{De2018breakdown,DiStefano2019,generalized2023akbari}. 
This gauge ambiguity has been well resolved in Ref.~\cite{DiStefano2019} by applying the minimal-replacement rule not only to the particle kinetic energy but also to the particle nonlocal potential. In this case, the resulting system Hamiltonian in the Coulomb gauge, i.e.,
\begin{equation}
	H_{\rm Rabi}^{\rm cg}=\omega_{\rm cav}a^{\dagger}a+\frac{\omega_{q}}{2}\left\{\sigma_{z}\cos\left[2\eta(a+a^{\dagger})\right]+\sigma_{y}\sin\left[2\eta(a+a^{\dagger})\right]\right\},
\end{equation}
where $\eta=g_{\rm dg}/\omega_{\rm cav}$, is gauge invariant and can produce the same eigenvalues as the Hamiltonian in the dipole gauge given in Eq.~(\ref{eq-Rabi-dipole-2level}).

When $g \ll \{\omega_q, \omega_{\rm cav}\}$, the RWA can be applied to Eq.~(\ref{eq:HRabi}) to yield the simplified model described by the JC Hamiltonian
\be
H_{\rm JC} = \omega_{\rm cav} a^\dag a + \frac{\omega_q}{2} \sz + g \mleft( a \sp + a^\dag \sm \mright).
\label{eq:HJC}
\ee
To use a simple system modeled by the Hamiltonian in \eqref{eq:HJC} to simulate the USC between the qubit and the cavity, two classical drives (at frequencies $\omega_1$ and $\omega_2$, and with strengths $\Omega_1$ and $\Omega_2$, respectively) are applied to the qubit. By choosing the drive strengths and frequencies appropriately, it turns out to be possible to recover, in a rotating frame, the dynamics of \eqref{eq:HRabi} with parameters renormalized, such that the USC of light and matter is realized.

In more detail, the corresponding derivation is as follows. Making the RWA also for the two drive terms added to \eqref{eq:HJC}, and then moving to a frame rotating at $\omega_1$, the Hamiltonian of the system becomes
\begin{align}
H = &\: \mleft( \omega_{\rm cav} - \omega_1 \mright) a^\dag a + \frac{\omega_q - \omega_1}{2} \sz + g \mleft( a \sp + a^\dag \sm \mright) \nonumber\\
& + \Omega_1 \sx + \Omega_2 \mleft[ e^{i \mleft( \omega_2 - \omega_1 \mright) t} \sm + e^{- i \mleft( \omega_2 - \omega_1 \mright) t} \sp \mright].
\end{align}
Next, we move to the interaction picture with respect to the drive term, $\Omega_1 \sx$, and rewrite the qubit terms in the rotated basis $\ket{\pm} = \mleft(\ket{g} \pm \ket{e} \mright) / \sqrt{2}$. This yields the transformed Hamiltonian
\begin{align}
H_{\rm tr} = &\: \mleft( \omega_{\rm cav} - \omega_1 \mright) a^\dag a - \frac{\omega_q - \omega_1}{2} \mleft( e^{i 2 \Omega_1 t} \ketbra{+}{-} + \text{H.c.} \mright) \nn\\
& + \frac{g}{2} \mleft[ \mleft( \ketbra{+}{+} - \ketbra{-}{-} + e^{i 2 \Omega_1 t} \ketbra{+}{-} - e^{- i 2 \Omega_1 t} \ketbra{-}{+} \mright) a + \text{H.c.} \mright] \nn\\
& + \frac{\Omega_2}{2} \mleft[ \mleft( \ketbra{+}{+} - \ketbra{-}{-} - e^{i 2 \Omega_1 t} \ketbra{+}{-} + e^{- i 2 \Omega_1 t} \ketbra{-}{+} \mright) e^{i \mleft( \omega_2 - \omega_1 \mright) t} + \text{H.c.} \mright].
\label{eq:Htr}
\end{align}
Assuming that $\Omega_1 \gg \{g/4 , \omega_q - \omega_1\}$ and that $\omega_2 - \omega_1 = 2 \Omega_1$, we apply the RWA to \eqref{eq:Htr}, keeping only the time-independent terms, and as a result, have 
\begin{empheq}[box =\widecolourbox]{equation}
H_{\rm eff} = \mleft( \omega_{\rm cav} - \omega_1 \mright) a^\dag a - \frac{\Omega_2}{2} \sz + \frac{g}{2} \mleft( a + a^\dag \mright) \sx.
\label{eq:Heff}
\end{empheq}
This is of the same form as \eqref{eq:HRabi}, but with an effective cavity frequency $\omega_{\rm cav} - \omega_1$ and an effective qubit frequency $\Omega_2$, both of which can be tuned by the two classical drives to values smaller than the effective coupling strength $g/2$.

It is important to note that the effective model in \eqref{eq:Heff} is obtained in a rotating frame. Experimentally, some quantities in this model, like the qubit population, would thus display rapid oscillations, which could make them hard to observe. To remedy this, Ref.~\cite{Ballester2012} proposes that after the two drives have been on for a time $t$, they are turned off non-adiabatically, then the qubit and the drive at $\omega_{1}$ are both detuned by some amount before the drive at $\omega_{1}$ is turned on again for the same time $t$, but with an opposite phase to before. This protocol allows one to recover the dynamics of the model in a rotating frame.



\paragraph{Variation for trapped ions}

Single ions can be trapped using radio-frequency fields and have some of their electronic degrees of freedom (a qubit) couple to their motion in the trap (a harmonic oscillator) through laser driving. The proposal in Ref.~\cite{Pedernales2015} considers the situation when two laser drives are applied, in which case the Hamiltonian for such a setup can be written as
\be
H = \omega_{\rm mot} a^\dag a + \frac{\omega_q}{2} \sz + \sum_{n = r, b} \Omega_n \sx \cos \mleft[ \eta \mleft( a + a^\dag \mright) - \omega_n t + \phi_n \mright],
\label{eq:HIon2Drives}
\ee
where $\omega_q$ is the transition frequency of the ionic qubit, $\omega_{\rm mot}$ is the frequency of the harmonic motional mode, $\Omega_n$ are the Rabi frequencies of the drives denoted by $r$ and $b$, $\omega_n$ are the drive frequencies, $\phi_n$ are the drive phases, and $\eta = k/\sqrt{2 m \omega_{\rm mot}}$ is the Lamb-Dicke parameter, with $k$ the component of the laser wave vector in the direction of the harmonic motion and $m$ the mass of the ion. Transforming to the interaction picture with respect to the bare qubit and the motional degree of freedom, the Hamiltonian becomes
\be
H_\text{int} = \sum_{n = r, b} \frac{\Omega_n}{2} \mleft\{\exp\{i \eta \mleft[ a(t) + a^\dag(t) \mright]\} \exp[i \mleft( \omega_q - \omega_n \mright) t] \sp + \text{H.c.} \mright\},
\label{eq:HintIon}
\ee
where $a(t) = a e^{i \omega_{\rm mot} t}$.

Next, we assume that the two drive frequencies are tuned to $\omega_{b/r} = \omega_q \pm \omega_{\rm mot} + \delta_{b/r}$, such that they are close to (detuned by $\delta_{b/r}$ from) the blue ($b$) and red ($r$) sidebands, respectively. We further assume that we are in the Lamb-Dicke regime, i.e., $\eta \sqrt{\expec{a^\dag a}} \ll 1$, and that the drives are resonant and weak enough ($\delta_{r,b} , \Omega_{r,b} \ll \omega_{\rm mot}$) not to excite any other sidebands. In this limit, also assuming equal drive strengths $\Omega \equiv \Omega_r = \Omega_b$, \eqref{eq:HintIon} simplifies to
\be
H_\text{int} = \frac{i \eta \Omega}{2} \sp \mleft( a e^{-i \delta_r t} + a^\dag e^{-i \delta_b t} \mright) + \text{H.c.}
\label{eq:HintIonSimplified}
\ee
Upon inspection, we see that this is the Hamiltonian in the interaction picture that would result from starting from the original Hamiltonian
\begin{empheq}[box =\widecolourbox]{equation}\label{eq-rabi-2}
H = \frac{1}{2} \mleft( \delta_b - \delta_r \mright) a^\dag a + \frac{1}{4} \mleft( \delta_r + \delta_b \mright) \sz + \frac{i \eta \Omega}{2} \mleft( \sp - \sm \mright) \mleft( a + a^\dag \mright).  
\end{empheq}
The coupling form of $ i\mleft( \sp - \sm \mright) \mleft( a + a^\dag \mright)$ is an equivalent version of the light-matter coupling form of the quantum Rabi Hamiltonian in the dipole gauge [see Eq.~(\ref{eq-Rabi-dipole-2level})]. Here, we can identify an effective harmonic-mode frequency $\mleft( \delta_b - \delta_r \mright)/2$, an effective qubit frequency $\mleft( \delta_r + \delta_b \mright)/2$, and an effective coupling strength $\eta \Omega/2$. These effective parameters can all be tuned by adjusting the drive frequencies and strengths, allowing the simulator to explore a wide range of parameters in the USC regime. As is pointed out in Ref.~\cite{Pedernales2015}, all transformations performed here commute with observables of interest (e.g., the number of excitations in the qubit and the harmonic mode), such that these observables can be directly measured in experiments. A later theory work~\cite{Puebla2017} suggests that some additional laser drives may need to be added to explore the particular parameter regime where a quantum phase transition can occur.



The setup of a bichromatically driven ion can be taken further to realize other versions of the quantum Rabi model. In Ref.~\cite{Felicetti2015}, it was shown that it also can simulate the two-photon Rabi model, i.e., by replacing $a$ and $a^\dag$ in the interaction in \eqref{eq:HRabi} with $a^2$ and $\mleft( a^\dag \mright)^2$, respectively, such that the photonic population only can be changed in steps of two. The key to realizing this simulation is to tune the two drives to the second sidebands instead of the first ones, i.e., to set the drive frequencies to $\omega_{b/r} = \omega_q \pm 2 \omega_{\rm mot} + \delta_{b/r}$. Using the same assumptions as when going from \eqref{eq:HintIon} to \eqref{eq:HintIonSimplified}, with these drive frequencies \eqref{eq:HintIon} simplifies to
\be
H_\text{int}^{(2)} = - \frac{\eta^2 \Omega}{4} \sp \mleft[ a^2 e^{-i \delta_r t} + \mleft( a^\dag \mright)^2 e^{-i \delta_b t} \mright] + \text{H.c.},
\ee
which we can recognize as being the interaction-picture version of the original Hamiltonian
\begin{empheq}[box =\widecolourbox]{equation}
H = \frac{1}{4} \mleft( \delta_b - \delta_r \mright) a^\dag a + \frac{1}{4} \mleft( \delta_r + \delta_b \mright) \sz - \frac{\eta^2 \Omega}{4} \sx \mleft[ a^2 + \mleft( a^\dag \mright)^2 \mright].  
\end{empheq}
In this case, we identify an effective harmonic-mode frequency $\mleft( \delta_b - \delta_r \mright)/4$, an effective qubit frequency $\mleft( \delta_r + \delta_b \mright)/2$, and an effective coupling strength $\eta^2 \Omega/4$. As before, we here have the freedom to adjust the drive frequencies and detunings to simulate a wide range of parameters that correspond to the USC regime of the two-photon quantum Rabi model.


Further generalizations of the scheme in Ref.~\cite{Pedernales2015} were proposed in Ref.~\cite{Aedo2018}. There, it was shown that the scheme can be extended to the Dicke model (the generalization of the quantum Rabi model to multiple qubits), as well as variations on that, like the biased Dicke model (which includes a term proportional to $\sx$ for each qubit) and the anisotropic Dicke model (where the rotating and counter-rotating parts of the interaction have different strengths).



\subsubsection{Experimental implementations}

\paragraph{Superconducting circuits}

\begin{figure}
	\centering
    \includegraphics[width=\textwidth]{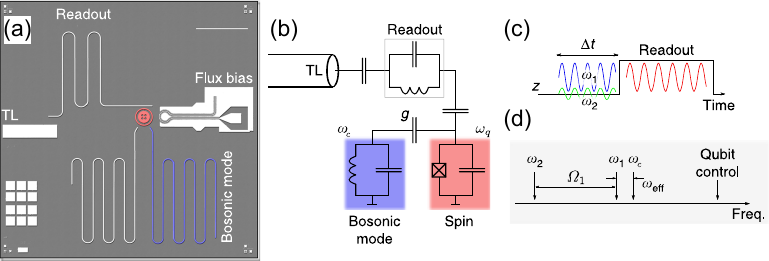}
	\caption{Experimental setup used in Ref.~\cite{Braumuller2017} for simulation of USC with superconducting circuits.
    (a) Optical micrograph of the chip used, with the transmon qubit highlighted in red and the $\lambda/2$ microstrip resonator constituting the harmonic mode highlighted in blue. The resonator next to the harmonic mode is not used in the experiment; it is detuned from the other components in frequency. The scale bar in the bottom left corner is \unit[1]{mm}.
    (b) Circuit diagram of the setup on the chip. The qubit and the resonator are capacitively coupled, and the qubit also couples capacitively to another resonator, which is probed through a transmission line (TL) to read out the qubit state.
    (c) Pulse sequence used to simulate the quantum Rabi Hamiltonian.
    (d) Illustration of the frequencies for the harmonic mode and the applied drives.
    Figures, with some modifications, from Ref.~\cite{Braumuller2017}, J.~Braum\"{u}ller et al., \href{https://doi.org/10.1038/s41467-017-00894-w}{Nat.~Commun.~\textbf{8}, 779 (2017)}, with permission.
    \label{fig:SetupBraumuller2017}}
\end{figure}

The first implementation of the proposal in Ref.~\cite{Ballester2012} came in 2017~\cite{Braumuller2017}, using superconducting circuits~\cite{gu2017microwave, Krantz2019, circuit2021blais}. The experimental setup, shown in \figref{fig:SetupBraumuller2017}, consisted of an LC oscillator (the harmonic mode) and a transmon qubit~\cite{Koch2007} (an anharmonic LC oscillator, made anharmonic by the nonlinear inductance of Josephson junctions). Using the notation from \eqref{eq:HJC}, this setup had a harmonic-mode frequency $\omega_{\rm cav} / 2 \pi= \unit[5.948]{GHz}$, a qubit frequency $\omega_q$ tuned close to $\omega_{\rm cav}$, and a bare coupling strength $g / 2 \pi = \unit[4.3]{MHz}$, i.e., $g/\omega_{\rm cav} \lesssim 0.001$.

Applying two drives to the qubits as explained in \secref{sec:TwoDrivesJCTheory} and illustrated in \figpanel{fig:SetupBraumuller2017}{c,d}, the experimental simulation could reach deep into the USC regime with a ratio $g_{\rm eff} / \omega_{\rm eff} \approx 0.6$ between the effective coupling strength $g_{\rm eff}$ and the effective resonance frequency $\omega_{\rm eff}$. In these conditions, the quantum revival characteristics of the USC regime were observed (they are due to a different mechanism, compared to the revivals that can occur in the JC model~\cite{Eberly1980}).

The maximum drive strength of the stronger of the applied drives is limited by the anharmonicity of the qubit, $\alpha / 2 \pi = \mleft( \omega_{21} - \omega_{10} \mright) / 2 \pi = \unit[0.36]{GHZ}$, where $\omega_{10}$ and $\omega_{21}$ are the transition frequencies for the first and second transitions of the qubit, respectively. This is because a too strong drive would excite higher levels of the qubit. However, this did not preclude simulating a higher $g_{\rm eff}$ or a lower $\omega_{\rm eff}$; the actual limitation to observing dynamics for higher ratios $g_{\rm eff} / \omega_{\rm eff}$ was the decay rates of the qubit and the harmonic mode.


\paragraph{Trapped ions}

The proposal of Ref.~\cite{Pedernales2015}, discussed in \secref{sec:TwoDrivesJCTheory}, was first implemented in an experiment with a trapped ion in 2018~\cite{Lv2018}. In that experiment, an $^{171}\text{Yb}^+$ ion was trapped and cooled down to its motional ground state. Using the notation from \eqref{eq:HIon2Drives}, the qubit transition frequency was $\omega_q / 2 \pi = \unit[12.642812]{GHz}$ (using a transition in the $S_{1/2}$ hyperfine manifold) and the harmonic motional mode had a resonance frequency of $\omega_{\rm mot} / 2 \pi = \unit[2.498]{MHz}$. The drive strength $\Omega$ of the bichromatic laser drive was chosen such that the effective coupling strength was held fixed at $g_{\rm eff} = \eta \Omega/2 = \unit[2\pi \times 12.5]{MHz}$ and the detuning from the red sideband was set to zero, i.e., $\delta_r = 0$. The effective resonance frequency of both the qubit and the harmonic mode then became $\omega_{\rm eff} = \delta_b/2$, and the detuning $\delta_b$ from the blue sideband was varied to achieve different ratios $g_{\rm eff} / \omega_{\rm eff}$ between 0.04 and 1.2, with the latter simulating the DSC regime.

In simulations starting from a state with the qubit excited and the harmonic mode in its ground state, the experiment showed how the dynamics qualitatively differ between the regime where the JC model holds, the regime of USC but not DSC, and the regime of DSC. In the DSC case, the effect of the counter-rotating terms is especially clear, with the number of phonons climbing above six. Additional experiments simulating the DSC regime demonstrated phonon-number wave packets bouncing back and forth, adiabatically prepared the ground state of the quantum Rabi Hamiltonian in the DSC regime (starting from the JC regime and slowly increasing the effective coupling strength), and studied the low-energy spectrum of the quantum Rabi Hamiltonian.

In 2021, another experimental group used the same setup with a trapped $^{171}\text{Yb}^+$ ion to study a quantum phase transition in the quantum Rabi model~\cite{Cai2021}, demonstrating a theoretical prediction from Ref.~\cite{Hwang2015}. By making the red and blue sideband detunings $\delta_r$ and $\delta_b$ almost the same, the regime with $\omega_q \gg \omega_{\rm mot}$ was simulated. Here, the drive strength $\Omega$ was then swept to simulate sweeping $g_{\rm eff}$ across a quantum critical point. The behaviors predicted from the quantum phase transition for both qubit and phonon populations were observed, with both populations increasing significantly when $g_{\rm eff}$ exceeded the critical value.

The next year, the same group took this setup further by simulating the Rabi--Hubbard model~\cite{Mei2022}. In that model, several sites, each governed by a quantum Rabi Hamiltonian, are coupled together through hopping between the bosonic modes of the respective sites. The quantum Rabi Hamiltonian was simulated as above, while the hopping interaction was given by the Coulomb interaction between ions. In the experiment, which used up to 16 ions, a quantum phase transition was observed and various dynamics of the system were studied at a scale beyond what a classical computer can simulate.


\subsection{Simulated ultrastrong interactions in single-mode-driven Jaynes--Cummings systems}
\label{sec:SimulationSingleDrive}

One of the most important hallmarks of the ultrastrong regime of light-matter coupling is the existence of nonlinear processes that do not conserve the total number of excitations in the system. In Ref.~\cite{Kockum2017}, a complete description of many such processes is provided, which includes, for instance, the conversion of a single photon into multiple atomic excitations~\cite{Garziano2016, Macri2022, Tomonaga2023} or multiple photons~\cite{Wang2024}, multiphoton oscillations~\cite{Garziano2015}, converting an atomic superposition state into an entangled photonic state~\cite{Macri2018}, frequency conversion~\cite{Kockum2017a}, and nonlinear photon-mediated interactions~\cite{Stassi2017}.

In order to implement the majority of these nonlinear processes, a crucial requirement is that the Hamiltonian should neither conserve the total number of excitations nor its parity. The quantum Rabi model, with a light-matter coupling term of the form $g(a+a^\dag)(\sigma^-+\sigma^+)$, does not conserve the total number of excitations due to the terms $g a \sigma^-$ and $g a^\dag \sigma^+$. However, these terms still keep the total parity constant, since they add or remove excitations by an even number. Therefore, most of the aforementioned processes, such as the simultaneous absorption of a single photon by several atoms~\cite{Garziano2016, Tomonaga2023}, require the implementation of a generalized Rabi model with the addition of another coupling term of the form $\sigma_z (a+a^\dag)$, which allows to increase or decrease excitations by an odd number. As a result, one can engineer non-linear processes characterized by the coherent oscillation between an initial state $\ket{i}$ and a final state $\ket{f}$ with a different total number of excitations, such as the simultaneous absorption of a single photon by two atoms, that would be represented by the two states $\ket{i} = \ket{1,g,g}$ and $\ket{f} = \ket{0,e,e}$. This can be achieved if the system parameters are tuned such that $\ket{i}$ and $\ket{f}$ are eigenstates of the bare Hamiltonian with the same energy, allowing one to obtain an effective Hamiltonian that generates the coherent conversion between the two states,
\begin{empheq}[box =\widecolourbox]{equation}
	H_\mathrm{eff} = g_\mathrm{eff} \mleft( \ketbra{f}{i} + \mathrm{H.c.} \mright).
	\label{eq:H-general-effective}
\end{empheq}

From the point of view of simulations, a recent proposal~\cite{SanchezMunoz2020a} showed that all the nonlinear processes emerging from the generalized quantum Rabi model --- including those breaking parity conservation --- can be simulated in JC systems well below the USC regime, using only a single coherent drive. This result is achieved by working in the dressed basis of the atoms driven by the classical field. Let us consider the Hamiltonian of a single driven atom coupled to a cavity mode in the weak-driving regime described by the JC model, and written in the rotating frame of the drive:
\begin{equation}
	H = \Delta_a a^\dag a + \Delta_\sigma \sigma^+\sigma + g \mleft( a^\dag \sigma^- + a \sigma^+ \mright)+ \Omega \mleft( \sigma^+ + \sigma^- \mright),
\end{equation}
where $\Omega$ is the drive amplitude. The part that describes the driven atom can be exactly diagonalized. Denoting the
ground and excited eigenstates of the undriven atom $\ket{g}$ and $\ket{e}$, respectively, the eigenstates of the driven atom, $\ket{\pm}$, can be obtained as a rotation of the original eigenstates around the $y$ axis:
\begin{align}
	\ket{+} = \;& \cos \theta \ket{g} + \sin \theta \ket{e} = \exp(i \sigma_y 2 \theta) \ket{g}, \nonumber \\
	\ket{-} = \;& \sin \theta \ket{g} - \cos \theta \ket{e} =  - \exp[i
		\sigma_y (2 \theta + \pi)] \ket{g},
	\label{eq:dressed-qubit}
\end{align}
where $\cos\theta = 1 / \sqrt{1 + \xi^{-2}}$, $\sin \theta
= 1 / \sqrt{1 + \xi^2}$, $\theta \in [0, \pi/2]$, $\xi
= \Omega / (\Delta_\sigma / 2 + R)$, and $R$ is the Rabi
frequency given by 
\begin{equation}
	R = \sqrt{\Omega^2 + (\Delta_\sigma / 2)^2}.
	\label{Rabi}
\end{equation}
One can then define a new atomic lowering operator in the dressed basis, $\tilde \sigma^- = \ketbra{-}{+}$, related to $\sigma^-$ by
\begin{equation}
	\sigma^- = 
	\sin^2\theta \tilde \sigma^- - \cos^2\theta \tilde \sigma^+ + \cos\theta \sin\theta \tilde
	\sigma_z,
\end{equation}
with $\tilde\sigma_z = 2 \tilde \sigma^+ \tilde \sigma^- - \mathbb{I}$. This allows one to obtain a new light-matter Hamiltonian between the cavity mode and the dressed atom, of the form
\begin{equation}
	H = \Delta_a  a^\dag a +R \tilde \sigma_z 
	+ g \mleft[ \mleft( \sin^2\theta \tilde \sigma^- - \cos^2\theta \tilde \sigma^+
	+ \sin\theta\cos\theta \tilde \sigma_z \mright) a^\dagger + \mathrm{H.c.} \mright].\quad \label{eq:H-rotated}
\end{equation}
Crucially, this light-matter Hamiltonian has terms of the types $\propto a\tilde\sigma^-$ and $\propto a^\dagger \tilde\sigma^+$, which violate the conservation of the total number of excitations, and $\propto a \tilde\sigma_z$ and $\propto a^\dagger \tilde\sigma_z$, which violate the conservation of parity. Therefore, a light-matter interaction of the form in \eqref{eq:H-rotated} has all the necessary ingredients to simulate the nonlinear processes that can be achieved with a generalized quantum Rabi model, described by effective Hamiltonians of the form of \eqref{eq:H-general-effective}. The matrix elements of the effective Hamiltonian can be obtained, for instance, using the method described in the previous section, based on \eqref{eq:Heff-second-order}. An alternative technique to obtain this type of effective Hamiltonians, which is particularly convenient when processes higher than second order are involved, is described in \appref{app:Derivations of effective Hamiltonian}, with details on how to achieve a second-quantized form of the effective Hamiltonians in \appref{sec:hamiltonians-second-quantization}.

\subsubsection{Example: Two atoms simultaneously excited by a single photon}

We now briefly discuss a particular example of the simulation of a particular nonlinear process by which two atoms are simultaneously excited by a single photon. This process is a characteristic of ultrastrongly coupled light-matter systems~\cite{Garziano2016, Garziano2020, Tomonaga2023}.
We consider two atoms coupled to a single cavity mode. The states of the bare basis, working in the dressed-qubit picture, are $\{\ket{\pm,\pm,n}\}$. For the nonlinear process where a single photon is simultaneously absorbed by two atoms, the initial and final states are
\begin{equation}
\ket{i} = \ket{-,-,n+1}; \quad \ket{f} = \ket{+,+,n}.
\end{equation}

The Hamiltonian, in the rotating frame of the driving, reads
\begin{equation}
H = \Delta_a a^\dag a + \Delta_\sigma \mleft( \sigma_1^\dag \sigma_1 + \sigma_2^\dag \sigma_2 \mright) + \Omega \mleft( \sigma_1 + \sigma_2 + \mathrm{H.c.} \mright) + g \mleft[ a \mleft( \sigma_1^\dag + \sigma_2^\dag \mright) +
\mathrm{H.c.} \mright] \, , 
\end{equation}
where $a$ is the bosonic annihilation operator of the cavity and
$\Delta_a = \omega_a - \omega_\mathrm L$
($\Delta_\sigma = \omega_\sigma - \omega_\mathrm L$) 
is the cavity (qubit) detuning from the drive frequency. In the dressed qubit basis, it then becomes:
\begin{multline}
H = \Delta_a a^\dag a + R \mleft( \tilde\sigma_1^\dag \tilde\sigma_1 + \tilde\sigma_2^\dag \tilde\sigma_2 \mright)\\
+ g \mleft\{ a \mleft[\sin^2\theta \mleft(\tilde\sigma_1 + \tilde \sigma_2\mright)-\cos^2\theta \mleft(\tilde\sigma^\dagger_1 + \tilde \sigma^\dagger_2\mright) +\sin\theta\cos\theta \mleft(\tilde \sigma_{1,z} + \tilde\sigma_{2,z}\mright)\mright] + \mathrm{H.c.} \mright\} \,.
\end{multline}
When the resonance condition $\Delta_a \approx 4R$ is satisfied, the nonlinear process $\ket{-,-,n+1} \leftrightarrow \ket{+,+,n}$ is enabled, which can be described by an effective Hamiltonian
\begin{empheq}[box =\widecolourbox]{equation}
H_\mathrm{eff} = \Delta_a a^\dag a + \sum_i (R + \lambda) \tilde \sigma_{z, i} + \chi a^\dag a \tilde \sigma_{z, i}
+ g_\mathrm{eff} \mleft( a^\dag \tilde\sigma_1 \tilde\sigma_2 + a \tilde\sigma_1^\dag \tilde\sigma_2^\dag \mright).
\end{empheq}
Here, $\lambda$ and $\xi$ are, respectively, a Lamb shift and a dispersive coupling rate. Their expression can be found in \appref{sec:examples-one-drive}, where we elaborate on three examples of different nonlinear processes in further detail, including the one discussed here. Moreover, using perturbation theory, the effective rate of this process is computed to be
\begin{equation}
g_\mathrm{eff} = \frac{g^3}{3 R^2}\mleft(\cos^3\theta \sin^3\theta + 3\cos\theta \sin^5\theta \mright).
\label{eq:geffIII}
\end{equation}
%


\subsection{Digital simulation}
\label{sec:DigitalSimulation}

The simulation protocols described so far in this section all fall into the category of analog quantum simulation, where a system is arranged (in the cases we have seen, by the use of external drives) to behave like the system we want to simulate. The other approach to simulating quantum systems is digital quantum simulation~\cite{Lloyd1996, Georgescu2014}, where quantum gates or other means are used to turn different Hamiltonian interactions on and off in a sequence (a Trotter decomposition~\cite{Trotter1959}). This yields the same time evolution as the total Hamiltonian we wish to simulate. In examples in this subsection, the proposed and implemented schemes could also be called digital-analog quantum simulations~\cite{Lamata2018}, where the Hamiltonians available in the simulator are interleaved with quantum gates on parts of the system.


\subsubsection{Theoretical proposal}

A proposal for a digital-analog quantum simulation of the quantum Rabi model was first presented in 2014 in Ref.~\cite{Mezzacapo2014}. Similar to the proposal of Ref.~\cite{Ballester2012}, discussed in \secref{sec:TwoDrivesJCTheory}, the starting point is a system with a qubit coupled to a resonator, which is well described by the JC Hamiltonian in Eq.~(\ref{eq:HJC}).
The key insight used is that the quantum Rabi Hamiltonian
\be
H_{\rm Rabi} = \omega_{Rc} a^\dag a + \frac{\omega_{Rq}}{2} \sz + g_R \mleft( a + a^\dag \mright) \sx
\label{eq:HRabiDigital}
\ee
can be divided up into $H_{\rm Rabi} = H_1 + H_2$, where
\begin{align}
H_1 =\: & \frac{\omega_{Rc}}{2} a^\dag a + \frac{\omega_{q1}}{2} \sz + g_R \mleft( a \sp + a^\dag \sm \mright), \label{eq:H1} \\
H_2 =\: & \frac{\omega_{Rc}}{2} a^\dag a - \frac{\omega_{q2}}{2} \sz + g_R \mleft( a \sm + a^\dag \sp \mright), \label{eq:H2}
\end{align}
with $\omega_{q1} - \omega_{q2} = \omega_{Rq}$.

In a frame rotating with frequency $\omega_{RF}$, the standard JC Hamiltonian in \eqref{eq:HJC} becomes
\be
\tilde{H}_{\rm JC} = \Delta_c a^\dag a + \frac{\Delta_q}{2} \sz + g \mleft( a \sp + a^\dag \sm \mright),
\label{eq:HJCDigitalRotating}
\ee
with $\Delta_{c/q} = \omega_{c/q} - \omega_{RF}$. This has the same form as $H_1$ in \eqref{eq:H1}. Applying a qubit rotation (a bit flip) to \eqref{eq:HJCDigitalRotating} yields
\be
e^{-i \pi \sx / 2} \tilde{H}_{\rm JC} e^{i \pi \sx / 2} = \Delta_c a^\dag a - \frac{\Delta_q}{2} \sz + g \mleft( a \sm + a^\dag \sp \mright),
\label{eq:HAJCDigitalRotating}
\ee
which has the same form as $H_2$ in \eqref{eq:H2}. By changing the qubit frequency before performing the bit flip, we can set the effective simulated qubit frequency $\omega_{Rq}$. The effective simulated resonator frequency becomes $\omega_{Rc} = 2 \Delta_c$ and the effective simulated coupling is unchanged, $g_R = g$.

The simulation then consists of short Trotter steps, where the system is switched between evolving in time with the Hamiltonian in \eqref{eq:HJCDigitalRotating} and the Hamiltonian in \eqref{eq:HAJCDigitalRotating}, by detuning and flipping the qubit. To lowest order in $t/s$, where $t$ is the total evolution time and $s$ is the number of Trotter steps, this approximates the time evolution of the full quantum Rabi Hamiltonian as~\cite{Trotter1959}
\begin{empheq}[box =\widecolourbox]{equation}
\exp \mleft( - i t H_{\rm Rabi} \mright) \approx \mleft[ \exp \mleft( - i t H_2 / s \mright) \exp \mleft( - i t H_1 / s \mright) \mright]^s + \mathcal{O} \mleft( \frac{t^2}{s} \mright).
\label{eq:Trotter}
\end{empheq}
However, a better Trotter approximation is to let each step consist of first applying $H_1$ for a quarter of the time, then $H_2$ for half the time, and finally $H_1$ again for a quarter of the time. This ensures a smaller Trotter error.

As is also pointed out in Ref.~\cite{Mezzacapo2014}, this scheme is readily extended to simulating the Dicke Hamiltonian~\cite{Dicke1954}
\be
H_{\rm D} = \omega_{Dc} a^\dag a + \sum_{j=1}^N \frac{\omega_{Dq}}{2} \sigma_z^{(j)} + \sum_{j=1}^N g_D \mleft( a + a^\dag \mright) \sigma_x^{(j)}
\label{eq:HDickeDigital}
\ee
for $N$ qubits interacting with a single harmonic mode, by using a system well described by the Tavis--Cummings Hamiltonian~\cite{tavis68a}
\be
H_{\rm TC} = \omega_{\rm cav} a^\dag a + \sum_{j=1}^N \frac{\omega_q}{2} \sigma_z^{(j)} + \sum_{j=1}^N g \mleft( a \sigma_+^{(j)} + a^\dag \sigma_-^{(j)} \mright).
\label{eq:HTCDigital}
\ee
All that is required is to flip and detune all the qubits in parallel in exactly the same way as one qubit is flipped and detuned in the simulation of the quantum Rabi model. Note that in order to satisfy the gauge principle, the Dicke Hamiltonian in the dipole gauge is (see Ref.~\cite{Garziano2020gauge} for the case in the Coulomb gauge)
\begin{equation}\label{eq-dicke-dipole}
		H_{\rm D}^{\rm dg}=\omega_{\rm cav}a^{\dagger}a+\omega_{q}S_{z}+2ig_{\rm dg}(a^{\dagger}-a)S_{x}+4\eta g_{\rm dg} S_{x}^2,
\end{equation}
where $S_{x}=\frac{1}{2}\sum_{j=1}^{N}\sigma_{x}^{(j)}$. It is seen that the Dicke Hamiltonian in Eq.~(\ref{eq:HDickeDigital}) can be viewed as a simpler variant of Eq.~(\ref{eq-dicke-dipole}) in the dipole gauge, with neglecting the diamagnetic term $\propto S_{x}^2$. Note that, however, although this diamagnetic term is safely negligible below the USC regime, it can become important in the USC regime~\cite{DiStefano2019,Garziano2020gauge,generalized2023akbari}. 


The ideas from Ref.~\cite{Mezzacapo2014} were further developed and generalized in Ref.~\cite{Lamata2017}, which considered simulating some variations of the Dicke model, and also provided more detailed expressions for the error in the Trotter expansion shown in \eqref{eq:Trotter}. For example, the Hamiltonian of a Fermi-Bose condensate can be mapped onto a generalized Dicke model with inhomogeneous qubit frequencies~\cite{Yuzbashyan2005}, which can be simulated just like the standard Dicke Hamiltonian by tuning the qubit frequencies individually. A biased Dicke model, where a term $\sum_{j=1}^N \Delta \sigma_x^{(j)}$ is added to \eqref{eq:HDickeDigital}, appears somewhat more complicated to simulate, requiring more gates applied to the qubits. For a pulsed Dicke model~\cite{Dasgupta2015}, where the coupling strength $g_D$ in \eqref{eq:HDickeDigital} is time-dependent, the simulation scheme of Ref.~\cite{Mezzacapo2014} can be modified to include tunable couplings $g$ (provided that the experimental setup allows that).



\subsubsection{Experimental implementation}

\begin{figure}
	\centering
    \includegraphics[width=0.8\textwidth]{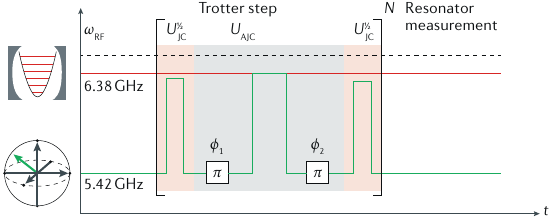}
	\caption{Digital quantum-simulation protocol implemented in Ref.~\cite{Langford2017}. The qubit (green line) starts at its sweet spot, detuned from the resonator frequency (red line). At the beginning of a Trotter step, the qubit frequency is tuned close to the resonator frequency. Next, the qubit is detuned again and flipped by a $\pi$ pulse (an additional phase $\phi_1$ is introduced by the choice of the rotation axis for the pulse), tuned back close to resonance (to a different frequency than the preceding time), detuned again and flipped back by a second $\pi$ pulse (introducing a second additional phase $\phi_2$), tuned close to resonance again (with the same frequency as the first time), and finally detuned back to the sweet spot. The first and last tuning close to resonance implements an effective JC Hamiltonian [\eqref{eq:HJCDigitalRotating}]; the middle tuning close to resonance (with the qubit flipped) implements an effective anti-JC Hamiltonian [\eqref{eq:HAJCDigitalRotating}]. The phase difference $\phi_1 - \phi_2$ sets the rotating frame in which the simulation takes place. 
    Figure from Ref.~\cite{Kockum2019}, A.~F.~Kockum et al., \href{https://doi.org/10.1038/s42254-018-0006-2}{Nat.~Rev.~Phys.~\textbf{1}, 19--40 (2019)} [adapted from Ref.~\cite{Langford2017}, N.~K.~Langford et al., \href{https://doi.org/10.1038/s41467-017-01061-x}{Nat.~Commun.~\textbf{8}, 1715 (2017)}], with permission. 
    \label{fig:SchemeLangford2017}}
\end{figure}

The proposal of Ref.~\cite{Mezzacapo2014} for digital quantum simulation of the quantum Rabi model was implemented experimentally in 2017~\cite{Langford2017}, using superconducting circuits. The main components of the setup were a resonator with a resonance frequency $\omega_{\rm cav} / 2\pi = \unit[6.381]{GHz}$ and a transmon qubit with a tunable frequency (sweet spot for protection from decoherence at $\omega_q / 2\pi = \unit[5.452]{GHz}$). The two were coupled capacitively with a strength $g / 2\pi = \unit[1.95]{MHz}$, so $g / \omega_{\rm cav} = 3 \times 10^{-4}$, far from the USC regime. In the simulation, an effective coupling strength up to $g_R / \omega_{Rc} \approx 1.8$ was realized.

The experimental protocol is sketched in \figref{fig:SchemeLangford2017}. To reduce Trotterization error, a second-order Trotter step as described below \eqref{eq:Trotter} was used. Each Trotter step took \unit[122]{ns}; one simulation run consisted of up to 90 Trotter steps. An improvement on the original protocol in Ref.~\cite{Mezzacapo2014} was that the rotation axes used to flip the qubit were chosen to yield phases that defined the frequency of the rotating frame in which the simulation takes place.

In the experiment, the simulation was used to observe the time evolution of both qubit and photon population from an initial state with the qubit excited and the resonator in its ground state. The phase-space dynamics of the resonator state were also mapped out through Wigner tomography of the resonator for different initial states and conditioned on the results of qubit measurements, demonstrating qubit-resonator entanglement. These insights into USC and DSC photonic states were helped by the fact that the resonator photons in this simulation setup always are real, not virtual. However, this feature of the simulator also meant that decoherence from the photon decay was a major limitation for the simulator performance.



\subsection{Other simulation methods}
\label{sec:OtherSimulation}

There are several approaches to simulating USC, which do not cleanly fall into the categories above of digital quantum simulation or analog quantum simulation through additional drives defining renormalized parameters, or which target other ultrastrongly coupled systems rather than those described by the quantum Rabi or Dicke models. In this subsection, we give an overview of these other approaches, starting from VQS to find the ground state of the multimode Dicke model in \secref{sec:VQS}, and continuing with schemes for analog simulation of the quantum Rabi model: using coupled waveguides in \secref{sec:CoupledWaveguides} and using two variations on setups with cold atoms in Secs.~\ref{sec:UltracoldAtoms} and \ref{sec:AtomicQDs}. We then conclude with a method for simulating USC between two resonators in \secref{sec:TwoResonatorsUSC} and schemes for the simulation of USC optomechanics in \secref{sec:SimOptomechUSC}.




\subsubsection{Variational quantum simulation}
\label{sec:VQS}

The broad class of quantum algorithms known as variational quantum algorithms (VQAs)~\cite{Cerezo2021} have been widely explored over the past few years, in the hope that they will be able to provide advantages when run on near-term quantum computers still subject to noise. In a VQA, the aim is to solve some problems encoded as minimizing a cost function. To find a solution, an ansatz in the form of a parameterized quantum circuit, applied to some initial quantum states, is chosen. By measuring properties of the output state after the circuit has been applied, information is gathered that lets a classical optimization algorithm update the parameters of the circuit to approach an optimum encoding the solution to the problem. In particular, VQAs can be applied to finding the ground or excited states of a given Hamiltonian in this manner, or even simulating the time evolution; this class of VQAs are known as VQS~\cite{Yuan2019, Cerezo2021}.

In 2020, Di Paolo et al.~\cite{DiPaolo2020} proposed the use of VQS to find the ground state of a system with an USC. The proposal considered a generalization of the quantum Rabi model [\eqref{eq:HRabi}] to $N$ qubits and $M$ harmonic modes, i.e., a multimode Dicke model, which is described by the Hamiltonian
\be
H = \sum_{i = 1}^N \frac{\omega_{qi}}{2} \sigma_i^z + \sum_{k = 1}^M \omega_k a^\dag_k a_k + \sum_{i = 1}^N \sum_{k = 1}^M g_{ik} \sigma_i^x \mleft( a_k + a_k^\dag \mright),
\label{eq:HMultiModeDicke}
\ee
where $\omega_{qi}$ is the transition frequency of qubit $i$, $\omega_k$ is the resonance frequency of mode $k$, and $g_{ik}$ is the strength of the coupling between qubit $i$ and mode $k$.

To treat the problem with VQS, the first step is to encode the bosonic modes in the qubits used in the VQS. The encoding chosen in Ref.~\cite{DiPaolo2020} truncated the Fock space of mode $k$ to have at most $n_k^{\rm max}$ photons and then encoded this state using $n_k^{\rm max} + 1$ qubits as $\ket{n_k} \rightarrow \ket{0_0 \ldots 0_{n_k - 1} 1_{n_k} 0_{n_k + 1} \ldots 0_{n_k^{\rm max}}}$. The annihilation operator of mode $k$ then maps to the qubit operators, such that
\be
a_k \rightarrow \sum_{n_k = 0}^{n_k^{\text{max} - 1}} \sqrt{n_k + 1} \sigma_{n_k}^+ \sigma_{n_k + 1}^-,
\ee
which facilitates simulation by only containing nearest-neighbour interactions between pairs of qubits.

The simulator is initialized in the state $\ket{\rm vac} = \otimes_{i = 1}^N \ket{0_{qi}} \otimes_{i = 1}^M \ket{0_k}$, where $\ket{0_k}$ is encoded in the qubits as shown above. To find the ground state $\ket{G}$ of the system, one applies an ansatz, a parameterized unitary operator $U(\boldsymbol{\theta})$, to $\ket{\rm vac}$ and searches for an optimal set of paramaters $\boldsymbol{\theta}^*$, such that
\be
\ket{G} \simeq U(\boldsymbol{\theta}^*) \ket{\rm vac}
\ee
and $\boldsymbol{\theta}^*$ minimizes the energy of the system [set by the Hamiltonian in \eqref{eq:HMultiModeDicke}]. In Ref.~\cite{DiPaolo2020}, the choice of $U(\boldsymbol{\theta})$ is based on the polaron transformation~\cite{Diaz-Camacho2016}, which approximately disentangles \eqref{eq:HMultiModeDicke}. For one qubit, this transformation is given by
\be
P = \prod_{k = 1}^M \exp \mleft[ \frac{g_k}{\omega_k + \omega_q'} \sx \mleft( a_k + a_k^\dag \mright) \mright],
\label{eq:PolaronTransformation}
\ee
where $\omega_q'$ is a renormalized qubit frequency. The transformation, applied as $P^\dag H P$, displaces the modes depending on the qubit state.

The final form of the variational ansatz is
\be
\prod_{i = 1}^N \prod_{k = 1}^M \prod_{s = 1}^{d_{ik}} \exp \mleft( \frac{f_{ik}^s}{d_{ik}} \sigma_i^x X_k^e \mright) \exp \mleft( \frac{f_{ik}^s}{d_{ik}} \sigma_i^x X_k^o \mright), 
\ee
where $X_k^e$ and $X_k^o$ act on the even and odd sites, respectively, in the qubit register encoding mode $k$, so as to form $X_k^e + X_k^o = a_k + a_k^\dag$. The variational parameters $f_{ik}^s$ are to be optimized in the VQS. They stem from the terms $\frac{g_k}{\omega_k + \omega_q'}$ in \eqref{eq:PolaronTransformation}. For the case of multiple qubits, an additional term is added before this ansatz to initialize the qubits in an entangled state instead of $\otimes_{i = 1}^N \ket{0_{qi}}$.

In Ref.~\cite{DiPaolo2020}, the ability of this variational ansatz to find the ground-state energy of small instances of \eqref{eq:HMultiModeDicke} was tested both in numerical simulations and in an experiment using three superconducting qubits on an IBM quantum processor. In the numerical simulations, the VQS came within a few percent (relative error) of the ground-state energy for the single- and two-mode quantum Rabi and Dicke models with a Fock-space truncation of $3$--$5$ photons per mode. Interestingly, the energy estimate from the VQS was better than the value given by doing the exact polaron transformation in \eqref{eq:PolaronTransformation}. In the experiment using superconducting qubits, the errors were larger, but there was qualitative agreement in how the ground-state energy changed as a function of $g/\omega_q$ when simulating the resonant quantum Rabi model [\eqref{eq:HRabi}] using two qubits to encode the mode. The errors were attributed to both the noise in the quantum processor and the performance of the classical optimization algorithm used in the VQS.



\subsubsection{Coupled waveguides}
\label{sec:CoupledWaveguides}

A scheme for analog simulation of the dynamics of the quantum Rabi model [\eqref{eq:HRabi}] proposed in 2011~\cite{Longhi2011} takes as its starting point the observation that the state of the system can be written as
\be
\ket{\Psi(t)} = \sum_{n = 0}^\infty A_n(t) \ket{g, n} + B_n(t) \ket{e, n},
\label{eq:RabiWfct}
\ee
where the first entry in the kets denotes the qubit state and $n$ denotes the photon number. Since the quantum Rabi Hamiltonian preserves the parity of the total number of excitations, the evolution of the coefficients in \eqref{eq:RabiWfct} decouples into two parity chains. If we define $c_n(t)$ to be $A_n(t)$ for even $n$ and $B_n(t)$ for odd $n$, its evolution is determined by
\be
i \frac{d c_n(t)}{d t} = \kappa_n c_{n+1} + \kappa_{n-1} c_{n-1} + \frac{(-1)^n}{2} \omega_q c_n + n \omega_{\rm cav} c_n,
\label{eq:RabiWfctCoefficients}
\ee
where $\kappa_n = g \sqrt{n + 1}$. The other parity chain has coefficients $f_n(t)$, which are $B_n(t)$ for even $n$ and $A_n(t)$ for odd $n$. The coefficients $f_n(t)$ obey the same evolution as in \eqref{eq:RabiWfctCoefficients}, but with $c_n(t)$ replaced by $f_n(t)$ everywhere.

The proposal of Ref.~\cite{Longhi2011}, implemented experimentally in 2012~\cite{Crespi2012}, is to construct an array of coupled photonic waveguides where the amplitudes of the propagating light in the waveguides evolve according to \eqref{eq:RabiWfctCoefficients}. In that setup, the propagation distance in the waveguide corresponds to the time, the coupling $g$ is set by the coupling between adjacent waveguides (which has to increase as $\sqrt{n+1}$ to make $g$ constant; this is achieved by setting the spacing between the waveguides), the qubit transition frequency is set by a modulation of the effective refractive index in the waveguides, and the resonator frequency is determined by the gradient of the refractive index.

In the experiment of Ref.~\cite{Crespi2012}, this proposal was implemented in 15 waveguides written with a femtosecond laser on a fused silica substrate. The relevant parameters were set such that $g / \omega_{\rm cav} = 0.65$, well in the USC regime, was realized. By measuring the light distribution in the waveguides, it was possible to observe the time evolution of the qubit and resonator populations in the simulation. To simulate a different set of parameter values, it was necessary to manufacture a new set of waveguides; no parameters were tunable in situ for this simulator.




\subsubsection{Ultracold atoms in optical lattices}
\label{sec:UltracoldAtoms}

Another proposed method for simulating the physics of the quantum Rabi model uses ultracold atoms in an optical lattice~\cite{Felicetti2017a}. In the limit of few atoms, the interactions between atoms are negligible and the Hamiltonian of a single atom in the lattice can be written as
\be
H = \frac{p^2}{2m} + \frac{V}{2} \cos \mleft( 4 k_0 x \mright) + \frac{m \omega_0^2}{2} x^2,
\label{eq:ColdAtomsH}
\ee
where $m$ is the mass of the atom, $p = - i \hbar \frac{\partial}{\partial x}$ is the momentum of the atom, $x$ is the position of the atom, $\omega_0$ is the frequency of the atomic motion in a harmonic trap induced by laser driving in the setup, $V$ is the depth of a periodic potential induced by laser driving, and $4 k_0$ is the wave vector of that potential (stemming from a four-photon interaction with a driving field having a wave vector $k_0$~\cite{Ritt2006}).

Assuming that the harmonic trap varies slowly on the scale of the periodic potential, a suitable basis for the system is the Bloch functions $\braket{x}{\phi_{n_b}(q)} = \phi_{n_b}(q, x) = \exp \mleft( i q x / \hbar \mright) u_{n_b} (x)$, where $q$ is the quasimomentum, $n_b$ is the band index, and $u_{n_b} (x)$ is a periodic function with the same periodicity as the periodic potential. Taking $u_{n_b} (x) = \exp \mleft( - 2 i k_0 x \mright) \exp \mleft( 4 i n_b k_0 x \mright)$, which yields a first Brillouin zone $q \in (- 2 \hbar k_0 , 2 \hbar k_0 ]$, in the Bloch basis the Hamiltonian in \eqref{eq:ColdAtomsH} becomes, for the two lowest-energy bands (i.e., $n_b = 0,1$),
\begin{empheq}[box =\widecolourbox]{equation}
H = \frac{1}{2m}
\begin{pmatrix}
q^2 + 4 \hbar k_0 q & 0 \\
0 & q^2 - 4 \hbar k_0 q
\end{pmatrix}
+ \frac{V}{4}
\begin{pmatrix}
0 & 1 \\
1 & 0
\end{pmatrix}
- \frac{m \hbar^2 \omega_0^2}{2} \frac{\partial^2}{\partial q^2} 
\begin{pmatrix}
1 & 0 \\
0 & 1
\end{pmatrix}
,
\label{eq:ColdAtomsHTwoBands}
\end{empheq}
assuming that the system state is fully contained in the first Brillouin zone. 

Rotating the qubit basis appropriately, the Hamiltonian in \eqref{eq:ColdAtomsHTwoBands} corresponds to the quantum Rabi Hamiltonian in \eqref{eq:HRabi} with a cavity frequency $\omega_{\rm cav} = \omega_0$, a qubit frequency $\omega_q = \frac{V}{2 \hbar}$, and a coupling strength $g = 2 k_0 \sqrt{\frac{\hbar \omega_0}{2 m}}$. In Ref.~\cite{Felicetti2017a}, it is shown that with typical parameters of ultracold rubidium atoms in an optical lattice, these values would correspond to $g / \omega_{\rm cav} \approx 10$, well into the regime of DSC, without much tunability possible, while the effective qubit frequency could be widely tuned in the range $\omega_q / \omega_{\rm cav} \in [0, 30]$.


The experimental implementation of the proposal from Ref.~\cite{Felicetti2017a} was published in 2023~\cite{Koch2023}. The experiment employed rubidium atoms in a setup with a trap frequency $\omega_0 / 2\pi$ in the range $[350, 750]\:\text{Hz}$ and an effective coupling strength $g / 2\pi \in [2290, 3090]\:\text{Hz}$, achieving $g / \omega_{\rm cav}$ up to 6.6. The effective qubit frequency $\omega_q / 2\pi$ could be tuned widely, from \unit[0]{Hz} to \unit[5850]{Hz}. These settings enabled the observation of the time evolution of both the bosonic and qubit excitations in simulations of the DSC regime. In a follow-up experiment with the same setup, collapse and revival of the excitation number in the DSC regime was observed~\cite{Hunanyan2023}.



\subsubsection{Atomic quantum dots coupled to superfluid Bose-Einstein condensates}
\label{sec:AtomicQDs}

An alternative way, compared to the proposal of Ref.~\cite{Felicetti2017a} in \secref{sec:UltracoldAtoms}, to use cold-atom systems for the simulation of the quantum Rabi model was put forward in Ref.~\cite{Felicetti2017}. In that proposal, a superfluid BEC in a shallow confining trap is made to interact with an atomic quantum dot~\cite{Recati2005}, i.e., a single atom (of the same species as the atoms in the BEC, but in a different hyperfine state) in a tight trap. The system is assumed to be in the collisional-blockade regime, meaning that the interaction $g_{dd}$ between atoms in the dot is much larger than interaction $g_{cc}$ between atoms in the BEC, such that the dot is never occupied by more than one atom, making it a two-level system. The BEC is assumed to be in the low-temperature superfluid regime, such that its quantum fluctuations can be described by a continuum of bosonic modes (phononic excitations). The coupling between the atomic quantum dot and the BEC is introduced through a Raman transition (effective Rabi frequency $\Omega$, detuning $\delta$) from external lasers.

The setup outlined above can be described by the Hamiltonian
\be
H = \sum_\mathbf{k} \omega_\mathbf{k} a^\dag_\mathbf{k} a_\mathbf{k} + \frac{\Omega_d}{2} \sz + \mleft[ - \delta' + \sum_\mathbf{k} g \mleft( a_\mathbf{k} + a^\dag_\mathbf{k} \mright) \mright] \frac{\sx}{2},
\ee
where the effective qubit frequency $\Omega_d$ is set by $\Omega$ and the number of atoms in the BEC, $\omega_\mathbf{k}$ are the BEC mode frequencies with the corresponding annihilation operators $a_\mathbf{k}$, and the coupling between the atomic quantum dot and these phononic modes is given by
\be
g = \sqrt{\frac{\omega_\mathbf{k}}{2 \hbar V g_{cc}}} \mleft( g_{cd} - g_{cc} \mright),
\ee
where $V$ is the volume of the BEC and $g_{cd}$ is the coupling between the atomic quantum dot and the atoms in the condensate. The parameter $\delta'$ depends on $g_{cd}$, $g_{cc}$, $\Omega$, and $\delta$~\cite{Recati2005}. Since the couplings $g_{cd}$, $g_{cc}$, and $g_{dd}$ can be tuned by external magnetic fields close to Feshbach resonances, it is possible to set $\delta' = 0$. By further having suitable boundary conditions for the BEC trap, the phononic modes can be spaced widely enough in frequency that only one of them effectively couples to the two-level system of the atomic quantum dot. We then recover the quantum Rabi Hamiltonian from \eqref{eq:HRabi}. 

In Ref.~\cite{Felicetti2017}, it is shown that for standard parameters of a setup with a rubidium BEC and a potassium atomic quantum dot, one can achieve an effective cavity frequency of a few hundred Hz (clearly lower than in most other simulation proposals discussed in this review) and an effective coupling $g$ that ranges from much less than that to much more than that, i.e., spanning from much less than USC all the way to DSC. It is also possible to add more atomic quantum dots to simulate the Dicke model instead of the quantum Rabi model.



\subsubsection{Ultrastrong coupling of two resonators via three-wave mixing}
\label{sec:TwoResonatorsUSC}

A close relative of the quantum Rabi and Dicke models is the Hopfield model~\cite{Hopfield1958}, which describes the interaction between two harmonic modes (not a harmonic mode and a qubit; however, the second harmonic mode can be an effective model for a collection of atoms) without having made the RWA. Omitting the diamagnetic term, the Hopfield Hamiltonian reads
\be
H_{\rm hpd} = \omega_a a^\dag a + \omega_b b^\dag b + G \mleft( a + a^\dag \mright) \mleft( b + b^\dag \mright),
\label{eq:Hopfield}
\ee
where $\omega_a$ and $\omega_b$ are the resonance frequencies of the modes $a$ and $b$, respectively, and $G$ is the strength of the coupling between them. Note that here, in order to satisfy the gauge principle, the Hopfield Hamiltonian in the dipole gauge is given by~\cite{Garziano2020gauge}
	\begin{equation}\label{eq-hopfield-dipole}
		H_{\rm hpd}^{\rm dg}=\omega_{a}a^{\dagger}a+\omega_{b}b^{\dagger}b+iG_{\rm dg}(a^\dagger-a)(b+b^{\dagger})+G^{\prime}_{\rm dg} (b+b^\dagger)^2,
	\end{equation}
and in the Coulomb gauge, by
	\begin{equation}\label{eq-hopfield-coulomb}
	H_{\rm hpd}^{\rm dg}=\omega_{a}a^{\dagger}a+\omega_{b}b^{\dagger}b-iG_{\rm cg}(b^\dagger-b)(a+a^{\dagger})+G^{\prime}_{\rm cg}(a+a^\dagger)^2,
\end{equation}
where $G_{\rm dg}$ and $G_{\rm cg}$ are the coupling strengths in the dipole and Coulomb gauges, respectively. Moreover, $G^{\prime}_{\rm dg}$ and $G^{\prime}_{\rm cg}$ are the corresponding diamagnetic amplitudes, which are related to the strengths $G_{\rm dg}$ and $G_{\rm cg}$, respectively. 
The Hopfield Hamiltonian in Eq.~(\ref{eq:Hopfield}) can be viewed as a simper variant in the dipole or Coulomb gauge, neglecting the diamagnetic term $\propto(b+b^\dagger)^2$ or $(a+a^\dagger)^2$. However, note that although these diamagnetic terms can be safely removed below the USC regime, they can become important in the USC regime~\cite{DiStefano2019,Garziano2020gauge,generalized2023akbari}.

\begin{figure}
	\centering
    \includegraphics[width=0.85\textwidth]{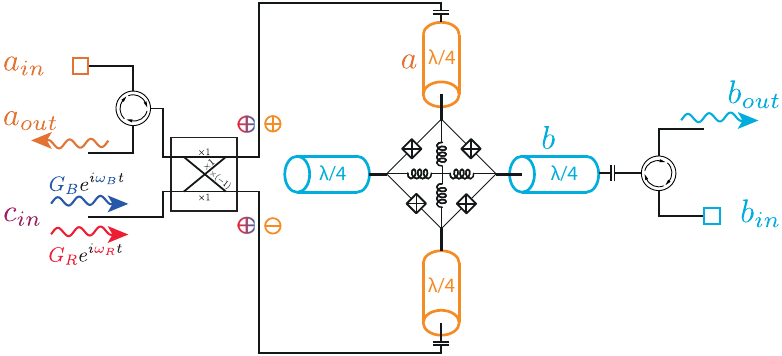}
	\caption{Setup of Ref.~\cite{Fedortchenko2017} for simulating USC between two resonator modes, $a$ (orange) and $b$ (blue). The ring of four Josephson junctions shorted by inductors in the middle of the two resonators realizes three-wave mixing between $a$, $b$, and a mode $c$ that can be driven from the left in the setup. The box on the left is a hybrid coupler which serves to couple modes $a$ and $c$ to different transmission lines.
    Figure from Ref.~\cite{Fedortchenko2017}, S.~Fedortchenko et al., \href{https://link.aps.org/doi/10.1103/PhysRevA.95.042313}{Phys.~Rev.~A \textbf{95}, 042313 (2017)}, with permission.
    \label{fig:SetupFedortchenko2017}}
\end{figure}

In 2017, Ref.~\cite{Fedortchenko2017} put forward a proposal for simulating \eqref{eq:Hopfield} in a setup with two superconducting resonators connected through a three-wave-mixing process generated by a Josephson mixer~\cite{Abdo2013}. By pumping the Josephson mixer with two tones, at frequencies $\omega_B = \omega_a + \omega_b + 2 \delta$ and $\omega_R = \omega_a - \omega_b$, respectively, where $\delta$ is a small detuning, the Hamiltonian for the setup (shown in \figref{fig:SetupFedortchenko2017}) becomes
\be
H_{3w} =  \omega_a a^\dag a + \omega_b b^\dag b + \sum_{j = B,R} \mleft[ \omega_j c_j^\dag c_j + \chi \mleft( c_j + c_j^\dag \mright) \mleft( a + a^\dag \mright) \mleft( b + b^\dag \mright) \mright],
\ee
where $c_R$ and $c_B$ are the annihilation operators for the two drive tones.

If the two drives are off resonance from the mixer, the stiff-pump approximation lets us treat $c_R$ and $c_B$ as two complex amplitudes. In the regime where $\abs{\delta} \ll \omega_a , \omega_b , \abs{\omega_a - \omega_b}$ and $\abs{\chi c_{B,R}} \lesssim \abs{\delta}$, applying the RWA results in an effective Hamiltonian
\begin{empheq}[box =\widecolourbox]{equation}
H_{\rm eff} = -\delta a^\dag a - \delta b^\dag b + G_B \mleft( a^\dag b^\dag + a b \mright) + G_R \mleft( a^\dag b + a b^\dag \mright), 
\end{empheq}
where $G_{B/R} = \chi c_{B/R}$, describing the system in a frame where the mode $a$ rotates at a frequency $\omega_a + \delta$, and the mode $b$ rotates at a frequency $\omega_b + \delta$. Choosing drive amplitudes such that $G_B = G_R=G$, this is of the same form as \eqref{eq:Hopfield} and allows tuning the ratio $G /|\delta|$ into the USC regime by selecting the detuning $\delta$ of the drive frequencies.

The proposal of Ref.~\cite{Fedortchenko2017} was implemented in an experiment in 2018~\cite{Markovic2018}. There, the mode frequencies were $\omega_a / 2\pi = \unit[8.477]{GHz}$ and $\omega_b / 2\pi = \unit[6.476]{GHz}$, much larger than the achievable interaction strength $\chi c_{B,R}$ from the three-wave mixing. However, the simulated effective frequency and coupling were of comparable size, with both a few tens of MHz. In this regime, Ref.~\cite{Markovic2018} demonstrated how the modes hybridize in the USC ground state, and measured both single- and two-mode squeezing of the emitted fields from the two resonators, as a result of the entanglement in the ground state.



\subsubsection{Simulation of ultrastrong optomechanics}
\label{sec:SimOptomechUSC}

The USC regime in cavity optomechanics refers to the case where the single-photon optomechanical coupling $g_0$ is comparable to the mechanical frequency $\omega_m$, i.e., $g_0 \sim \omega_m$. However, as mentioned in \secref{Amplified photon-mechanical interactions in cavity optomechanic}, the single-photon coupling $g_0$ is typically extremely weak, and reaching ultrastrong optomechanical interaction has remained challenging so far, despite considerable theoretical~\cite{rimberg2014cavity, nation2016ultrastong, shevchuk2017strong, kounalakis2020flux, Manninen2022} and experimental~\cite{fogliano2021mapping} effort aimed at this goal. Therefore, there have been several methods proposed to simulate ultrastrong optomechanical interaction with current technologies~\cite{johansson2014optomechanical,kim2015circuit,liao2020generalized}. 

Among these possible simulation methods, it was suggested in Ref.~\cite{liao2020generalized} that two bosonic modes, with annihilation operators $a$ and $b$, are coupled with a cross-Kerr interaction of strength $\chi$, i.e.,
\begin{equation}
H_{\rm Kerr} = \chi a^\dag a b^\dag b,
\end{equation}
similar to Eq.~(\ref{HamiltonianKerr}) and Eq.~(\ref{HamiltonianKerrPrime}), and one of these bosonic modes, say $b$, is driven by a detuned, strong driving of frequency $\omega_d$. Upon introducing a displacement transformation, $b = \delta b + \beta_b$, with $\beta_b$ being the field amplitude induced by the driving, it follows that
\begin{empheq}[box =\widecolourbox]{equation}
H_{\rm Kerr} \rightarrow -\beta_b \chi a^\dag a \mleft( \delta b^\dag + \delta b \mright).
\label{eq-simulated-ultrastrong-optomechanical-coupling-01}
\end{empheq}
Here, $\beta_b$ has been assumed to be real for simplicity and the residual cross-Kerr interaction has been neglected.

The right-hand side of \eqref{eq-simulated-ultrastrong-optomechanical-coupling-01} is of the same form as the optomechanical interaction Hamiltonian given in \eqref{eq:optomechanical interaction}. The effective optomechanical coupling strength is now $\beta_b \chi$ and the effective frequency of the mode $\delta b$, which plays the role of the mechanical resonator, is the detuning $\Delta_b = \omega_b - \omega_d$, where $\omega_b$ is the bare frequency of the mode $b$. Clearly, the ratio $\beta_b\chi/\Delta_{b}$ can be made comparable to (or even much larger than) unity by tuning the drive amplitude and frequency, such that an ultrastrong optomechanical interaction can be well simulated.

In an analogous way, it has been theoretically shown that a Fredkin-type interaction of three bosonic modes can also be used to simulate ultrastrong optomechanical interaction~\cite{yin2022all}, and that with the squeezing of the mode $b$, even a quadratic optomechanical coupling in the USC regime can be simulated~\cite{Zhou2021}. 

%
%
%
%
%
\section{Summary and perspectives}
\label{sec:SummaryPerspectives}

In this review, we have described theoretical and experimental quantum simulation and amplification methods developed in the past decade or so for increasing the coupling between light and matter, from strong to ultrastrong and deep strong, i.e., when the coupling strength becomes comparable to the frequencies of photons and characteristic atomic transitions.

It should be stressed that the methods described here are not limited to photon-mediated interactions, but have also been applied to amplify interactions mediated by phonons or magnons. In fact, the first experimental demonstrations of quantum amplified boson-mediated interactions were reported for phonons~\cite{burd2020quantum}, and only later for photons~\cite{villiers2022dynamically}. We believe that these described methods can also be generalized and applied to amplify the interactions mediated by exciton-polaritons, plasmons, magneto-plasmons, rotons, or other types of collective bosonic interactions.

In the past two decades, we have witnessed impressive theoretical and experimental progress demonstrating how the ultrastrong coupling of light and matter facilitates the generation or emulation of complex quantum phenomena that were hitherto beyond reach. Multiple experimental platforms, which comprise superconducting circuits, quantum wells, molecules, organic light-emitting diodes, two-dimensional electron gases, plasmonic nanoparticle crystals, or YIG (yttrium-iron-garnet) crystals, have been developed for probing a wide array of physical phenomena. From non-perturbative light-matter interactions to the emergence of novel quantum phases, such platforms have unraveled a host of intriguing phenomena with far-reaching fundamental and practical implications.

Despite this impressive progress, both theoretical and experimental, we hold the perspective that the exploration of the ultrastrong and deep-strong coupling regimes remains in its nascent stages, with a vast realm of novel effects and  potential applications for quantum technologies awaiting their emergence. It should be stressed that such applications are not limited to quantum information processing, quantum metrology and sensing, advanced materials (like cavity-mediated electron-photon superconductors) and nanotechnology, but also include possible applications for quantum chemistry~\cite{levine2009quantum}, quantum biology~\cite{lambert2013quantum}, or for testing supersymmetric (SUSY) theories. We believe that, by applying the described or other methods for amplifying light-matter interactions, such fundamental and practical objectives can be accomplished with greater ease and speed.

Let us conclude this review with the following opinion of Carlo Rubbia about the progress in quantum physics in general: ``I think Nature is smarter than physicists. We should have the courage to say: `Let Nature tell us what is going on.' Our experience of the past has demonstrated that in the world of the infinitely small, it is extremely silly to make predictions as to where the next physics discovery will come from and what it will be. In a variety of ways, this world will always surprise us all. We have to leave all this spectrum of possibilities open and just enjoy this extremely fascinating science.'' This opinion is not limited to elementary-particle physics, but it is also very relevant for the emerging field of the ultrastrong light-matter quantum interactions. Our perspective is that, with quantum amplified light-matter interactions, Nature can reveal a great deal about what is going on and how to make use of it.

\section*{Acknowledgements}
We thank Daniel H.~Slichter and Marius Villiers for their valuable comments.
W.Q.~acknowledges support of the National Natural Science Foundation of China (NSFC) (Grants No.~0401260012 and No.~62131002). A.M.~is supported by the Polish National Science Centre (NCN) under the Maestro Grant No.~DEC-2019/34/A/ST2/00081.
A.F.K.~acknowledges support from the Japan Society for the Promotion of Science through a BRIDGE Fellowship, the Swedish Research Council (grant number 2019-03696), the Swedish Foundation for Strategic Research, the Horizon Europe programme HORIZON-CL4-2022-QUANTUM-01-SGA via the project 101113946 OpenSuperQPlus100, and from the Knut and Alice Wallenberg Foundation through the Wallenberg Centre for Quantum Technology (WACQT). 
C.S.M.~acknowledges support of a fellowship from “la Caixa” Foundation (ID 100010434) and from the European Union’s Horizon 2020 research and innovation programme under the Marie Skłodowska-Curie Grant Agreement No.~847648, with fellowship code LCF/BQ/PI20/11760026, and financial support from the MCINN project PID2021-126964OB-I00 (QENIGMA) and the Proyecto Sinérgico CAM 2020 Y2020/TCS-6545 (NanoQuCo-CM).
F.N.~is supported in part by: 
Nippon Telegraph and Telephone Corporation (NTT) Research, the Japan Science and Technology Agency (JST) [via the Quantum Leap Flagship Program (Q-LEAP), and the Moonshot R\&D Grant Number JPMJMS2061], the Asian Office of Aerospace Research and Development (AOARD) (via Grant No.~FA2386-20-1-4069), and the Office of Naval Research (ONR) Global  (via Grant No.~N62909-23-1-2074).

\setcounter{figure}{0}
\appendix
\label{sec:Appendix}


\section{Eliminating squeezing-induced noise}
\label{app:Derivations of eliminating the noise induced by squeezing the cavity with a squeezed vacuum reservoir}

For simplicity, we consider a two-photon-driven cavity coupled to a squeezed vacuum bath with a squeezing parameter $r_{e}$ and a reference phase $\theta_{e}$. To begin, we work within a frame rotating at $H_{\rm rot}=\omega_{\rm 2ph}a^{\dagger}a/2+H_{B}$, where $a$ is the annihilation operator for the cavity mode, $\omega_{\rm 2ph}$ is the frequency of the parametric driving of the cavity mode, and $H_{B}=\sum_{l}\nu_{l}t^{\dagger}\mleft(\nu_{l}\mright)t\mleft(\nu_{l}\mright)$ is the free Hamiltonian of the bath, with $t\mleft(\nu_{l}\mright)$ being the annihilation operator for the bath mode of frequency $\nu_{l}$. In this frame, the full Hamiltonian is
\begin{equation}
H_F = H_{\rm DPA} + H_I ,
\end{equation}
where $H_{\rm DPA}$ is the Hamiltonian responsible for the parameter driving, which is given in Eq.~(\ref{eq:parametric driving Hamiltonian}), and 
\begin{equation}
H_I = \sum_{l}\lambda\mleft(\nu_{l}\mright)\mleft[t\mleft(\nu_{l}\mright)a^{\dagger}+{\rm H.c.}\mright]
\end{equation}
represents the coupling of the cavity mode to the bath, with the frequency-dependent coupling strength $\lambda\mleft(\nu_{l}\mright)$. 

To proceed, we introduce the squeezed cavity mode $a_{\rm sq}$ of frequency $\omega_{\rm sq}$ using the Bogoliubov transformation given in \eqref{eq:Bogoliubov transformation}. Then, we again switch into the frame rotating at $\omega_{\rm sq}a_{\rm sq}^{\dagger}a_{\rm sq}$; in this frame, the coupling of the cavity mode to the bath is transformed to
\begin{equation}\label{eq:cavity-bath coupling in the rotating frame}
H_{I}\mleft(t\mright)=a\mleft(t\mright)T^{\dagger}\mleft(t\mright)+{\rm H.c.}
\end{equation}
Here, we have defined
\begin{align}
a\mleft(t\mright) = \;&\exp\mleft(-i\omega_{\rm 2ph}t/2\mright)\exp\mleft(i\omega_{\rm sq}a_{\rm sq}^{\dag}a_{\rm sq}\mright)a\exp\mleft(-i\omega_{\rm sq}a_{\rm sq}^{\dag}a_{\rm sq}\mright),\\
T\mleft(t\mright) = \;&\sum_{l}\lambda\mleft(\nu_{l}\mright)t\mleft(\nu_{l}\mright)\exp\mleft(-i\nu_{l}t\mright).
\end{align}

Following the standard procedure in Ref.~\cite{scully1997book} and then returning to the frame rotating at $H_{\rm rot}$, we obtain the master equation expressed, in terms of the $a_{s}$ mode, as
\begin{align}\label{eq:master equation with thermal and two-photon noise}
\dot{\rho_{c}}=\;&-i\mleft[\omega_{\rm sq}a_{\rm sq}^{\dagger}a_{\rm sq},\rho_{c}\mright]\nonumber\\
&+\kappa\mleft(N_{\rm sq}+1\mright)\mathcal{L}\mleft(a_{\rm sq}\mright)\rho_{c}+\kappa N_{\rm sq}\mathcal{L}\mleft(a^{\dag}_{\rm sq}\mright)\rho_c \nonumber\\
&-\kappa M_{\rm sq}\mathcal{L}^{\prime}\mleft(a_{\rm sq}\mright)\rho_{c}-\kappa M^{*}_{\rm sq}\mathcal{L}^{\prime}\mleft(a^{\dag}_{\rm sq}\mright)\rho_{c},
\end{align}
where 
\begin{align}
\label{effective-thermal-noise}
N_{\rm sq} = \;& \cosh^{2}\mleft(r\mright)\sinh^{2}\mleft(r_{e}\mright)+\sinh^{2}\mleft(r\mright)\cosh^{2}\mleft(r_{e}\mright) + \frac{1}{2}\sinh\mleft(2r\mright)\sinh\mleft(2r_{e}\mright)\cos\mleft(\theta_{e}-\theta_{\rm 2ph}\mright),\\
\label{effective-tow-photon-correlation}
M_{\rm sq} = \;& \exp\mleft(-i\theta_{\rm 2ph}\mright)\mleft[\sinh\mleft(r\mright)\cosh\mleft(r_{e}\mright)+\exp\mleft[-i\mleft(\theta_{e}-\theta_{\rm 2ph}\mright)\mright]
\cosh\mleft(r\mright)\sinh\mleft(r_{e}\mright)\mright]\nonumber\\
&\times\mleft[\cosh\mleft(r\mright)\cosh\mleft(r_{e}\mright)+\exp\mleft[i\mleft(\theta_{e}-\theta_{\rm 2ph}\mright)\mright]\sinh\mleft(r_{e}\mright)
\sinh\mleft(r\mright)\mright],
\end{align}
and the Lindblad superoperators are defined by
\begin{align}
\mathcal{L}\mleft(o\mright)\rho = \;& o\rho o^{\dag}-\frac{1}{2}o^{\dag}o\rho-\frac{1}{2}\rho o^{\dag}o, \\
\mathcal{L}'\mleft(o\mright)\rho = \;& o\rho o-\frac{1}{2}oo\rho-\frac{1}{2}\rho oo.
\end{align}

The decay rate $\kappa$ of the cavity mode in Eq.~(\ref{eq:master equation with thermal and two-photon noise}) is expressed as 
\begin{equation}\label{eq:expression of the cavity decay rate}
\kappa=2\pi d\mleft(\omega_{\rm 2ph}/2\mright)\lambda^{2}\mleft(\omega_{\rm 2ph}/2\mright),
\end{equation}
where $d\mleft(\omega_{\rm 2ph}/2\mright)$ is the density of states for the squeezed vacuum bath at frequency $\omega_{\rm 2ph}$/2. Here, we have assumed that the central frequency of the squeezed vacuum bath is equal to half the two-photon driving frequency (i.e., $\omega_{\rm 2ph}$/2). In addition, we have also made the approximation
\begin{align}
d\mleft(\omega_{\rm 2ph}/2\pm\omega_{s}\mright)\approx d\mleft(\omega_{\rm 2ph}/2\mright).
\end{align}
This is justified as the frequency $\omega_{s}$ of the squeezed cavity mode is typically much smaller than the frequency $\omega_{\rm 2ph}$ of the parametric driving.
 
From Eqs.~(\ref{effective-thermal-noise}) and (\ref{effective-tow-photon-correlation}), we can, by setting $r_{e}=0$, see the effects of the thermal noise and the two-photon correlation noise caused by the detuned parametric driving. However, when choosing $r_{e}=r$ and $\theta_{e}-\theta=\pm n\pi$ ($n=1,3,5,\cdots$), we have
\begin{equation}
N_{\rm sq}=M_{\rm sq}=0,
\end{equation}
such that the master equation in Eq.~(\ref{eq:master equation with thermal and two-photon noise}) is simplified to 
\begin{equation}
\dot{\rho_{c}} = -i\mleft[\omega_{\rm sq}a_{\rm sq}^{\dagger}a_{\rm sq},\rho_{c}\mright] + \kappa\mathcal{L}\mleft(a_{s}\mright)\rho_{c}.
\end{equation}
It is seen that the squeezing-induced noises, arising both from the parametric driving and the squeezed vacuum reservoir, are completely suppressed as desired, and as a result, that the squeezed cavity mode is equivalently coupled to a vacuum reservoir.


\section{Effective Hamiltonians from perturbation theory}
\label{app:Derivations of effective Hamiltonian}

In this appendix, we describe a perturbative method that allows one to obtain the effective Hamiltonian in a relative straightforward manner. The main advantage of this method is that it can be cast in a very simple form, very similar to second-order perturbation theory, but can encapsulate effects beyond the second-order theory. 

Let us start by considering the following Hamiltonian in a matrix form:
\begin{equation}
H = 
\begin{pmatrix} 
H_A & \lambda H_{AB} \\ 
\lambda H_{AB}^\dag & H_B
\end{pmatrix}.
\label{eq:hamiltonian-initial-app}
\end{equation}
Written in this form, the separation between the two sectors of the Hilbert space, $A$ and $B$, is made explicit. Here, $A$ is the subspace of the Hilbert space consisting of $N_A$ states $\{\ket{a_1}, \ket{a_2}, \ldots \ket{a_{N_A}} \}$, and $B$ is another subspace, consisting of $N_B$ states $\{\ket{b_1}, \ket{b_2}, \ldots \ket{b_{N_B}} \}$. The Hamiltonians within $A$ and $B$ are $H_A$ and $H_B$, respectively, with matrix elements  given by $H_{A,ij} = \brakket{a_i}{H}{a_j}$ and $H_{B,ij} = \brakket{b_i}{H}{b_j}$. These two subspaces do not have independent dynamics; they are coupled by the ``interaction'' Hamiltonian $\lambda\hat H_{AB}$, where $\lambda H_{AB,ij} = \brakket{a_i}{H}{b_j}$. The parameter $\lambda$ has been included in this definition to act as a perturbation parameter that is considered small in the following. Defining the projectors
\begin{align}
P_A = \;& \sum_{i=1}^{N_A} \ketbra{a_i}{a_i}, \\
P_B = \;& \sum_{i=1}^{N_B} \ketbra{b_i}{b_i},
\end{align}
we can write
\begin{align}
H_A = P_A H P_A , \\
H_B = P_B H P_B , \\
\lambda H_{AB} = P_A H P_B.
\end{align}
The Hamiltonian $H_A$ is thus an $N_A\times N_A$ matrix, $H_B$ is an $N_B\times N_B$ matrix, and $H_{AB}$ is an $N_A\times N_B$ matrix.

\begin{figure}
	\centering	\includegraphics[width=0.4\textwidth]{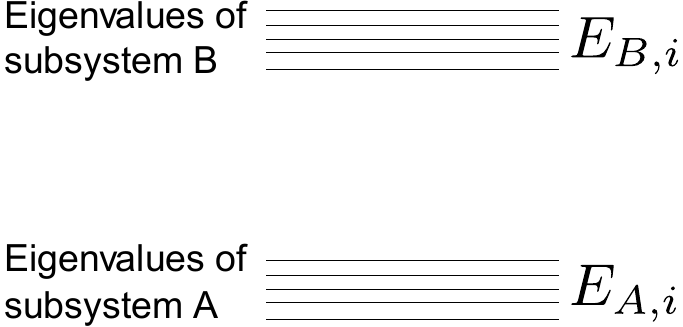}
	\caption{Requirements for applying adiabatic elimination. The energy levels, $E_{A, i}$ and $E_{B, i}$, of the subsystems $A$ and $B$ need to be distributed in well-separated manifolds. At the same time, the elements of coupling $A$ and $B$ need also to be much smaller than this difference. }\label{fig_cluster_eigenvalues}
\end{figure}

Our main goal is to obtain an effective Hamiltonian that describes the evolution of subsystem $A$ under the influence of subsystem $B$, without the need of explicitly describing subsystem $B$. This is only possible if the timescales of the intrinsic dynamics of the two subspaces are very different---this is why it is called adiabatic elimination---and still remain different even after considering the coupling between them. In energetic terms, this requires a ``clustering'' of energy levels, so that these energy levels can be split into some well-separated manifolds, as discussed, e.g., in Ref.~\cite{CohenTannoudji1998} and illustrated in Fig.~\ref{fig_cluster_eigenvalues}. In particular, by denoting the eigenvalues of $H_A$ and $H_B$ as $\{E_{A,i}\}$ and $\{E_{B,i}\}$, respectively, we require
\begin{equation}
\mleft| E_{A,i}-E_{A,j} \mright| \ll \mleft| E_{A,i}-E_{B,j} \mright|
\end{equation}
and
\begin{equation}
\lambda \mleft| \brakket{a_i}{H_{AB}}{b_j} \mright| \ll \mleft| E_{A,i} - E_{B,j} \mright|.
\end{equation}

Let us now discuss how this elimination can be performed in a simple way. For this, we write the eigenvalue equation for the Hamiltonian in \eqref{eq:hamiltonian-initial-app} as
\begin{equation}
\begin{pmatrix} 
H_A  & \lambda H_{AB} \\ 
\lambda H_{AB}^\dagger &  H_B 
\end{pmatrix} 
\begin{pmatrix} 
\varphi_A \\ 
\varphi_B 
\end{pmatrix} 
= E 
\begin{pmatrix} 
\varphi_A \\ 
\varphi_B 
\end{pmatrix} ,
\label{eq:starting-eigenvalue-eq}
\end{equation}
where $\varphi_A$ is an $N_A\times 1$ vector and $\varphi_B$ an $N_B\times 1$ vector. Explicitly performing the matrix multiplication yields two coupled equations for $\varphi_A$ and $\varphi_B$:
\begin{subequations}
	\begin{align}
	\label{eq:system1}
	(E-H_A)\varphi_A &= \lambda H_{AB}\, \varphi_B,\\
	\label{eq:system2} (E-H_B)\varphi_B &= \lambda H_{AB}^\dagger \,\varphi_A.
	\end{align}
\end{subequations}
Solving \eqref{eq:system2}, we obtain
\begin{equation}
\varphi_B = \frac{1}{E- H_B} \lambda H^\dagger_{AB}\, \varphi_A.
\end{equation}
Substituting this result into \eqref{eq:system1}, we obtain an eigenvalue equation for $\varphi_A$, equivalent to \eqref{eq:starting-eigenvalue-eq}, but in terms of an $N_A\times N_A$ Hamiltonian $H_\mathrm{eff}$, which only acts on elements of the subspace $A$:
\begin{equation}
(E-H_\mathrm{eff})\varphi_A = 0,
\label{eq:new-eigenvalue_equation}
\end{equation}
where
\begin{equation}
H_\mathrm{eff} = H_A + \lambda^2 H_{AB}\frac{1}{E-H_B} H^\dagger_{AB}.
\end{equation}
So far, this is an exact result, and one can see that the apparent simplicity of \eqref{eq:new-eigenvalue_equation} is hiding a considerable degree of complexity residing in the denominator of $H_\mathrm{eff}$ in the presence of $E$. This equation is therefore not so straightforward to solve. 

The key to obtain a truly effective Hamiltonian for the dynamics in the subspace $A$ is to apply our previous assumptions about the clustering of energy levels, which allows us to treat the Hamiltonian $\lambda H_{AB}$ as a perturbation. The unperturbed energy levels of the subspace $A$, i.e., the eigenvalues of $H_A$, are all clustered around an energy level that we call $E_0$, and they are well separated from unperturbed energy levels of the subspace $B$. All the terms of the coupling Hamiltonian $\lambda H_{AB}$ are small in comparison to this separation. Therefore, the corresponding perturbed energy levels of $A$ can have the energy $E=E_0 +O(\lambda)$. Considering that \eqref{eq:new-eigenvalue_equation} is an eigenvalue equation for one of those perturbed energy levels, we can, up to lowest order in the interaction parameter $\lambda$, write $H_{\mathrm{eff}}$ as
\begin{equation}
H_\mathrm{eff} \equiv H_A + \lambda^2 H_{AB}\frac{1}{E_0- H_B} H^\dagger_{AB}.
\label{eq:h_eff}
\end{equation}
This can now be understood as en effective Hamiltonian for the subspace $A$. This method of computing $H_\mathrm{eff}$, which has the appealing form of second-order perturbation theory, has the significant property of being able to encapsulate any higher-order process that occurs within the eliminated subspace $B$. This aspect is discussed in more detail in the main text.


\section{Effective Hamiltonians in second-quantized form}
\label{sec:hamiltonians-second-quantization}

The effective Hamiltonian of \eqref{eq:h_eff} is defined in the subspace $A$. In most instances, this result can be generalized and written in terms of the general bosonic and atomic annihilation and creation operators, giving a clearer understanding of nonlinear processes. To show how this can be done, let us consider a general situation in which we have $N_{\mathrm{cav}}$ cavities and $N_\mathrm{a}$ atoms (or qubits), and the initial and final states are some tensor products in the dressed-qubit basis:
\begin{align}
\ket{i} =\;& \bigotimes_{k=1}^{N_{\mathrm{cav}}} \ket{n_k} \bigotimes_{j=1}^{N_{\mathrm{a}}} \ket{s_j}, \nonumber\\
\ket{f} =\;& \bigotimes_{k=1}^{N_{\mathrm{cav}}} \ket{n'_k} \bigotimes_{j=1}^{N_{\mathrm{a}}} \ket{s'_j},
\end{align}
where $n_k,n'_k \in \mathbb{N}$ denote the photon numbers and $s_j, s'_j \in \{+,-\}$. We assume $n_k \neq n'_k$ and $s_j \neq s'_j$ --- otherwise, the corresponding cavity or atom does not change during the whole process, and then it can be factored out.
Let us also define the change of the photon number in cavity $k$ as $\Delta n_k = n'_k-n_k$ (which characterizes the process),  a transition operator $C_k$ for cavity $k$ as
\begin{equation}
	C_k \equiv \begin{cases}
		a_k & \text{if $\Delta n_k <0$} , \\
		a_k^\dag & \text{if $\Delta n_k >0$},
	\end{cases} 
\end{equation}
and a transition operator $A_j$ for atom $j$ as
\begin{equation}
	A_j \equiv \begin{cases}
		\sigma_j^+ & \text{if $s'_j = +$} , \\
		\sigma_j^- & \text{if $s'_j = -$},
	\end{cases} 
\end{equation}
so that $\ket{f} \propto \prod_{k,j}C_k^{|\Delta n_k|} A_j \ket{i}$.

As a result, we can obtain a general expression for the effective Hamiltonian in terms of the creation and annihilation operators,  provided we can find a set of parameters $\chi_{i,j}$, $\lambda_j$, and $g_{\mathrm{eff}}$, which allows one to rewrite the matrix $H_\mathrm{eff}$ in the form
\begin{equation}
	H_\mathrm{eff} = H_A + 
        \begin{pmatrix}
		\sum_{k,j}(\chi_{k,j}n_k s_j + s_j \lambda_j) + \alpha & g_\mathrm{eff}^* \prod_k  \sqrt{(\max[n_k,n_k'])_{|\Delta n_k|}} \\
		g_\mathrm{eff} \prod_k  \sqrt{(\max[n_k,n_k'])_{|\Delta n_k|}} & \sum_{k,j} (\chi_{k,j}n'_k s'_j  + s'_j \lambda_j) + \alpha
	    \end{pmatrix}.
	\label{eq:h_eff_general}
\end{equation}
Here, note that $\alpha$ is an overall energy shift that can be ignored. We can then rewrite \eqref{eq:h_eff_general} as
\begin{equation}
	H_\mathrm{eff} = H_A  + \sum_{k,j}\mleft[ \chi_{k,j}\, a^\dagger_k a_k \tilde\sigma_{j,z}  + \lambda_j \tilde\sigma_{j,z}  +g_\mathrm{eff} \mleft( \prod_{k,j} C_k^{|\Delta n_k|}A_j + \text{H.c} \mright) \mright].
\end{equation}
From this expression, it is clear that $\chi_{k,j}$ describes an effective dispersive coupling between cavity $k$ and atom $j$; $\lambda_j$ is a Lamb shift of the natural frequency of atom $j$ and $g_\mathrm{eff}$ is the rate of the coherent exchange between $\ket{i}$ and $\ket{f}$.


\setcounter{figure}{0}

\section{Examples of simulating ultrastrong light-matter interactions in single-mode-driven Jaynes--Cummings systems}
\label{sec:examples-one-drive}

\begin{figure}
	\centering	
    \includegraphics[width=0.75\textwidth]{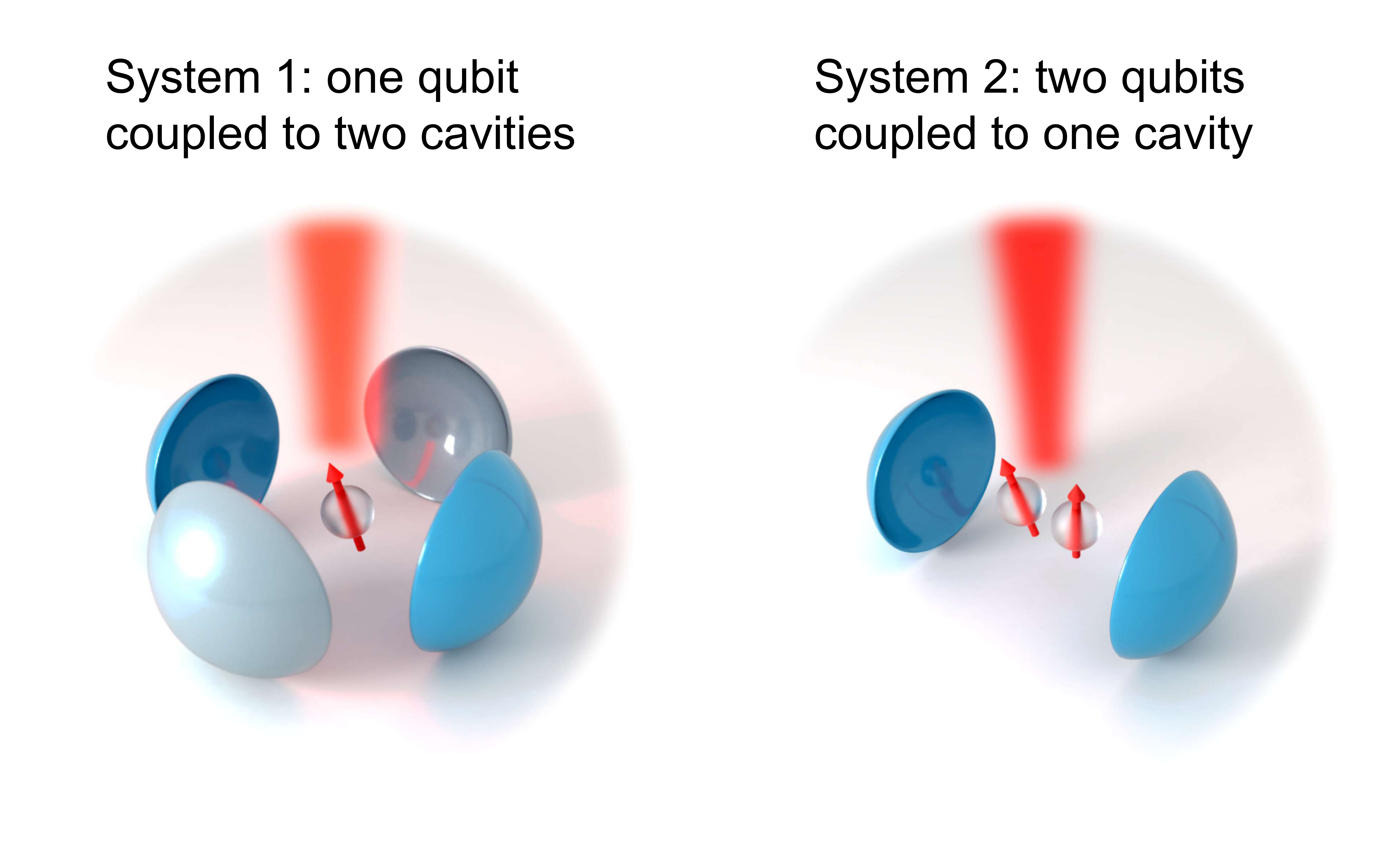}
	\caption{Cavity-QED setups used to simulate ultrastrong light-matter interactions with the addition of a single atomic drive. In system 1, a single qubit (or atom) is assumed to be coupled to two cavity modes, and in system 2, two qubits are assumed to be coupled to a single cavity mode.}
    \label{fig:sketch-one-drive}
\end{figure}

In this section, we provide details of the derivation of the effective Hamiltonians, which  describe the nonlinear characteristic of ultrastrong light-matter interactions simulated in JC-type systems with the addition of a single drive, as discussed in \secref{sec:SimulationSingleDrive} of the main text. The three cases that we consider here are based on two examples of the cavity-QED systems depicted in Fig.~\ref{fig:sketch-one-drive}: a single qubit coupled to two cavities, and two qubits coupled to a single cavity. The three nonlinear processes that we discuss are: 
(i) two photons simultaneously emitted by a single atom; 
(ii) frequency conversion of photons; and 
(iii) two atoms simultaneously excited by a single photon. These three cases are examples of nonlinear processes that have been discussed in the context of ultrastrongly coupled light-matter systems, e.g., in Refs.~\cite{Garziano2016, Stassi2017, Kockum2017a}.


\subsection{Example I: Two photons simultaneously emitted by a single atom}

\begin{figure}[t]
	\centering	\includegraphics[width=\textwidth]{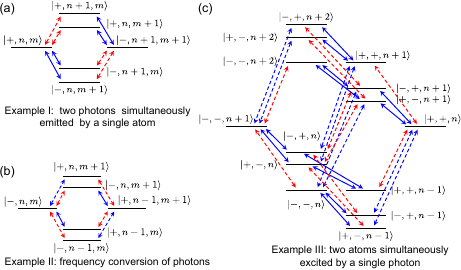}
	\caption{Intermediate transition processes that mediate the effective interactions 
    (a) between the two states $\ket{+,n,m}$ and $\ket{-,n+1,m+1}$ in example I, 
    (b) between the two states $\ket{+,n,m+1}$ and $\ket{-,n+1,m}$ in example II, and 
    (c) between the two states $\ket{-,-,n+1}$ and $\ket{+,+,n}$ in example III. Blue solid arrows denote transitions that conserve the number of excitations, red dashed arrows denote transitions that change the number of excitations by one, and blue dashed arrows denote transitions that change the number of excitations by two. Transitions marked by blue arrows conserve parity; transitions denoted by red arrows do not. See also Refs.~\cite{Garziano2016, Kockum2017a, Kockum2017}.}
    \label{fig:eigenvalues-one-drive2}
\end{figure}

\paragraph{System}

The setup considered is system 1 displayed in Fig.~\ref{fig:sketch-one-drive}: a single atom---described as a qubit---is coupled to two cavity modes. The states of the bare basis in the dressed-qubit picture are $\{ \ket{\pm,n,m} \}$, where $-$ and $+$ stand for the dressed-qubit states of \eqref{eq:dressed-qubit}, and $n,m=0,1,\ldots,$ are the numbers of photons in the two cavity modes, respectively.
	
\paragraph{Nonlinear process}

We consider the nonlinear process where a single dressed atom simultaneously emits two photons, and vice versa. The initial and final states are thus
\begin{equation}
	\ket{i} = \ket{+,n,m}; \quad \ket{f} = \ket{-,n+1,m+1}.
\end{equation}

\paragraph{Intermediate states}

For any choice of $\ket{i}$ and $\ket{f}$---i.e., any value of $m$ and $n$---the subspace of intermediate states involved in the process (to lowest order) consists of the following 12 states:
\begin{multline}
	\{ \ket{\pm,n+1,m}, \ket{\pm,n,m+1}, \ket{\pm,n+2,m+1}, \\
	\ket{\pm,n+1,m+2}, \ket{\pm,n,m-1}, \ket{\pm,n-1,m} \}.  
\end{multline}
Among these 12 states, there are four that contribute, as intermediate states, to the interaction between $\ket{i}$ and $\ket{f}$, as shown in \figpanel{fig:eigenvalues-one-drive2}{a}. The other eight states, which are not plotted in \figpanel{fig:eigenvalues-one-drive2}{a}, contribute to the Lamb shifts and the dispersive couplings, which need to be accounted for since these are of the same order in $g$ as the effective coupling.
	
\paragraph{System Hamiltonian}

The Hamiltonian, in the rotating frame of the drive, is given by
\begin{equation}
	H = \Delta_1 a_1^\dag a_1 + \Delta_2 a_2^\dag a_2 + \Delta_\sigma \sigma^\dag \sigma + \Omega \mleft( \sigma + \sigma^\dag \mright) + g \mleft[ \sigma \mleft( a_1^\dag + a_2^\dag \mright) +
	\mathrm{H.c.} \mright], \label{eq:hamiltonian_two_photons}
\end{equation}
with $a_{1,2}$ the bosonic annihilation operators of the cavity modes, $\Delta_1$, $\Delta_2$, and $\Delta_\sigma$ the frequency detunings of the cavities and qubit with the drive (i.e., $\Delta_x = \omega_x - \omega_\mathrm L$), $\Omega$ the amplitude of the driving field, and $g$ the coupling rate between the cavities and the qubit (considered to be equal for simplicity). In the dressed-qubit basis, we obtain
\begin{equation}
	H = \Delta_1 a_1^\dag a_1+\Delta_2 a_2^\dag a_2 + R \tilde \sigma_z 
	+ g \mleft[ \mleft( \sin^2\theta \tilde \sigma - \cos^2\theta \tilde \sigma^\dag
	+ \sin\theta\cos\theta \tilde \sigma_z \mright) \mleft( a_1^\dag + a_2^\dag
	\mright) + \mathrm{H.c.} \mright]. \label{eq:H-dfg}
\end{equation}

\paragraph{Approximate resonance conditions}

The nonlinear process $\ket{+,m,n} \leftrightarrow \ket{-,n+1,m+1}$ is enabled when the following resonance conditions are met:
\begin{equation}
	\Delta_1 + \Delta_2 \approx 2R, \quad  \Delta_1 \neq \Delta_2 \neq
	\mleft( \pm R, \pm 2R \mright). \label{eq:resonance-condition1}
\end{equation}
The second condition is imposed in order to detune the first-order processes (e.g., $\tilde \sigma a_1^\dag + \text{H.c.}$ if $\Delta_1 = 2 R$), and the competing second-order processes [e.g., $\mleft(\tilde \sigma {a_1^\dag}^2 + \text{H.c.}\mright)$ for $\Delta_1 = R$] that can excite degenerate photon pairs within a single cavity~\cite{SanchezMunoz2014,SanchezMunoz2015, SanchezMunoz2018}.

\paragraph{Effective Hamiltonian}

The resulting effective Hamiltonian, using \eqref{eq:h_eff}, takes the form
\begin{equation}
	H_\mathrm{eff}^\mathrm{I} = H_A + 
    \begin{pmatrix}
		\chi_1 n + \chi_2 m + \lambda + \alpha  & g^\mathrm{I}_\mathrm{eff}\sqrt{(n+1)(m+1)} \\
		g^\mathrm{I}_\mathrm{eff}\sqrt{(n+1)(m+1)}  & - \chi_1 (n+1) - \chi_2 (m+1) - \lambda + \alpha
	\end{pmatrix},
	\label{eq:heff_I_matrix}
\end{equation}
which, following the procedure outlined above, can be written as
\begin{equation}
	H_\mathrm{eff}^\mathrm{I} = \Delta_1 a_1^\dag a_1 + \Delta_2 a_2^\dag a_2
	+ \mleft(R + \lambda \mright) \tilde \sigma_z + \mleft( \chi_1 a^\dag_1 a_1 + \chi_2 a^\dag_2 a_2 \mright) \tilde \sigma_z + g_\mathrm{eff}^\mathrm{I} \mleft( a_1^\dag a_2^\dag \tilde \sigma + \mathrm{H.c.} \mright).
\end{equation}
Here, $g_\mathrm{eff}^\mathrm{I}$ refers to the effective two-photon coupling rate; it is given by
\begin{equation}
	g_\mathrm{eff}^\mathrm{I} = g^2 \cos\theta \sin^3\theta[R f (1 - f)]^{-1},
	\label{eq:geffI}
\end{equation}
where we have introduced a parameter $f \in (0,1)$ such that
\begin{equation}
	\Delta_1 = 2 f R,\quad \Delta_2 =2 (1 - f) R,
	\label{Delta}
\end{equation}
ensuring that the resonance condition is automatically fulfilled. Moreover, the Lamb shift of the qubit is
\begin{equation}
	\lambda = \frac{g^2}{2} \mleft[\cos^4\theta \mleft(
	\Delta^{-1}_{1,+}+\Delta^{-1}_{2,+} \mright) - \sin^4\theta
	\mleft(\Delta^{-1}_{1,-}+ \Delta^{-1}_{2,-} \mright) \mright],
\end{equation}
and the dispersive coupling rates are
\begin{equation}
	\chi_i = g^2 \mleft( \frac{\cos^4\theta}{\Delta_{i,+}}- \frac{\sin^4\theta}{\Delta_{i,-}}\mright), \, i\in\{1,2\},
\end{equation}
with $\Delta_{i,\pm}\equiv \Delta_i \pm 2 R$. 

\paragraph{Exact resonance conditions}

\begin{figure}
	\centering
    \includegraphics[width=\textwidth]{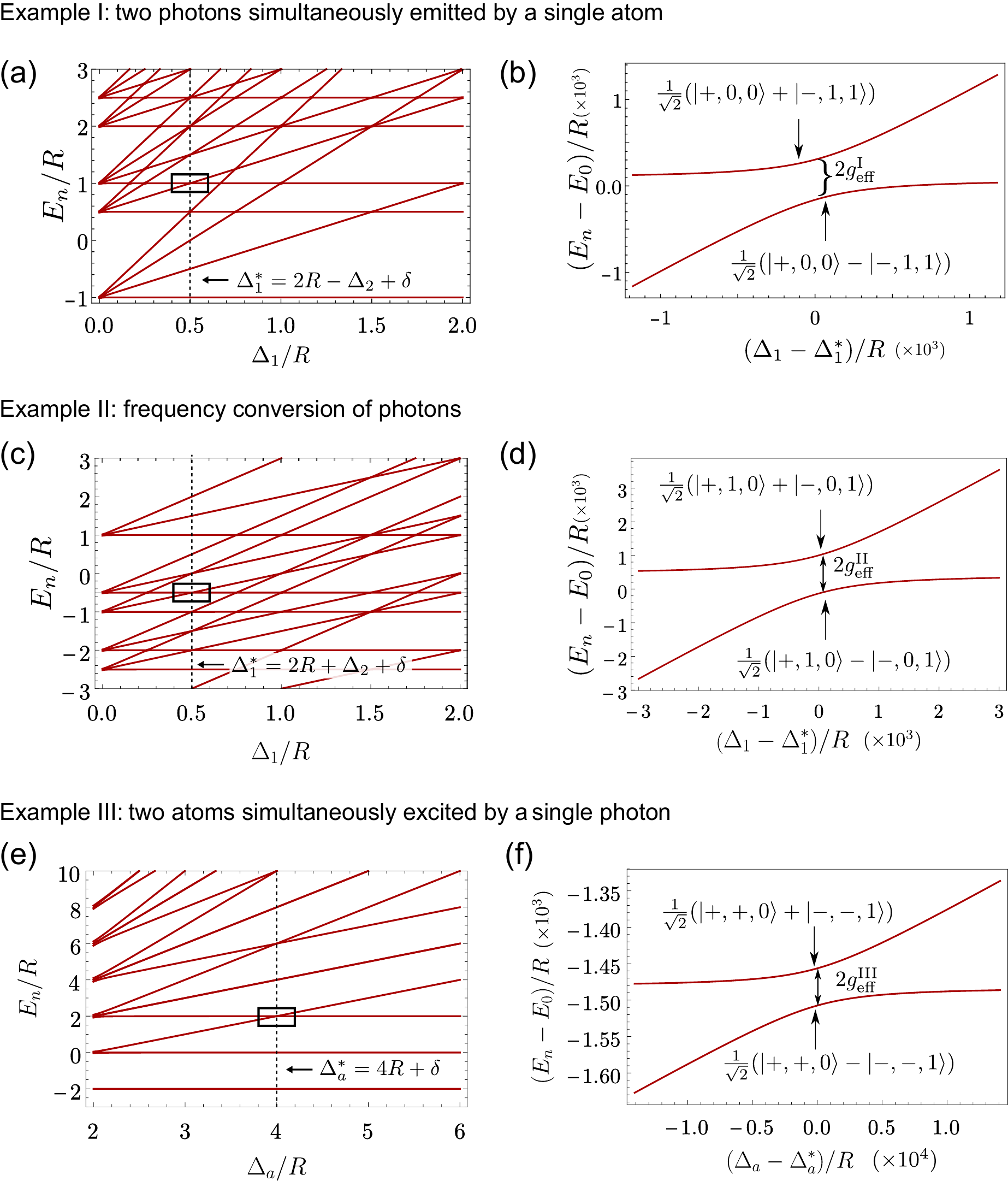}
	\caption{(a), (c), (e) Lowest energy levels of the system Hamiltonians in examples I, II, and III, respectively, of simulating ultrastrong light-matter interactions. Here, $E_n$ refers to the eigenenergy of the $n$th eigenstate; $\Delta_{1}^*$ and $\Delta_a^*$ denote the values of the detunings $\Delta_{1}$ and $\Delta_a$, respectively, at which the exact resonance conditions are fulfilled. 
    (b), (d), (f) Enlarged view of the boxed regions in (a), (c), (e), respectively, each showing an avoided level crossing around their bare eigenenergy $E_0$. These avoided level crossings originate from the state hybridization due to the high-order nonlinear interactions shown in Fig.~\ref{fig:eigenvalues-one-drive2}.}
    \label{fig:eigenvalues-one-drive1}
\end{figure}

In order to obtain the full Rabi oscillations between the states $\ket{i}$ and $\ket{f}$, the diagonal elements of $H_\mathrm{eff}^\mathrm{I}$ should be equal. While the approximate resonance condition, given in \eqref{eq:resonance-condition1}, allows us to justify the derivation of $H_\mathrm{eff}^\text{I}$, it does not exactly meet the condition for the equal diagonal elements of $H_\mathrm{eff}^\mathrm{I}$. Thus, it needs to be fine-tuned. In order to do this, we introduce a small correction $\delta \ll R$ to the value of $\Delta_1$, such that 	
\begin{equation}
\Delta_1 = 2fR + \delta=2R-\Delta_{2}+\delta,
\end{equation} 
and then find the $\delta$ that makes the diagonal elements of $H_\mathrm{eff}^\text{I}$ equal. The resulting $\delta$ is 
\begin{equation}
	\delta = 2\lambda + \chi_1(2n+1) + \chi_2(2m+1).
	\label{eq:delta_resonance_condition_caseI}
\end{equation}
The lowest energy levels and the avoided level crossing arising under the exact resonance condition are plotted in Figs.~\figpanelNoPrefix{fig:eigenvalues-one-drive1}{a} and \figpanelNoPrefix{fig:eigenvalues-one-drive1}{b}, respectively.

\subsection{Example II: Frequency conversion of photons}

\paragraph{System}

The setup considered here is again system 1 in \figref{fig:sketch-one-drive} as in example I: a single atom---described as a qubit---is coupled to two cavity modes. The states of the bare basis, working in the dressed-qubit picture, are again $\{ \ket{\pm,n,m} \}$.
	
\paragraph{Nonlinear process}

We consider the nonlinear process where a photon is converted into a photon of different energy. The initial and final states are, thus, given by:
\begin{equation}
	\ket{i} = \ket{+,n,m+1}; \quad \ket{f} = \ket{-,n+1,m}.
\end{equation}

\paragraph{Intermediate states}

For any choice of $\ket{i}$ and $\ket{f}$---i.e., any value of $m$ and $n$---the subspace of intermediate states involved in the process (to lowest order) consists of the following 12 states:
\begin{multline}
	\{ \ket{\pm,n+1,m+1}, \ket{\pm,n,m}, \ket{\pm,n+2,m}, \\ \ket{\pm,n+1,m-1}, \ket{\pm,n,m+2}, \ket{\pm,n-1,m} \}
\end{multline}
As in the previous example, there are four states that contribute, as intermediate states, to the interaction between $\ket{i}$ and $\ket{f}$, as shown in \figpanel{fig:eigenvalues-one-drive2}{b}. The other eight states, which are not shown, yield the Lamb and dispersive shifts.
	
\paragraph{System Hamiltonian}

The Hamiltonian is the same as Eq.~(\ref{eq:hamiltonian_two_photons}) in example I.
	
\paragraph{Approximate resonance conditions}

The nonlinear process $\ket{+,n,m+1} \leftrightarrow \ket{-,n+1,m}$ is enabled when the following resonance conditions are met:
\begin{equation}
	\Delta_1 \approx  2 R + \Delta_2, \quad  \Delta_1 \neq \Delta_2
	\neq \pm R.
    \label{eq:resonance-condition2}
\end{equation}
As in example I, the second condition guarantees that the second-order processes, introducing photon pairs into the cavities, are out of resonance.
	
\paragraph{Effective Hamiltonian}

The resulting effective Hamiltonian, using \eqref{eq:h_eff}, takes the form
\begin{equation}
	H_\mathrm{eff}^\mathrm{II} =  H_A +
    \begin{pmatrix}
	\chi_1 n + \chi_2 (m+1) + \lambda + \alpha & g^\mathrm{II}_\mathrm{eff}\sqrt{(n+1)(m+1)} \\
	g^\mathrm{II}_\mathrm{eff}\sqrt{(n+1)(m+1)} & - \chi_1 (n+1) - \chi_2 - \lambda + \alpha
	\end{pmatrix},
\end{equation}
which, following the procedure outlined above, can be written as
\begin{equation}\label{eq-effH-example2}
	H_\mathrm{eff}^\mathrm{II}= \Delta_1 a_1^\dag a_1 + \Delta_2 a_2^\dag a_2
	+ \mleft(R + \lambda \mright) \tilde \sigma_z  + \mleft( \chi_1 a^\dag_1 a_1 + \chi_2 a^\dag_2 a_2 \mright)  \tilde \sigma_z + g_\mathrm{eff}^\mathrm{II} \mleft( a_1^\dag a_2 \tilde \sigma + \mathrm{H.c.} \mright).
\end{equation}
Here, $g_\mathrm{eff}^\mathrm{II}$ refers to the effective rate of the nonlinear process; it is given by  
\begin{equation}
	g_\mathrm{eff}^\mathrm{II} =g^2 \frac{(f - 1) \cos^3\theta \sin\theta + f \cos\theta \sin^3\theta} {R f (f - 1)} .
	\label{eq:geffII}
\end{equation}
In Eq.~(\ref{eq-effH-example2}), we have defined
\begin{equation}
	\Delta_1 = 2 f R, \quad \Delta_2 = 2(f-1) R,
	\label{N}
\end{equation}
for $f \in (0,1)$, such that the resonance conditions are automatically fulfilled.
Moreover, the Lamb shift of the qubit is
\begin{equation}
	\lambda = \frac{g^2}{2} \frac{\cos^4\theta
	(4R+\Delta_1+\Delta_2)}{(2R+\Delta_1)(2R+\Delta_2)}
	+\frac{g^2}{2}\frac{\sin^4\theta(4R-\Delta_1-\Delta_2)}{(2R-\Delta_1)(2R-\Delta_2)}
\end{equation}
and the dispersive coupling rates are 
\begin{equation}
	\chi_i = g^2 \mleft(\frac{\cos^4\theta}{2R+\Delta_i}+\frac{\sin^4\theta}{2R-\Delta_i} \mright), \quad i\in\{1,2\}.
\end{equation}

\paragraph{Exact resonance conditions}

Following the same discussion as in example I, and setting 
\begin{equation}
\Delta_1 = 2fR + \delta=2R+\Delta_{2}+\delta,
\end{equation} 
one can see that the $\delta$ for the exact resonance condition has the same form as given in \eqref{eq:delta_resonance_condition_caseI}. The lowest energy levels and the avoided level crossing arising under the exact resonance condition are plotted in Figs.~\figpanelNoPrefix{fig:eigenvalues-one-drive1}{c} and \figpanelNoPrefix{fig:eigenvalues-one-drive1}{d}, respectively.


\subsection{Example III: Two atoms simultaneously excited by a single photon}

Here we discuss the nonlinear effect originally described in the USC regime in Ref.~\cite{Garziano2016} and recently experimentally realized in Ref.~\cite{Tomonaga2023}; that is, two atoms can be simultaneously excited by a single photon, and vice versa. 

\paragraph{System}

The setup corresponds to system 2 in Fig.~\ref{fig:sketch-one-drive}: two atoms are coupled to a single cavity mode. The states of the bare basis, working in the dressed-qubit picture, are again $\{ \ket{\pm,\pm,n} \}$.
	
\paragraph{Nonlinear process}

We consider the nonlinear process where a single photon is simultaneously absorbed by two atoms. The initial and final states are, thus, given by:
\begin{equation}
	\ket{i} = \ket{-,-,n+1}; \quad \ket{f} = \ket{+,+,n}.
\end{equation}

\paragraph{Intermediate states}

For any choice of $\ket{i}$ and $\ket{f}$---i.e., any value of $n$---the subspace of intermediate states involved in the process (to lowest order) consists of the following  12 states:
\begin{multline}
	\{ \ket{+,-,n+1}, \ket{-,+,n+1}, \ket{+,+,n+1}, \ket{+,-,n}, \ket{-,+,n}, \ket{-,-,n}, \\
	\ket{+,-,n+2}, \ket{-,+,n+2}, \ket{-,-,n+2}, \ket{-,+,n-1}, \ket{+,-,n-1}, \ket{+,+,n-1} \}.
\end{multline}
These states mediate the interaction between $\ket{i}$ and $\ket{f}$, up to third order, as shown in \figpanel{fig:eigenvalues-one-drive2}{c}. Note that there is no second-order process coupling $\ket{i}$ and $\ket{f}$. 
	
\paragraph{System Hamiltonian}

The Hamiltonian, in the rotating frame of the driving, reads
\begin{equation}
	H = \Delta_a a^\dag a + \Delta_\sigma \mleft( \sigma_1^\dag \sigma_1 + \sigma_2^\dag \sigma_2 \mright) + \Omega \mleft( \sigma_1 + \sigma_2 + \mathrm{H.c.} \mright) + g \mleft[ a \mleft( \sigma_1^\dag + \sigma_2^\dag \mright) + \mathrm{H.c.} \mright] \, ,
\end{equation}
where $a$ is the bosonic annihilation operator of the cavity and
$\Delta_a=\omega_a-\omega_\mathrm L$
($\Delta_\sigma=\omega_\sigma-\omega_\mathrm L$)
is the cavity (qubit) detuning from the drive frequency. In the dressed qubit basis, it reads:
\begin{multline}
	H = \Delta_a a^\dag a + R \mleft( \tilde\sigma_1^\dag \tilde\sigma_1 + \tilde\sigma_2^\dag \tilde\sigma_2 \mright) \\
	+ g \mleft\{ a \mleft[\sin^2\theta \mleft(\tilde\sigma_1 + \tilde \sigma_2\mright)-\cos^2\theta \mleft(\tilde\sigma^\dagger_1 + \tilde \sigma^\dagger_2\mright) +\sin\theta\cos\theta \mleft(\tilde \sigma_{1,z} + \tilde\sigma_{2,z}\mright)\mright] +
	\mathrm{H.c.} \mright\} \,.
\end{multline}
	
\paragraph{Approximate resonance condition}

The nonlinear process $\ket{-,-,n+1} \leftrightarrow \ket{+,+,n}$ is enabled when the following resonance condition is satisfied:
\begin{equation}
	\Delta_a \approx 4R.
    \label{eq:resonance-condition3}
\end{equation}

\paragraph{Effective Hamiltonian}

The resulting effective Hamiltonian, using \eqref{eq:h_eff}, takes the form
\begin{equation}
	H_\mathrm{eff}^\mathrm{III} =  H_A + 
    \begin{pmatrix}
	- \chi_1 (n+1) - \chi_2 - \lambda + \alpha & g^\mathrm{III}_\mathrm{eff}\sqrt{n+1} \\
	g^\mathrm{III}_\mathrm{eff}\sqrt{n+1} & 2\chi n +2 \lambda + \alpha  
	\end{pmatrix},
\end{equation}
which, following the procedure outlined above, can be written as
\begin{equation}
	H_\mathrm{eff}^\mathrm{III} = \Delta_a a^\dag a + \sum_i (R + \lambda) \tilde \sigma_{z, i} + \chi a^\dag a \tilde \sigma_{z, i}
	+ g_\mathrm{eff}^\mathrm{III} \mleft( a^\dag \tilde\sigma_1 \tilde\sigma_2
	+ a \tilde\sigma_1^\dag \tilde\sigma_2^\dag \mright).
\end{equation}
Here, $g_\mathrm{eff}^\mathrm{III}$ is the effective rate of the nonlinear process; it is computed to be
\begin{equation}
	g_\mathrm{eff}^\mathrm{III} = \frac{g^3}{3 R^2}\mleft(\cos^3\theta \sin^3\theta + 3\cos\theta \sin^5\theta \mright).
\end{equation}
It is clear from the cubic dependence on $g$ that, in contrast to the two previous examples, this is a third-order process; the intermediate transitions are shown in \figpanel{fig:eigenvalues-one-drive2}{c}. On the other hand, both the Lamb shifts of the atoms and the dispersive coupling between the atoms and the cavity are dominated by the second-order processes and, thus, exhibit a quadratic dependence on $g$:
\begin{align}
		\lambda = \;& \frac{g^2}{2} \frac{(2 R - \Delta_a) \cos^4\theta + (2 R + \Delta_a) \sin^4\theta}{4 R^2 - \Delta_a^2}, \\
		\chi = \;& 2\lambda.
\end{align} 

\paragraph{Exact resonance conditions}

Following the same approach as in the previous examples, the fine-tuning of the resonance condition is done by defining a small energy detuning $\delta \ll R$, such that 
\begin{equation}
	\Delta_a = 4R + \delta.
\end{equation}
The detuning is then found to be
\begin{equation}
	\delta = 4(n+1)\chi.
\end{equation}
The lowest energy levels and the avoided level crossing arising under the exact resonance condition are plotted in Figs.~\figpanelNoPrefix{fig:eigenvalues-one-drive1}{e} and \figpanelNoPrefix{fig:eigenvalues-one-drive1}{f}, respectively.

\addcontentsline{toc}{section}{References}
\nocite{apsrev41Control}


%

\end{document}